\documentclass[letterpaper,11pt]{article}

\usepackage{sidecap}
\usepackage{graphicx}
\usepackage{slashed}
\usepackage{amsmath}
\usepackage{amssymb}
\usepackage{epsfig}
\usepackage{bm}
\usepackage{enumitem}

\usepackage{color}

\newcommand{\gsim}{{\;\raise0.3ex\hbox{$>$\kern-0.75em\raise-1.1ex\hbox{$\sim$}}\;}}
\newcommand{\beq}{\begin{equation}}
\newcommand{\eeq}{\end{equation}}
\newcommand{\bea}{\begin{eqnarray}}
\newcommand{\eea}{\end{eqnarray}}
\newcommand{\bit}{\begin{itemize}}
\newcommand{\eit}{\end{itemize}}

\usepackage{color}
\definecolor{darkgreen}{rgb}{0,0.5,0}
\definecolor{violet}{rgb}{0.5,0.5,1}

\def\beq{\begin{equation}}
\def\eeq{\end{equation}}
\def\bea{\begin{eqnarray}}
\def\eea{\end{eqnarray}}

\def\bit{\begin{itemize}}
\def\eit{\end{itemize}}

\def\l{\left}
\def\r{\right}

\def\ra{\rightarrow}

\def\baa{\begin{array}}
\def\eaa{\end{array}}

\def\d{\partial}

\def\simgt{\mathrel{\lower2.5pt\vbox{\lineskip=0pt\baselineskip=0pt
           \hbox{$>$}\hbox{$\sim$}}}}
\def\simlt{\mathrel{\lower2.5pt\vbox{\lineskip=0pt\baselineskip=0pt
           \hbox{$<$}\hbox{$\sim$}}}}

\def\ds{\displaystyle}

\begin{document} 

\baselineskip=14pt


\thispagestyle{empty}
\vspace{5pt}
\font\cmss=cmss10 \font\cmsss=cmss10 at 7pt

\begin{flushright}
\today \\
CERN-PH-TH-2014-238\\
UMD-PP-014-020 \\
\end{flushright}
\hfill

\begin{center}
{\LARGE \textbf
{Warped Dipole Completed, \\
\vspace{3pt}
with a Tower of Higgs Bosons
}} 
\end{center}

\vspace{8pt}

\begin{center}
{\large Kaustubh Agashe$\, ^{a}$, Aleksandr Azatov$\, ^{b}$, Yanou Cui$\, ^{a}$$\, ^{c}$, 
Lisa Randall$\, ^{d}$, Minho Son$\, ^{e}$} \\
\vspace{15pt}
$^{a}$\textit{Maryland Center for Fundamental Physics,
     Department of Physics,
     University of Maryland,
     College Park, MD 20742, U.~S.~A.} 
      \\     
     $^{b}$\textit{CERN, Theory Division, CH-1211 Geneva 23, Switzerland} 
      \\ 
      $^{c}$\textit{Perimeter Institute for Theoretical Physics, Waterloo, Ontario, Canada N2L 2Y5}
      \\
      $^{d}$\textit{Department of Physics, Harvard University, Cambridge, MA, 02138, U.~S.~A.} 
      \\ 
      $^{e}$\textit{
      Institut de Th\' eorie des Ph\' enom\` enes Physiques, EPFL, CH-1015 Lausanne, Switzerland} 

\end{center}

\vspace{5pt}

\begin{center}
\textbf{Abstract}
\end{center}
\vspace{5pt} {\small \noindent

In the context of warped extra-dimensional models which address both the Planck-weak- and flavor-hierarchies of the Standard Model (SM), it has been argued that certain observables can be calculated within the 5D effective field theory only with the Higgs field propagating in the bulk of the extra dimension, just like other SM fields. The related studies also suggested an interesting form of decoupling of the heavy Kaluza-Klein (KK) \emph{fermion} states in the warped 5D SM in the limit where the profile of the SM Higgs approaches the IR brane. We demonstrate that a similar phenomenon occurs when we include the mandatory \emph{KK excitations of the SM Higgs} in loop diagrams giving dipole operators for SM fermions, where the earlier work only considered the SM Higgs (zero mode). In particular, in the limit of a quasi IR-localized SM Higgs, the effect from summing over KK Higgs modes is unsuppressed (yet finite), in contrast to the naive expectation that KK Higgs modes decouple as their masses become large. In this case, a wide range of KK Higgs modes have quasi-degenerate masses and enhanced couplings to fermions relative to those of the SM Higgs, which contribute to the above remarkable result. In addition, we find that the total contribution from KK Higgs modes in general can be comparable to that from the SM Higgs alone. It is also interesting that KK Higgs couplings to KK fermions of the same chirality as the corresponding SM modes have an unsuppressed overall contribution, in contrast to the result from the earlier studies involving the SM Higgs. Our studies suggest that KK Higgs bosons are generally an indispensable part of the warped 5D SM, and their phenomenology such as signals at the LHC are worth 
further investigation.}

\vspace{6pt}
\begin{center}
 \let\thefootnote\relax\footnotetext{*Email addresses: kagashe@umd.edu,~Aleksandr.Azatov@cern.ch,~ycui@perimeterinstitute.ca,\newline~~~~~~randall@physics.harvard.edu,~minho.son@epfl.ch}
\end{center}
\vfill\eject
\noindent
\newpage
\enlargethispage{2\baselineskip}
\tableofcontents
\setcounter{page}{1}
\newpage


\section{Introduction}
\label{sec:intro}

The Randall-Sundrum (RS1) warped extra-dimensional framework \cite{Randall:1999ee}, coupled with a suitable radius stabilization mechanism (for example,
\cite{Goldberger:1999uk}), provides a solution to the Planck-weak hierarchy problem of the standard model (SM). It requires the Higgs field to be localized on the TeV/infrared (IR) brane
of the extra dimension. In the original model, it was assumed that the rest of the SM, i.e., gauge and fermion fields, were also
TeV-brane-localized. However, with SM gauge and fermion fields propagating in the extra dimension \cite{Goldberger:1999wh,
Grossman:1999ra, Gherghetta:2000qt}, it was soon realized that the same framework can also address the flavor hierarchy in the SM \cite{Grossman:1999ra, Gherghetta:2000qt, Huber:2000ie}. In this ``SM in the bulk" version of the warped extra-dimensional framework, there are contributions to various SM precision tests from massive Kaluza-Klein (KK) excitations of SM particles, which in the four-dimensional (4D) effective theory are essentially the manifestation of SM fields propagating in the extra dimension. However, custodial symmetries \cite{Agashe:2003zs} ameliorate the resulting constraints from electroweak precision tests (EWPT), such that a KK mass scale as low as $\sim 3$ (or a few) TeV might be allowed \cite{Carena:2006bn}. As far as consistency with flavor changing neutral currents (FCNC's) and CP-violating processes is concerned, in spite of a built-in analog of GIM mechanism of the SM \cite{Gherghetta:2000qt, Huber:2000ie, Agashe:2004cp}, a KK mass scale of at least $\sim 10$ TeV seems necessary \cite{Csaki:2008zd}. However, a few TeV mass scale might still be allowed if the model is supplemented by appropriate flavor symmetries (see~\cite{Barbieri:2012tu, Konig:2014iqa} for recent work in a ``simplified" version of the 5D model).

In this paper, we consider contributions to dipole operators of the SM fermions in this framework, which induce various radiative processes involving either the photon or the gluon. In turn, they arise from loops of KK particles and the resulting sizes give interesting constraints or signals for this framework. Some of the most stringent bounds on this framework from flavor/CP violation -- both in the lepton sector, e.g. $\mu \rightarrow e \gamma$ and in the quark sector, e.g. neutron electric dipole moment (EDM) --  originate from dipole operators. More specifically, for dipole operators leading to flavor- and/or CP-violation, it turns out that the dominant contribution comes from loops with Higgs boson modes and KK fermions. Henceforth, we focus solely on these effects. In passing, we would like to mention that loops with gauge and fermion KK modes tend to be ``aligned" (in generation space) with SM fermion Yukawa couplings/masses term. Hence, such effects do not contribute to the above types of processes, but are still relevant for $\left( g - 2 \right)_{ \mu }$, for example.

As already indicated above, detailed computations of dipole operators arising from the Higgs-KK fermion loops have been performed  
before. 
We contextualize our contribution here by first giving a brief recap of the literature as follows.
\begin{itemize}

\item
Naive dimensional analysis (NDA) estimates show that for a strictly brane-localized Higgs,
the dipole effect from a 5D cutoff  is actually comparable to the lowest KK mode's contribution
\cite{Agashe:2004cp, Agashe:2006iy}.
However, these references also showed that such UV sensitivity is suppressed 
for the alternative case of a Higgs field propagating in the extra dimension
\cite{Luty:2004ye}.
\end{itemize}
At first sight such a bulk Higgs might seem to be a radical departure from the localization on the TeV brane
as considered in the original models. However, the Planck-weak hierarchy problem is still solved as long as the profile of the Higgs VEV peaks near the TeV brane. By a mild tuning, it is also possible to obtain a physical mode of this 5D Higgs field which is much lighter than the typical KK scale and which has approximately the same profile as the Higgs VEV. This could then be identified with the SM-like Higgs boson of mass 126 GeV discovered at the LHC. The localization of the SM Higgs boson and the Higgs VEV is controlled by a 5D mass parameter in such a way that one can even take the TeV brane-localized limit. A bulk Higgs is thus a more general possibility than the brane-localized one, with the former encompassing the latter.

Explicit calculations of dipole operators have been performed for a bulk Higgs (even if eventually the brane-localized limit is taken). However, while the 5D Higgs field also manifests itself as KK excitations of the SM Higgs boson, the earlier work has considered only the SM Higgs in the loop, along with KK fermions. From a 5D viewpoint, the inclusion of the KK Higgs bosons is mandatory for consistency with 5D covariance. The main goal of this paper is to conduct a comprehensive study of the effects from KK Higgs bosons on dipole operator calculations.

Naively speaking, the KK Higgs contribution decouples in the brane-localized limit, since it turns out that its mass is roughly 
proportional to the 5D mass parameter, which becomes larger as the Higgs profile gets narrower. However, the previous dipole calculations (and some others involving fermion-Higgs couplings \cite{Azatov:2009na, Azatov:2010pf}) for such models involve further subtleties of significance beyond the NDA expectations, especially in the brane-localized limit, including the realization that the (very) heavy modes are still relevant in some cases. Therefore, there is a potential for similar ``surprises" in the KK Higgs calculation as well; indeed we will find that this is the case. In order to set the stage for our new analysis including the KK Higgs modes, it is then necessary to first give a more extensive summary of the various related results from earlier literature as follows (we do it roughly in time order).

\begin{itemize}
\item
NDA estimates for the contribution to these dipole operators from the SM Higgs boson-KK fermion loops first 
performed in \cite{Agashe:2004cp} gave
\begin{eqnarray}
{\cal L}_{ 4D } & \ni &   m_{ \rm SM } \frac{ e \left( y_{ \rm KK } \right) ^2}{ 16 \pi^2 M_{ \rm KK }^2 } \overline{ \psi_L } \sigma_{ \mu \nu } \psi_R F^{ \mu \nu }~, 
\label{NDA1}
\end{eqnarray}
where $\psi$ is a SM fermion, $F^{ \mu \nu }$ is the photon field strength, $M_{ \rm KK}$ is the typical lightest mass scale of the KK excitations and $y_{ \rm KK }$ is the Yukawa coupling of the SM Higgs boson to the two KK fermions.

\item The first actual calculation of these effects (at one-loop order) was done in~\cite{Agashe:2006iy} in 2006 (followed by essentially a similar one in 2009 \cite{Gedalia:2009ws}). It only considered the contribution from the lowest/fixed KK level. 

Note also that, for later use the 4D loop momentum cutoff was taken to be infinity from the start, since the loops can be shown to be 
convergent, corresponding to a higher dimensional operator in the 4D effective theory.
At this point, it is convenient to differentiate two kinds of couplings of the Higgs boson to the KK fermions (which are taken to be 
in the weak/gauge eigenstate basis, i.e., treating the Higgs VEV as an insertion). Namely, the Higgs boson can couple a left ($L$) chirality $SU(2)$ doublet fermion to a right ($R$) chirality singlet, which is the same assignment of chiralities as in the SM  and thus has been dubbed \emph{``correct" chirality} coupling in the literature on this subject; whereas, a separate coupling involves the opposite choice, i.e., \emph{``wrong'' chirality}. Only for the massive KK fermions do both types of couplings exist, since the KK modes are vector-like. References \cite{Agashe:2006iy, Gedalia:2009ws} then showed that:  

\bit
\item
The contribution of the correct chirality KK fermions has an extra suppression $\sim 
\left( m_{ h } / M_{ \rm KK } \right)^2$ relative to the above NDA estimate ($m_{h }$ here is the SM Higgs boson mass). We will neglect this contribution. However, note that no symmetry argument was found for this feature, so we expect other previously neglected contributions will not necessarily be similarly suppressed.

\item %

The contribution of the wrong chirality does not have such a factor, but it is instead suppressed (again, for fixed KK level) as we make the 
profile of the SM Higgs boson narrower, for the following reason. The profile of the wrong chirality KK fermion always vanishes exactly at the TeV brane, which is precisely the location of the SM Higgs boson in  the brane-localized limit  and thus the wrong chirality coupling is negligible in this case (while the correct chirality is not). Naively speaking, it seems then that the dipole operator vanishes for the brane-localized limit of the SM Higgs boson.

\item These references then focused instead on the case of a more spread-in-the-bulk Higgs boson, but which is still peaked near the TeV brane. 
For this case, a dipole operator of size similar to the above NDA estimate of Eq.~\ref{NDA1} was found. Again, this effect is from the wrong chirality, but the point is that this coupling is not small for such a profile of the SM Higgs boson.
\eit

\item \cite{Csaki:2010aj} in 2010 (and follow-up in 2012) studied these dipole operators from an alternate angle. First, they used strictly brane-localized (i.e., $\delta$-function-like) Higgs throughout (cf. such a limit obtained from a bulk Higgs discussed above). This implies that the wrong chirality coupling does not come into play at all. More relevantly, they took a 5D covariant approach, where one appropriately coordinates the 4D loop momentum cutoff with that of the KK sum (clearly,~\cite{Agashe:2006iy} in 2006 did not follow this prescription). Using the 5D propagators, but also sketching the equivalent KK picture, \cite{Csaki:2010aj} went on to show that 
\begin{itemize}

\item
the correct chirality of KK fermions does contribute a similar size to the above NDA estimate for such a brane-localized Higgs.

\end{itemize}

As we outline later on, this dipole effect is finite. Nonetheless we argue that it is UV-sensitive, since the KK modes up to the 5D cutoff seem to be relevant for it.

\item \cite{Delaunay:2012cz} in 2012 returned to the bulk Higgs calculation (again, only with the SM Higgs boson in the loop). This work established more firmly the ``need" for wrong chirality as advocated in~\cite{Agashe:2006iy} (but still no symmetry argument!). They took the brane-localized limit more carefully (for the wrong chirality effect), in particular, performing the KK sum (which was not done in~\cite{Agashe:2006iy}).

\begin{itemize}

\item
They showed that the summed-up effect of wrong chirality of the KK fermions is actually unsuppressed even in this 
brane-localized limit (unlike what was stated in \cite{Agashe:2006iy}), which can be dubbed a ``non-decoupling" effect.
\end{itemize}

Namely, an individual KK level contribution is suppressed by the Higgs profile's width in this limit (as in~\cite{Agashe:2006iy}), in turn, due to the dependence on the wrong chirality coupling. However, this coupling simultaneously grows with the KK level, which tends to compensate the expected suppression due to the increase of the KK fermion mass, in such a manner that the effect is of roughly similar size for a large range of KK levels, namely up to the mass comparable to the inverse of the Higgs profile width. Then the KK sum does indeed give a contribution of size of the above NDA estimate\footnote{Other instances of such (apparent) non-decoupling of the effects of heavy KK modes have been discussed previously: in the tree-level coupling of SM fermions to Higgs boson \cite{Azatov:2009na} and in gluon couplings to Higgs boson \cite{Azatov:2010pf}.}. We emphasize that the result in~\cite{Delaunay:2012cz} applies when Higgs width is at least as large as the (inverse of) the 5D cutoff, and that still leaves room for it to be (much) smaller than the width of a typical KK mode: the authors called it ``quasi IR-localized" (it has also been dubbed ``narrow bulk Higgs" \cite{Malm:2013jia}). Similar results have been obtained by~\cite{Beneke:2012ie} over the past two years, but using 5D propagators (for fermions only) instead. Of course, as the Higgs profile's width actually approaches the inverse of the 5D cutoff (which can be thought of as the ``width" of the TeV brane itself), the $O(1)$ factors in the above KK result are not quite reliable, since KK modes up to the 5D cutoff should be included. This finding is consistent with the UV sensitivity of a (strictly) brane-localized Higgs which was obtained simply using NDA estimates. 

\end{itemize}

\begin{table}
\begin{center}

\scalebox{1}{
\begin{tabular}{ | c || c | c | c || c | }
\hline 
Refs/year $\rightarrow$ & \cite{Agashe:2006iy} in 2006 & \cite{Csaki:2010aj} in 2010 & \cite{Delaunay:2012cz, Beneke:2012ie} in 2012 
& new in 2014 \\
Features $\downarrow$ & & & & (this paper) \tabularnewline
\hline
\hline
Higgs mode & SM & SM & SM & KK \tabularnewline
& & & & (thus 5D covariant) \tabularnewline
\hline
KK fermion & wrong &  correct & wrong & correct (and wrong) \tabularnewline
chirality & chirality & chirality & chirality & chirality \tabularnewline
\hline
Higgs profile & bulk & 

brane-localized & bulk  & bulk  \tabularnewline
considered & &  & (including narrow) & (including narrow)
\tabularnewline
\hline
KK modes & only 1$^{\rm st}$ mode & KK sum & KK sum & KK sum \tabularnewline
included & included &  &  &  \tabularnewline
\hline
Size vs.~NDA & 1 & 1 & 1  & 1 \tabularnewline 
 & & & (even for narrow) & (even for narrow
) \tabularnewline
\hline
\end{tabular}
} 
\caption{\label{table:review} A summary of various features (top to bottom) of dipole effects considered in the literature
before (middle three columns) vs.~our contribution (last column). ``correct" (``wrong'') in the row labelled ``KK fermion chirality'' refers to whether the assigned chirality is same as (opposite to) that of the SM fermions. 
}

\end{center}

\end{table}

\noindent We summarize these past works 
in Table \ref{table:review}. \\

In spite of all this body of work on dipole operators, we felt that a puzzle still remained: for a bulk Higgs (including its quasi IR-localized limit), why is the dominant contribution arising from the wrong chirality (again, this seemed to be an accident), as discussed in~\cite{Agashe:2006iy,Delaunay:2012cz,Beneke:2012ie} above? On the other hand, if we simply start with a $\delta$-function brane-localized Higgs boson, the correct chirality seems to be enough, as in~\cite{Csaki:2010aj}.
\\

In light of the dominant dipole effect in~\cite{Csaki:2010aj} coming from the correct chirality whereas all the other analyses get the dominant effect from the wrong chirality, we wanted to check if any effects have been omitted. Our main contribution is to include for the first time the effects from KK excitations of the SM Higgs boson. In addition to rendering the dipole result complete and 5D covariant, the details of the KK Higgs effect are also very interesting. We give a preview of our main findings as follows.

\begin{itemize}

\item
The KK Higgs contribution has a significant part coming from the correct chirality, in contrast to the SM Higgs boson effect which is dominated by the wrong one. In other words, the suppression factor for the correct chirality contribution with the light SM Higgs boson, namely $\sim m_{h }^2 / M_{ \rm KK }^2$, is clearly absent for the KK Higgs (it becomes $O(1)$ with mass $\sim M_{ \rm KK}$ instead).
\end{itemize}

Moreover, we show that:

\begin{itemize}
\item
The summed-up of the KK Higgs effect in general is parametrically comparable to the NDA estimate in Eq.~\ref{NDA2} (and hence to the wrong chirality one from the SM Higgs), although our numerical results show that it is accidentally somewhat smaller.
\end{itemize}

Most strikingly, we demonstrate that:

\begin{itemize}
\item
The summed-up of KK Higgs effect retains the above size even in the limit where the bulk Higgs profile becomes very narrow. This result looks counter-intuitive, since (as already mentioned earlier) the KK Higgs naively decouples in this case. Roughly speaking, this unexpected result arises as follows. The suppression of any individual KK level's contribution by the KK Higgs mass in the brane-localized limit is partially compensated by the Yukawa couplings of the KK Higgs being enhanced compared to those of the SM Higgs. Furthermore, there is a large range of KK Higgs modes with nearly degenerate masses, so that when we sum over the whole tower of KK Higgs bosons, the effect is unsuppressed.

\end{itemize}

Our work on the KK Higgs effect is also included in Table \ref{table:review}. Indeed, our finding of the apparent ``non-decoupling'' behavior of the summed-up KK Higgs effect bears some resemblance to earlier results mentioned above \cite{Azatov:2009na, Azatov:2010pf, Delaunay:2012cz}, but those involved the multiplicity of KK fermion modes only (vs.~focus on Higgs here). Also, given that our result derives from the KK fermions with the correct chirality and the KK Higgs (again, inclusion of the latter is required by 5D covariance), it is in spirit similar to (i.e., shares features with) the approach of~\cite{Csaki:2010aj}. However, we show in detail that it is still a different effect from that in~\cite{Csaki:2010aj}, since we can argue that the contribution in~\cite{Csaki:2010aj} is actually relevant only in the brane-localized limit, whereas the KK Higgs effect is important even for the bulk case. While our KK Higgs computation does not change the order of magnitude result for the dipole operator, it is important to include it for better precision. In this paper we do not re-compute the signals associated with specific processes. Independent of its practical implications, the KK Higgs effect is also of theoretical interest as mentioned above (namely, respecting 5D covariance; the particular chirality structure and the apparent ``non-decoupling'' feature) which is the main focus of the current work. \\

Here is the outline of the rest of the paper. We begin in Section~\ref{model} with a description of the model, mainly to explain our notation. In Section~\ref{sec:NDAI}, we present a ``cartoon" of the profiles, masses and couplings of the KK modes arising from the 5D model. Based on this picture, we then discuss the semi-analytic estimates for dipole operators in Section~\ref{sec:NDAII}. In Section~\ref{toy}, we present a ``simplified'' model which is amenable to a semi-analytic (actual) calculation, 
followed by detailed numerical results in the full 5D model in Section~\ref{numerical}. We conclude in Section~\ref{conclude}. More technical details and formulae are relegated to the appendices.

\section{The model}
\label{model}

The metric is given by
\begin{eqnarray}\label{eq:yzcoordinate}
\begin{split}
ds^2 & = \quad \frac{1}{ \left( k z \right)^2 } \l (\eta_{ \mu \nu } dx^{ \mu } dx^{ \nu } - d z^2 \r ) \hspace{0.2in} \hbox{equivalently, with} \; k z  = \exp \left( k y \right)~, \\
& = \quad  \exp \left( -2 k y \right) \eta_{ \mu \nu } dx^{ \mu } dx^{ \nu } - d y^2,
\end{split}
\end{eqnarray}
where $k$ is the AdS curvature scale and $z =R= 1 / k$ ($y = 0$) and $z =R'= \exp \left( k \pi r_c \right) / k$ ($y = \pi r_c$ ) correspond to the UV and IR branes (often called Planck and TeV branes), respectively. $r_c$ denotes the size of the extra-dimension. The KK masses are quantized in units of $k \exp \left( - k \pi r_c \right)$, denoted by $M_{ \rm KK }$ henceforth, which sets the mass scale of the first KK mode.

All the SM fields (including the Higgs boson) are assumed to propagate in the bulk. We neglect brane-localized kinetic terms for bulk fermions, gauge and Higgs fields. We also take the EW gauge symmetry to be simply $SU(2)_L \times U(1)_Y$ in the bulk, i.e., neither the custodial symmetric extensions \cite{Agashe:2003zs} nor the one which provides extra protection for the Higgs potential (aka Higgs boson being an $A_5$, i.e., part of 5D gauge field, dual to it being a pseudo-Nambu-Goldstone boson in the interpretation based on purely 4D strong dynamics) \cite{{Agashe:2004rs}}.
We make the above assumptions mainly for simplicity, but also because we do not expect our results for dipole operators to change significantly even in the presence of such extensions.

The Higgs field is described by 
\bea
 S_{\rm Higgs} = \int dz d^4x \left(\frac{R}{z}\right)^3
\left[ |{\cal{D}}_M H|^2 - \frac{\mu^2}{z^2} |H|^2  \right] -
V_{UV}(H)\delta(z-R) - V_{IR}(H)\delta (z-R')~,
\eea
where $V_{UV}$ ($V_{IR}$) corresponds to the potential localized at the UV (IR) brane. 
By a suitable (but not fine-tuned) choice of UV and IR brane potentials, this can give a Higgs VEV profile which is 
localized near the IR brane, as needed to solve the hierarchy problem:
\bea
h(z)\sim z^{2+\beta},
\label{v(z)}
\eea
where $\beta=\sqrt{4+\mu^2}$ (equivalent to the 5D Higgs mass parameter) controls the localization of the profile. The brane Higgs scenario can be recovered by an appropriate limit, $\beta\ra \infty.$ We then perform a KK decomposition of the 5D Higgs field. This KK decomposition gives the masses and profiles in the extra dimension of the various 4D physical Higgs boson modes. The full details of the KK decomposition for the bulk Higgs are given in Appendix~\ref{higgsRS}, with the qualitative features being discussed in the Section~\ref{higgs:qual}. Here we just mention that a mild tuning -- of order $\sim \left( v / M_{ \rm KK }\right)^2$ -- gives a mode which is lighter than the typical KK scale (often referred to as the zero-mode) and which is then identified with the observed SM Higgs boson. Its profile is approximately the same as that of the VEV in Eq.~\ref{v(z)} up to corrections of order $\sim \left( v / M_{ \rm KK } \right)^2$. In addition, there are KK Higgs modes with masses quantized in units of the typical KK scale: their profiles also peak near the IR brane, but with a degree of localization which can be very different from that of the zero-mode (more details will be shown later).

The bulk fermion fields are described by
\bea
S_{\rm Fermion} & = & \int d^4 x d z \l(\frac{R}{z}\r)^5 \Big [ \frac{i}{2}\l(\bar Q \Gamma^A {\cal D}_A Q-{\cal D}_A \bar Q \Gamma^A Q \r)+\frac{c_q}{R}\bar{Q} Q+(Q,\, c_q\Leftrightarrow U,\, c_u \; \hbox{and} \; D,\, c_d ) \nonumber \\
& & \hspace{3cm} + Y^{ u }_{5D} \bar Q H U + Y^{ d } _{5D} \bar Q H D \Big ]~,
\eea
where $Q$, $U/D$ are the five-dimensional Dirac fermions ($SU(2)_L$ doublet and singlets respectively) and their corresponding 5D masses are $c_q$, $c_{ u/d }$ (in units of the AdS curvature scale, $k = 1/R$). Here we focus on the quark sector for simplicity, but an analogous analysis applies for the lepton sector. The masses and profiles of the 4D modes are obtained via KK decomposition. The details of the KK decomposition for the bulk fermion are given in Appendix \ref{fermionRS}, with a sketch outlined in Section~\ref{fermion:qual}. We will discuss briefly only the zero-mode fermions arising from this compactification here, obtained by imposing the appropriate $Z_2$ boundary conditions. These modes are to be identified with the chiral $SU(2)$ doublet (singlet) SM fermions. The behavior of the fermion zero modes are very different from that of the heavy KK modes. In particular, the profile of the SM left chiral fermions is given by
\bea
\label{fc}
q_L^0(z) &=& f(c_q)\frac{{R'}^{-\frac{1}{2}+c_q}}{ R^{2}} z^{2-c_q}~,
\eea
where the $f(c)=\sqrt{\frac{1-2c}{1-(R/R')^{1-2c}}}$.  The profiles of SM right chiral fermions are obtained by the same equation with $c\ra -c$. Thus, similarly to the Higgs zero-mode/VEV, the localization of the SM fermion profile is controlled by the five-dimensional mass parameter $c$. However, the crucial point is that in the case $c_q< 0.5$ the profile of the zero-mode fermion is localized near the IR brane, whereas for $c_q > 0.5$ it is near the UV brane. In contrast, KK fermion modes (like all KK modes) are always localized near the IR brane.

Here we are setting the Higgs VEV to be zero when dealing with the fermion fields, and we treat the Higgs VEV as an insertion. The fermion zero mode and KK modes undergo mass mixing after EWSB which corresponds to a higher order correction to our results (it is suppressed by powers of $v / M_{ \rm KK }$)\footnote{One could have instead worked directly with the resulting mass eigenstates (i.e., included effect of EWSB from the beginning in the mode decomposition). This latter approach should of course be equivalent to the one that we actually use.} .

In the resulting 4D effective theory, these modes (zero and massive KK) are then used as part of loop diagrams in order to calculate the dipole operators. The KK mode contributions to dipole operators are dictated not just by their masses, but also by their couplings between the particles. These couplings depend on the overlap in the extra dimension of the profiles of the particles involved. These overlap integrals are done numerically: the exact formulae are not so enlightening and thus are given in Appendix~\ref{app:couplings}, with estimates being discussed in Section~\ref{couplingestimate}. As an alternative to the KK approach involving a sum over modes, one can compute the same dipole operator by a 5D approach (independent of whether Higgs VEV is treated as an insertion or not), using 5D propagators where the KK sum is implicitly done to begin with.

\section{Semi-analytic estimates I: profiles, masses and couplings}
\label{sec:NDAI}

Armed with the masses and couplings of KK modes (from the appendices), it is rather straightforward to perform the full calculation of the dipole operator in the 5D model. However, such a procedure tends to be mostly numerical and so we defer it to Section~\ref{numerical}. In the intervening sections, we perform an approximate, semi-analytic study, which will be more insightful and indicate to us what results to expect from the full analysis. We begin with making naive dimensional analysis (NDA)-type estimates for the all parts of the dipole operator calculation involving both the 4D loop and the genuine 5D effects, such as couplings, masses of KK modes and their KK sum. In this section, we outline a cartoon of the profiles, their couplings, and masses of the KK particles. This is a rough sketch of the exact results given in the Appendices (or in~\cite{Davoudiasl:2000wi, Gherghetta:2000qt, Agashe:2004cp, Gherghetta:2006ha}, for example). In the next section, we will use these couplings and masses in order to provide semi-analytic estimates for the relevant effects on dipole operators. Such estimates, based on an NDA approach, although not accurate, provide the quickest understanding of and intuition for the full results.

We now start the process of NDA estimates of profiles and masses of KK modes. We will use the $y$ coordinate for this purpose, although it is simple to switch to the $z$ coordinate instead (see Eq.~\ref{eq:yzcoordinate}). Regarding the mass scales used in our NDA estimates: for $O(1)$ bulk masses, we will approximate the lightest KK mass by the standard unit
\beq
M_{ \rm KK }\equiv \frac{1}{R'}=k \exp \left( - k \pi r_c \right).\label{eq:mKK}
\eeq
The actual lightest KK mass is typically an O(1) factor different from $M_{ \rm KK }$, mostly depending on its spin and bulk mass. For example, the mass of the lightest gauge KK mode is actually $\approx 2.45 \; M_{ \rm KK }$. We will neglect such factors in this section.

As far as all the profiles (whether zero or KK modes) are concerned, we choose them to include the warp factor in such a way that the overlap integrals (relevant for computation of the normalization and couplings) do not have the explicit warp factor dependences (i.e., \`a la flat extra dimension). The profiles are normalized to 1, with above convention for the warp factors,
\begin{eqnarray}
\int^{ \pi r_c }_0 d y \left( \hbox{profile} \right)^2 & = & 1~.
\end{eqnarray}
%

%

\subsection{Fermion field profiles and masses}
\label{fermion:qual}

These depend on the 5D fermion mass $c$ (in units of $k$). In our estimates, we assume $c \sim O(1)$. The exact formulae are in Appendix~\ref{fermionRS}.

\subsubsection{Zero-mode: SM fermion}

The profile of the zero-mode (which is massless before EWSB) is very sensitive (exponentially for a certain range) to the $c$ parameter (see Eq.~\ref{fc}). Small variations in $c$ can result in localization either near the Planck brane, which is suitable for 1st- and 2nd-generation fermions with small Yukawa couplings to the SM Higgs boson, or near the TeV brane as for the top quark. For simplicity in our estimates we consider a (quasi-)flat profile for the zero-mode (strictly flat corresponds to $c = 1/2$) for both chiralities.

%
The zero-mode fermion profile is explicitly
given by
\begin{eqnarray}
f_{ \rm SM } ( y ) & \simeq& \ds\frac{1}{ \sqrt{ \pi r_c } } \quad \hbox{for} \; 0 \leq y \leq \pi r_c~.
\label{zerofermionprofile}
\end{eqnarray}
In the following, we will use the notation $f$ in profile and mass to denote a fermion.

\subsubsection{KK fermion modes}

The masses and profiles of KK fermions are not sensitive to $c$ in contrast with the zero mode. We neglect $c$ dependence, assuming $c \sim O(1)$.
The masses are quantized in units of $\sim M_{KK}$ and they are approximately given by ($n$ being the mode-number): 
\begin{eqnarray}
m^f_{ \rm KK } & \simeq& n \; M_{ \rm KK }~.
\end{eqnarray}
The profile is localized within $\sim 1 / k $ away from the TeV brane and the $n^{ \rm th } \left( \gg 1 \right)$ mode has $\sim n$ oscillations (roughly uniformly spread) inside this width. Here and henceforth, `width' refers to that of the profile in the extra dimension (not to be confused with decay width).
As a rough approximation, we simply take it to be
\begin{eqnarray}
f^{ \pm }_{ \rm KK } ( y ) & \sim &
\left\{ 
\begin{array} {l}
\sqrt{k} \hspace{0.2in} \hbox{for} \; \pi r_c -  O \left( \ds\frac{1}{k} \right) \leq y \leq \pi r_c, \; \hbox{with} \; \sim n \; \hbox{nodes} \\
\hspace{0.14in} 0 \hspace{0.2in} \hbox{for} \; 0 \leq y \leq \pi r_c -  O \left( \ds\frac{1}{k} \right)
\end{array}
\right.
\end{eqnarray}
where (and henceforth) we use alternatively $\pm$ symbols for notational simplicity in equations. ``$+$"  and ``$-$" denotes correct or wrong chiralities of KK fermions, namely, $SU(2)$ doublet $L$ (plus singlet $R$) and doublet $R$ (plus singlet $L$), respectively.

However, the wrong chirality profile vanishes exactly at the TeV brane, i.e., behaving like 
$  \propto
%
%
\sin \left\{ n \Big[  1 - e^{ k \left( y - \pi r_c \right) } \Big]  \right\}$
close to it (see discussion around Eq.~(C11) in appendix of~\cite{Azatov:2009na}).
%
%
Within the width of the SM Higgs, given by $\sim 1 / \left( \beta k \right)$ (which is relevant for couplings: see details below), we have (assuming $\beta \gg 1$)
\begin{eqnarray}\label{wrongprofile1}
f^{ - } _{ \rm KK } ( y ) & \sim & \sqrt{k}\; \frac{n}{ \beta } \hspace{0.2in} \hbox{for} \; \pi r_c -  O \left( \frac{1}{ \beta k} 
\right) \leq y \leq \pi r_c~.
\end{eqnarray}
Eq.~\ref{wrongprofile1} implies that the profile is suppressed and not oscillating for the case of (large) fermion mode-number $n$ while being still $\lesssim\beta$ (this is not the case for $n \gg \beta$).


\subsection{Higgs field profiles and masses}
\label{higgs:qual}

Just like for the fermion case above, these depend on the 5D mass of the Higgs field, in units of $k$ 
(denoted by $\beta$). Note that in the literature $\beta$ often denotes
$\sqrt{ 4 + \left( \hbox{5D Higgs mass} / k \right)^2 }$. 
However, we will be especially interested in the $\beta \gg 1$ case,
in which case the two definitions are equivalent and so henceforth we will neglect this difference. The exact formulae are in Appendix~\ref{higgsRS}.

\subsubsection{Zero mode: SM Higgs}

By a suitable choice of parameters such as $\beta$ and the TeV brane-localized Higgs potential, one can obtain a mode which is much lighter than $M_{ \rm KK }$ and which will be identified with the SM Higgs boson with the usual VEV. This mode will aquire a mixing with the massive modes after EWSB which is typically small as is discussed in Appendix~\ref{higgsRS}. Its profile (both for the VEV and the physical Higgs boson within our insertion approximation) is monotonic and peaked near the TeV brane, that too localized within $\sim 1 / \left( \beta k \right)$ of it (see Eq.~\ref{v(z)}). It can be approximately given by
\begin{eqnarray}
\phi^{ \rm light } ( y ) & \sim & 
\left\{ 
\begin{array} {l} 
\sqrt{ \beta k } \hspace{0.2in} \hbox{for} \; \pi r_c -  O \left( \ds\frac{1}{ \beta k} \right) \leq y \leq \pi r_c \\
\hspace{0.15in} 0 \hspace{0.28in} \hbox{for} \; 0 \leq y \leq \pi r_c -  O \left( \ds\frac{1}{ \beta k} \right)
\end{array}
\right.
\label{zeroHiggsprofile}
\end{eqnarray}
We will use $\phi$ to denote Higgs mode in general. Based on the above profile, we see that 
\begin{itemize}
\item
a quasi-localized \cite{Delaunay:2012cz} (or narrow ~\cite{Malm:2013jia}) bulk Higgs corresponds to the choice $\beta \gg 1$,
but still $\beta \lesssim\Lambda / k$, where $\Lambda$ is the cutoff of the 5D non-
gauge/Yukawa theory. 
\end{itemize}
(Henceforth, we will call it the `narrow limit' of a bulk Higgs.)
A reason for not taking even larger $\beta$, corresponding to a width smaller than $\sim 1 /  \Lambda$, is that inclusion of higher-dimensional operators will effectively give a width to even a (supposedly) brane-localized Higgs of $\sim  1 / \Lambda$ (see discussion around Eq.~C13 in appendix of \cite{Azatov:2009na}). In any case, a 5D mass for Higgs field larger than cutoff might not make sense to begin with.

We can then define 
\begin{itemize}

\item
the brane-localized Higgs limit of the bulk Higgs to be $\beta \rightarrow \Lambda / k $ (cf. $\delta$-function localization would correspond to $\beta \rightarrow \infty$).

\end{itemize}

\subsubsection{KK Higgs modes}

The masses are quantized in units of $\sim M_{KK}$. Unlike the case of the KK fermion, the 1st mode is much heavier than the typical KK scale
in the limit $\beta \gg 1$, namely for a narrow bulk Higgs:
%
%
\begin{eqnarray}
\label{KKHiggsmassapprox}
m^{\phi }_{ \rm KK } & \sim & \left( \beta + n \right) M_{ \rm KK }~.
\end{eqnarray}
We would like to highlight the above degeneracy of the KK Higgs modes up to $\sim\beta^{\rm th}$ mode which implies that masses of the first $\sim \beta$ number of modes ($n = 1$ to $n \sim \beta$) are $\sim \beta\; M_{ \rm KK }$. This degeneracy is one crucial property that leads to our new result. 

The profiles roughly look like
\begin{eqnarray}
\phi^{ \rm heavy } ( y ) & \sim & 
\left\{ 
\begin{array} {l} 
\sqrt{k} \hspace{0.2in} \hbox{for} \; \pi r_c -  O \left( \ds\frac{1}{ k} \right) \leq y \leq \pi r_c, \; \hbox{with} \; \sim n \; \hbox{nodes} \\
\hspace{0.14in} 0 \hspace{0.2in} \hbox{for} \; 0 \leq y \leq \pi r_c -  O \left( \ds\frac{1}{ k} \right)
\end{array}
\right.
\label{KKHiggsprofile}
\end{eqnarray}

Note that the width of the KK Higgs (and the number of nodes - roughly uniformly spread - within it) is
similar to that of KK fermions. In particular, the KK Higgs width is much larger than that of the SM Higgs for the case $\beta \gg 1$.

\subsection{ Couplings of various fermion- and Higgs- modes}
\label{couplingestimate}

Loop contributions of KK excitations depend on their masses and couplings, the latter being determined by their profiles. Based on the above choice of the inclusion of the warp factor in profiles (and taking into account that they are already normalized), it is clear that the coupling of two fermions and one Higgs modes is given by 
\begin{eqnarray}
\label{mastercoupling}
y^{ \rm  \phi }_{ f \; f^{ \prime} } &= \ds\int^{ \pi r_c }_0 d y \left( Y \sqrt{k} \right) f ( y ) f^{ \prime } (y ) \phi ( y )~,  
\end{eqnarray}
where $Y \sqrt{k}$ is the 5D Yukawa coupling (recall it has mass dimension $-1/2$ which implies that $Y$ here is dimensionless). Throughout this paper, we will use the superscript in the Yukawa coupling $y$ to indicate the Higgs mode, and two subscripts in $y$ to indicate fermion modes. The index $\phi$ can be either light or heavy which refers to the SM or KK Higgs. Similarly, $f$ and $f^{ \prime }$ can refer to either the SM or KK fermion.

We emphasize here that relations between couplings are crucial for estimating the final result, especially for doing the KK sum and, in this process, for understanding the dependence of the result on the Higgs boson width, which is set by the 5D mass of the Higgs field, $\beta$. In particular, as already mentioned,  we would like to study the brane-localized limit ($\beta \gg 1$). So, we prefer not to leave these couplings as free parameters in this section, i.e., we insist on estimating their sizes (even if crudely).\\

\noindent 
Our NDA estimates for couplings between various modes that will be discussed in detail in subsequent sections are summarized in Table~\ref{table:couplings} for the convenience. The exact formulae for couplings and the wave functions, corresponding to our NDA estimate in the approximation are given in Appendices~\ref{higgsRS}-\ref{app:couplings}. 

\subsubsection{SM Yukawa coupling and SM fermion mass}
\label{SMYukawa}

Plugging in the relevant profiles, Eqs.~\ref{zeroHiggsprofile} and \ref{zerofermionprofile}, 
into Eq.~\ref{mastercoupling}, it is straightforward to see that 
\begin{eqnarray}
\label{SMYukapprox}
\begin{split}
y^{ \rm light }_{ \rm SM \;  \rm SM } & \simeq \quad \frac{Y}{ \sqrt{ \beta } } \frac{1}{ k \pi r_c }~,\\
& \equiv \quad y_{ \rm SM }~,
\end{split}
\end{eqnarray}
where $y_{ \rm SM }$ denotes the SM Yukawa coupling (see Eq.~\ref{SMYukexact} for the exact formula being valid for any $c$ parameter). SM fermion mass is approximately, up to mixing of zero and KK fermion modes after EWSB, given by
\begin{eqnarray}
m_{ \rm SM } & \approx & y_{ \rm SM } v~.
\end{eqnarray}

Note that for fixed SM fermion profiles (whether flat or not), this Yukawa coupling decreases as we take the brane-localized Higgs limit ($\beta \gg 1$), if we also keep the 5D Yukawa coupling ($Y$) constant in this process. One has to keep the Yukawa coupling of zero-mode fermions at the SM value. To this end, one could compensate for this effect by either (i) localizing SM fermions closer to the TeV brane or (ii) rescaling the 5D Yukawa coupling appropriately (i.e., roughly by $\sqrt{ \beta }$). However, various precision tests would disfavor the former option and rescaling is the standard practice in the literature (starting with \cite{Agashe:2006iy}).\footnote{We will return to this issue when we present estimates for other couplings and when we show the results of the full 5D calculation (see an exact treatment of this issue at end of Appendix~\ref{app:couplings}).} Here, we instead keep explicit the factor of $\sqrt{ \beta }$ as above (instead of absorbing it in the 5D Yukawa coupling), just for clarity and -- more importantly -- for contrasting with the couplings of the KK Higgs (see below).

\subsubsection{SM-KK fermions to SM Higgs}
\label{SMKKYuklight}

Here, we focus on fermion mode-number, $n \lesssim\beta$, so that it does not oscillate (at least, not significantly) within the SM Higgs width. As we will argue later, higher fermion mode-numbers are not really relevant for estimates of dipole operators. Following a similar procedure to above, we then have:
\begin{eqnarray}
\label{SMKKYukapproxlight}
\begin{split}
y^{ \rm light }_{ \rm SM \; \rm KK } \; \left( \hbox{for} \; n \lesssim\beta \right) & \sim \quad \frac{Y}{ \sqrt{ \beta } } \frac{1}{ \sqrt{ k \pi r_c } }~, \\
& \sim \quad y_{ \rm SM } \sqrt{ k \pi r_c }~.
\end{split}
\end{eqnarray}
 In particular, 
KK-number conservation is badly violated for these fermion modes with $n \lesssim\beta$, i.e., the overlap of the corresponding profiles is not suppressed, since the two fermion profiles are roughly monotonic with in Higgs width. On the other hand, profiles of KK fermion modes with $n \gg \beta$ will oscillate within the Higgs width so that the corresponding overlap integral will be (highly) suppressed, resulting in a negligible coupling.

\subsubsection{KK-KK fermions to SM Higgs}
\label{KKYuklight}

The coupling of two KK fermions with correct chirality (denoted by ``$+$'') to SM Higgs is estimated to be
\begin{eqnarray}
\label{KKYukapproxcorrectlight}
\begin{split}
y^{ \rm light, + }_{ \rm KK \; \rm KK } 
\left( \hbox{for} \; n, \; p \lesssim\beta \right)  
& \sim \quad \frac{ Y }{ \sqrt{ \beta }  }~, \\
& \sim \quad y_{ \rm SM } \left( k \pi r_c \right)~.
\end{split}
\end{eqnarray}
For convenience of later estimates, we take the couplings in Eq.~\ref{KKYukapproxcorrectlight} as the standard unit for KK Yukawa coupling:
\beq
 y_{ \rm KK }\equiv y_{ \rm SM } \left( k \pi r_c \right)~.
\eeq
Just like the above coupling, for $n, p \lesssim\beta$, we do not have KK number conservation here, but we will recover it
for $n, \; p \gg \beta$, i.e., coupling will be significant, i.e., $\sim y_{ \rm KK }$, only for $n \sim p$ in this case, and finally, coupling will be negligible for $n \gg \beta$, but $p \lesssim \beta$ (or vice versa).

For wrong chiralities (denoted by ``$-$'') with mode numbers, $n, \; p \lesssim\beta$, we get (in particular, using Eq.~\ref{wrongprofile1})
\begin{eqnarray}
\label{KKYukapproxwronglight}
\begin{split}
y^{ \rm light, - }_{ KK \; KK } \left( \hbox{for} \; n, \; p \lesssim\beta \right) & \sim & \frac{ n \; p }{ \beta^2 } \frac{ \tilde{Y}}{ \sqrt{ \beta } }~, \\
& \sim & \frac{ n  \; p }{ \beta^2 } y_{ \rm KK }~.
\end{split}
\end{eqnarray}
The couplings will be similar to the correct chirality ones when mode-numbers exceed $\beta$.
In general, 5D covariance requires $Y = \tilde{Y}$, but we would like to keep them as separate parameters,
just as reminders of the chiralities involved.\\

The Yukawa coupling in Eq.~\ref{KKYukapproxwronglight} is suppressed (compared to the correct chirality) by the Higgs width in the brane-localized limit, i.e., $\beta \gg 1$ (as had already been anticipated in the introduction), but ``enhanced" by fermion mode-number. This feature is crucial for the non-decoupling effect of the wrong chirality fermions in the brane-localized limit (see below). Moreover, for fixed 5D Yukawa coupling, the KK Yukawa seems to decrease as we increase $\beta$. However, as discussed in Section~\ref{SMYukawa}, in practice, we need to rescale $Y$ by $\sqrt{ \beta }$ in order to keep the SM Yukawa coupling fixed (for fixed SM fermion profiles) as we take $\beta \gg 1$, in such a manner that the KK Yukawa also stays (roughly) constant in this limit.

\subsubsection{SM-KK fermions to KK Higgs}
\label{SMKKYukheavy}

Taking into account that the KK Higgs profile looks quite different from the SM one (compare 
Eqs.~\ref{zeroHiggsprofile} and~\ref{KKHiggsprofile}), we get 
for the coupling of the KK fermion to the KK Higgs boson
\begin{eqnarray}
\label{SMKKYukapproxheavy}
%
%
y^{ \rm heavy }_{ \rm SM \; \rm KK } & \sim & 
\left\{ 
\begin{array} {l} 
\ds\frac{ Y }{ \sqrt{ k \pi r_c } } \sim  \sqrt{ \beta } \left( y_{ \rm SM } \sqrt{ k \pi r_c } \right)  \quad \hbox{for similar mode-numbers} \vspace{-2.5mm} \\ 
\hspace{5.2cm} \hbox{of KK Higgs and fermion} \vspace{2mm} \\
\quad 0  \hspace{4.5cm} \hbox{otherwise}
\end{array}
\right.
%
%
%
\end{eqnarray}
where $\sqrt{\beta}$ is orginated by the width of the heavy Higgs being larger than the SM one, as was indicated in Eqs.~\ref{zeroHiggsprofile} and~\ref{KKHiggsprofile}.\\

The enhancement by $\sqrt{\beta}$ in the Yukawa coupling of the KK Higgs (which is more pronounced in the brane-localized limit of $\beta \gg 1$) relative to that of the SM Higgs is another crucial property leading to our new result.  Unlike for the case of the SM Higgs, there is (approximate) KK number conservation, since the KK profiles of fermion and Higgs must oscillate similarly within their widths (being $\sim 1 / k$ for both modes) in order for their overlap not to be suppressed.

\subsubsection{KK-KK fermions to KK Higgs}

For both wrong and correct chiralities of KK fermions, we get the following coupling to the heavy Higgs:
\begin{eqnarray}
\label{KKYukapproxheavy}
y^{ \rm heavy, \; \pm }_{ \rm KK KK } & \sim & 
\left\{ 
\begin{array} {l} 
Y \sim \ds\sqrt{ \beta } \; y_{ \rm KK } \quad \hbox{for KK-number conserving combination} \vspace{-2.5mm} \\
\\
 0 \hspace{2.5cm} \hbox{otherwise}
\end{array}
\right.
%
%
\end{eqnarray}
The point is that, even if the wrong chirality vanishes exactly at the TeV brane, it is obviously unsuppressed away from it (within its width of $\sim  1 / k$), where the heavy Higgs also lives, so that the wrong chirality coupling is similar in size to the correct chirality one. In particular, wrong chirality coupling of the KK Higgs is not suppressed as $\beta$ increases (i.e., the width of the SM Higgs decreases), unlike the similar coupling of the SM Higgs (see Eq.~\ref{KKYukapproxwronglight}). In fact, it is enhanced compared to the correct chirality coupling of the SM Higgs, $y_{ \rm KK }$ (of course, the correct chirality coupling of the heavy Higgs to two KK fermions also has a similar enhancement by $\sim\sqrt{\beta}$, compared to $y_{ \rm KK }$).

Finally, note that this coupling also features approximate KK number conservation (based on oscillating profiles,
just like the one above). It involves something like $n^{ \rm th }$ KK Higgs coupled to the $p^{ \rm th }$ and
$( p + n )^{\rm th }$ KK fermions.
For the exact formula, see Eq.~\ref{KKYukexact} by setting $n_H \geq 1$.

\subsubsection{SM-SM fermions to KK Higgs}

Although this coupling will not be used in our estimates since its effects are suppressed, we give it here for the sake of completeness:
\begin{eqnarray}
\label{SMYukapproxheavy}
\begin{split}
y^{ \rm heavy }_{ SM \; SM } & \sim \quad \frac{Y}{ k \pi r_c }\quad \hbox{only for 1st few KK Higgs}~, \\
& \sim \quad \sqrt{ \beta } \; y_{ \rm SM }~.
\end{split}
\end{eqnarray}

\begin{table}
\begin{center}
\scalebox{0.90}{
\begin{tabular}{|c||c|c|c|c|}
\hline 
Couplings of & & & &
\tabularnewline
fermion modes $\rightarrow$ & SM-SM & SM-KK  & KK-KK  (correct) & KK-KK (wrong)
\tabularnewline
to Higgs modes $\downarrow$ & & $\left( {\rm KK\ mass} \sim n\, M_{ \rm KK } \right)$ & &  
\tabularnewline
\hline
\hline
& & & &
\tabularnewline 
SM & $ \ds\frac{Y}{ \sqrt{ \beta } } \ds\frac{1}{ k \pi r_c } \equiv y_{ \rm SM } $ & $ y_{ \rm SM } \sqrt{ k \pi r_c } $ & $ y_{ \rm SM } \left( k \pi r_c \right)  \equiv y_{ \rm KK } $ &  $ \ds\frac{ n  \; p }{ \beta^2 } y_{ \rm KK } $
\tabularnewline
: KK \# not conserved & & & & (for $n, \; p \lesssim\beta$)
\tabularnewline
\hline
& & & &
\tabularnewline
KK  & $\sqrt{ \beta } \; y_{ \rm SM }$ & $ \sqrt{ \beta } \left( y_{ \rm SM } \sqrt{ k \pi r_c } \right) $  & $ \sqrt{ \beta } \; y_{ \rm KK } $ &  
$ \sqrt{ \beta } \; y_{ \rm KK } $
\tabularnewline
(mass\;  $\sim \left( \beta + n \right) M_{ \rm KK } )$ & & & & 
\tabularnewline
: KK \# conserved & & & & 
\tabularnewline
\hline 
\end{tabular}
}
\caption{\label{table:couplings} NDA estimates for couplings between various modes: fermion ones (assuming $n, \; p \lesssim\beta$) are indicated from left to right, whereas Higgs are top and bottom.  $\beta$ is the 5D mass of the Higgs field and sets the profile of the SM Higgs and the masses of KK Higgs modes. $Y$ is the dimensionless 5D Yukawa coupling in (appropriate) units of the AdS curvature scale. ``correct" and ``wrong" at the top of last two columns refer to the chirality of the KK fermion. It is convenient (as done here) to express other couplings in terms of those of the SM Higgs to two SM fermions ($y_{ \rm SM }$) or to 2 KK fermions of correct chirality ($y_{ \rm KK }$). For simplicity, the SM fermion profile is taken to be flat: in general, $1 / \sqrt{ k \pi r_c }$ factor should be replaced by the profile evaluated at the TeV brane.}
\end{center}
\end{table}

%
\section{Semi-analytic estimates II: coefficient of dipole operator}
\label{sec:NDAII}

We focus on the 4D Lagrangian for the chromomagnetic dipole operator
\begin{eqnarray}
{\cal L}_{ 4D } & \supset & m_{ \rm SM }  \frac{ g_{ \rm QCD } C_{\rm dipole} }{ 16 \pi^2 M_{ \rm KK }^2 } \overline{ \psi_L }T^a \sigma^{ \mu \nu } \psi_R G^{a}_{\mu \nu }~, 
\label{dipoledef}
\end{eqnarray}
and estimate the coefficient $C_{ \rm dipole }$ arising from KK fermion and Higgs boson modes in loops. We will only consider contributions which can be similar in size to the NDA estimate of Eq.~\ref{NDA1}. A similar estimate applies for photon field strength, with $g_{ \rm QCD } \rightarrow e$, and our analysis here can be easily applied to the electromagnetic dipole operator.
\\

Before proceeding with this section, we point out a few important properties that simplify our estimate. The NDA estimates show that dipole operators for a bulk Higgs are UV-insensitive whereas this is not the case for the brane limit, i.e., $\beta \sim$ 5D-cutoff. We expect that KK modes with mode-number well above $\beta$ will give a suppressed effect, i.e., decouple, which is partly due
to the heaviness of these modes. In addition, the structure of the couplings of the SM Higgs
to such KK fermion modes is different from that in the case of smaller fermion mode numbers. For example, KK number conservation is
recovered for these heavy KK fermions as discussed in Section~\ref{couplingestimate}. Hence, for simplicity, we restrict the sum over KK modes only up to mode-numbers $\sim \beta$ in our estimates. However, in our numerical computation, the KK sum is performed up to mode-numbers well above $\beta$.

We will only consider diagrams with KK fermion modes as internal lines (inside and outside the loop), since SM fermion modes will give contributions which are suppressed by the associated SM Yukawa couplings. As we will see, each contribution based on our NDA estimates here matches well with the corresponding exact loop-function, unless there is an accidental suppression or enhancement, which of course NDA cannot quite capture.

\subsection{SM Higgs in the Loop}

This part is mostly a review of earlier work, but it sets the stage for the newer results on the KK Higgs we present later on.

\begin{figure}
\begin{center}
\includegraphics[scale=0.75]{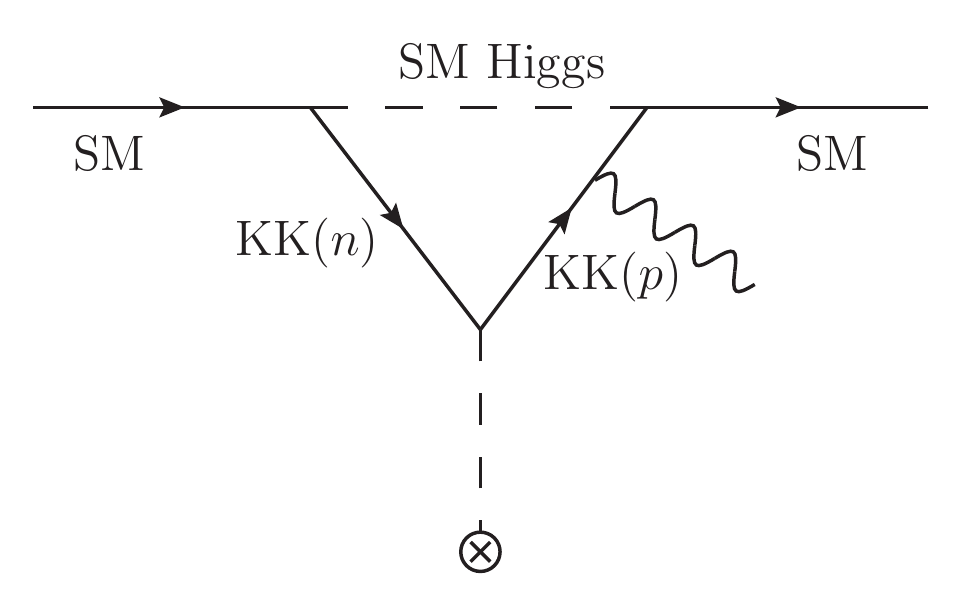}\quad
\includegraphics[scale=0.75]{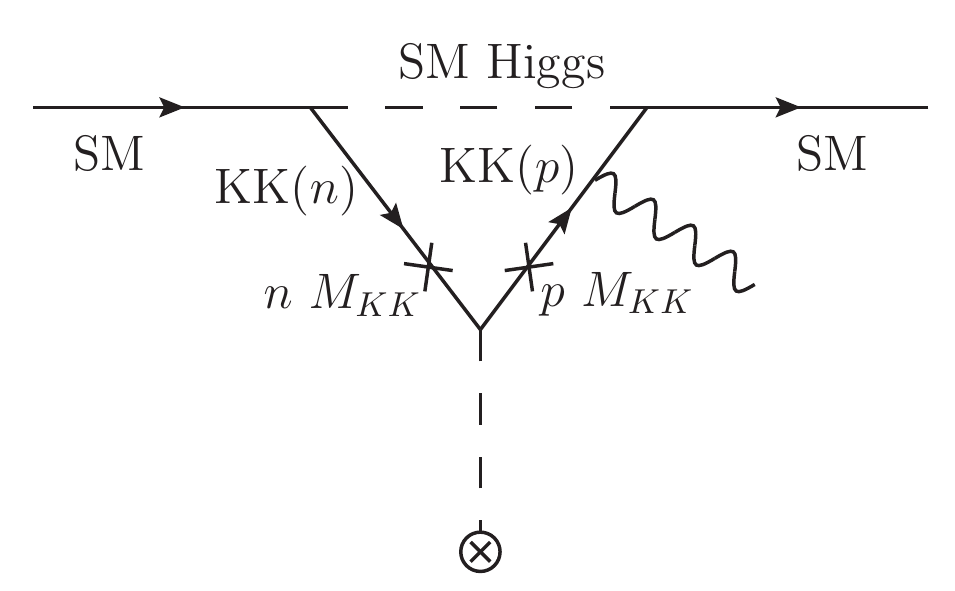}
\caption{Left: NDA estimate for SM Higgs-KK fermion loop. Right: Wrong chirality with SM Higgs and Higgs VEV insertion inside loop. The ``$\times$'''s on the KK fermion line denote chirality flips. For both diagrams, the ``$\otimes$" on the scalar line denotes a Higgs VEV insertion. There is a similar diagram with a gluon being emitted on the ``other" side of the Higgs VEV insertion.}
\label{fig:old_correct:old_wrong1}
\end{center}
\end{figure}

\subsubsection{Correct chirality}

\label{correctNDA}

The relevant diagram is shown on the left side of Fig.~\ref{fig:old_correct:old_wrong1}. We use the couplings given in the section above (and summarized in table \ref{table:couplings}) at the vertices, and masses given there in the propagators. For estimating the size of the loop momentum integral here (and henceforth), we basically invoke simple dimensional analysis or power counting, namely, the operator in Eq.~\ref{dipoledef} is dimension-6 and thus we should get $ 1 /  ( \hbox{mass} )^2$. Equivalently, we can consider powers of loop momenta in propagators and in the integration measure, including that we have to extract one power of (external) gluon momentum. Assuming that the masses of the two KK fermions are comparable, we expect the overall mass dimension for NDA estimates to be given roughly by adding the masses of all particles in the loop in quadrature\footnote{Of course, in the general case of a (large) hierarchy between the two KK fermion masses in this loop, there can be a logarithm of the ratio of these two KK masses: we neglect such a factor here, for simplicity and because as we will see soon, there is an accidental suppression factor of $m_h^2/m_{\rm KK}^2$ present.}. Note that here we are not considering chirality flips on KK fermion lines. This 4D loop diagram is convergent (Eq.~\ref{dipoledef} being a higher-dimensional operator) and so we take the 4D loop momentum cutoff to infinity to begin with (whether this procedure is consistent with 5D covariance is discussed later in Appendix~\ref{cornell}, and it does not affect the results here with finite $\beta$).

It is then rather straightforward to see that we expect 
(for fixed KK fermion mode-numbers, denoted by $n$ and $p$, as shown in figure):
\begin{eqnarray}\label{eq:NDA:light:np}
\begin{split}
\l ( C^{ \rm light, \; NDA}_{ \rm dipole } \r )_{(n, \; p)} & \sim \quad \frac{1}{ n^2 + p^2 } \frac{ \left( y^{ \rm light }_{ \rm SM \; KK } \right)^2 y^{ \rm light, + }_{ \rm KK \; KK } }
{ y_{ \rm SM } }~,\\
& \sim \quad \frac{ y_{ \rm KK }^2 }{ n^2 + p^2 }~, 
\end{split}
\end{eqnarray}
where $\left( n^2 + p^2 \right)$ in the denominator arises from the KK fermion masses (we neglect the SM Higgs mass). The loop factor of $16 \pi^2$ and overall KK mass scale has already been factored out in the definition of $C_{ \rm dipole }$ in Eq.~\ref{dipoledef}. $y_{ \rm SM }$ in the first line of Eq.~\ref{eq:NDA:light:np} is due to $m_{ \rm SM }$ out front in Eq~\ref{dipoledef}.
Finally, as mentioned in Sections~\ref{SMKKYuklight} and \ref{KKYuklight}, KK number is not conserved (even approximately) at the SM Higgs vertices, i.e., $n$ and $p$ are allowed to be large and quite different.

The KK sum then gives:
\begin{eqnarray}
\begin{split}
C^{ \rm light,\; NDA}_{ \rm dipole } & \sim \quad \sum^{ n \sim \beta }_{ n = 1 } 
 \sum^{ p \sim \beta }_{ p = 1 } \frac{ y_{ \rm KK }^2 }{  n^2 + p^2 }~,\\
& \sim \quad y_{ \rm KK }^2 \log \left( \beta \right)~.
\end{split}
\label{NDA2}
\end{eqnarray}
which is essentially the same as the estimate given in the introduction, i.e., Eq.~\ref{NDA1}, modulo the log-factor.
Naively one would expect that the above estimate is ``log-divergent" in the brane-localized limit, i.e., $\beta \rightarrow \Lambda / k$.

However, explicit calculations \cite{Agashe:2006iy, Gedalia:2009ws, Delaunay:2012cz} show that,
\begin{itemize}

\item
even though there is no symmetry argument for it,  the correct chirality effect is actually suppressed by a factor $\sim \left( m_{h} / M_{ \rm KK } \right)^2$ compared to the above estimate\footnote{This is (strictly speaking) valid for each KK level, but the KK sum does not change the result, since it is now $\propto \sim 1 / M_{ \rm KK }^4$ vs. $\sim 1 / M_{ \rm KK }^2$ in the NDA estimate above}.

\end{itemize}
There are actually two  diagrams here, namely, with a gluon attached to the left or the right of the Higgs VEV insertion: they are not shown separately in Fig.~\ref{fig:old_correct:old_wrong1} for simplicity (see instead Fig.~\ref{fig:insertioninside}). It turns out that each is separately suppressed (again, as far as we know, accidentally): see  Eqs.~\ref{Iaest} and \ref{eq:inside:correct} and discussion around them for the actual loop function. This suppression applies to the physical Higgs boson loop by itself (and similarly for the associated would-be Nambu-Goldstone bosons, equivalently the longitudinal $W/Z$). Also, as an aside, diagrams with a Higgs VEV insertion on an external leg (outside the loop) are also suppressed for the correct chirality case: see Section~\ref{toy} for more details.

We will return to the correct chirality contribution in Section~\ref{correctKKHiggs}, where we consider it instead for a KK Higgs in the loop and in Section~\ref{cornell}, where we go back to the SM Higgs, but being more careful about cutoff on 4D loop momentum (and KK sum).

\subsubsection{Wrong chirality}
\label{SM.in.out}

In this case, the diagrams with a Higgs VEV insertion inside and outside the loop both need to be considered, and they end up contributing similarly. We discuss these two contributions separately in order.

The relevant diagram with a Higgs VEV insertion inside the loop is seen in the right side of Fig.~\ref{fig:old_correct:old_wrong1}.
Using profiles and masses from above (but being careful with chiralities), it is easy to estimate this effect. As before, we start with fixed KK fermion modes (the superscript ``int" denotes a Higgs VEV insertion inside the loop):
\begin{eqnarray}\label{eq:NDAII:light:wrong:int}
\begin{split}
\l ( C^{ \rm light,\; wrong,\;  int}_{ \rm dipole } \r )_{(n, \; p)} \sim & \quad
 \frac{ n \; p }{ n^2 p^2} \frac{ \left( y^{ \rm light}_{ \rm SM \; KK } \right)^2 y^{ \rm light, - }_{ \rm KK \; KK } }{ y_{ \rm SM } }~,\\
 \sim & \quad \frac{ n^2 \; p^2 }{ n^2 p^2} \frac{ y_{ \rm KK }^2 }{ \beta^2 } \frac{ \tilde{Y} }{Y}~.
\end{split}
\end{eqnarray}
We must use the correct chirality for the external, zero-mode couplings, but the wrong chirality coupling for Higgs VEV insertion involving only KK fermions. This requires chirality flips on KK fermion propagators, giving factors of KK masses (i.e., mode-numbers, since $M_{ \rm KK }$ has been factored out) in the numerator of the first line in Eq.~\ref{eq:NDAII:light:wrong:int}. Simple power-counting then suggests four powers of KK mass in the denominator here; that this factor is $n^2 p^2$ in this case is based on the estimate that (in the general case of the two KK fermion masses being hierarchical) the largest contribution to the loop integral comes from a gluon attached to the lighter of these modes, as can be inferred from the diagram at the right side of Fig.~\ref{fig:old_correct:old_wrong1}. The above net estimate is confirmed by the exact loop function given in Eq.~\ref{eq:inside:wrong}. In the 2nd line in Eq.~\ref{eq:NDAII:light:wrong:int}, we have used the previous result that the wrong chirality coupling increases with KK fermion mode-number. 

Of course, the wrong chirality effect is naively (rather per KK level) still proportional to the Higgs width (due to the wrong chirality vanishing at the TeV brane), and is thus negligible in the narrow bulk Higgs limit ($\beta \gg 1$). However, we see that this contribution is roughly independent of (rather, not quite decoupling with) KK fermion mode-number, for instance, if we increase both $n$ and $p$. This feature comes from the growth of the wrong chirality coupling with mode-number compensating the KK mass suppression of the loop integral, cf. NDA estimate above (where heavier KK modes were indeed suppressed, as per the naive expectation).

Consequently, we find that 
\begin{itemize}

\item

the double sum over KK fermion modes compensates the above suppression due to the Higgs width

\end{itemize}
giving 
\begin{eqnarray}
\begin{split}
C^{ \rm light,\; wrong,\; int}_{ \rm dipole } & \sim \quad \frac{ y_{ \rm KK }^2 }{ \beta^2 } \frac{ \tilde{Y} }{ Y } \sum^{ n \sim \beta }_{ n = 1 } 
 \sum^{ p \sim \beta }_{ p = 1 }  1~, \\
& \sim \quad y_{ \rm KK }^2 \frac{ \tilde{Y} }{ Y }~,
\end{split}
\end{eqnarray}
which is indeed similar to the NDA estimate above (assuming $Y \sim \tilde{Y}$). This total contribution is roughly constant even as we take $\beta \gg1$ (due to the above mentioned non-decoupling feature). Recall that $y_{ \rm KK}$ remains roughly fixed in this process, contrary to the naive impression from the estimate in Eq.~\ref{KKYukapproxcorrectlight}, in turn, due to the rescaling of $Y$ that was mentioned earlier (see discussion in Section~\ref{KKYuklight}).

If we are sufficiently away from the narrow bulk Higgs limit, e.g., we consider $\beta \sim O(1)$, then the wrong chirality coupling is unsuppressed to begin with, which implies that even the 1st KK level contribution is sizeable. In other words, it is only in the narrow bulk Higgs limit that we stumble upon the apparent ``non-decoupling'' effect. However, in the brane-localized limit as $\beta \rightarrow$ 5D cutoff (in units of curvature scale), the result is UV-sensitive (even if finite, cf.~NDA estimate in Eq.~\ref{NDA2}), since KK modes up to the 5D cutoff give significant contribution.\\

\begin{figure}
\begin{center}
\includegraphics[scale=0.75]{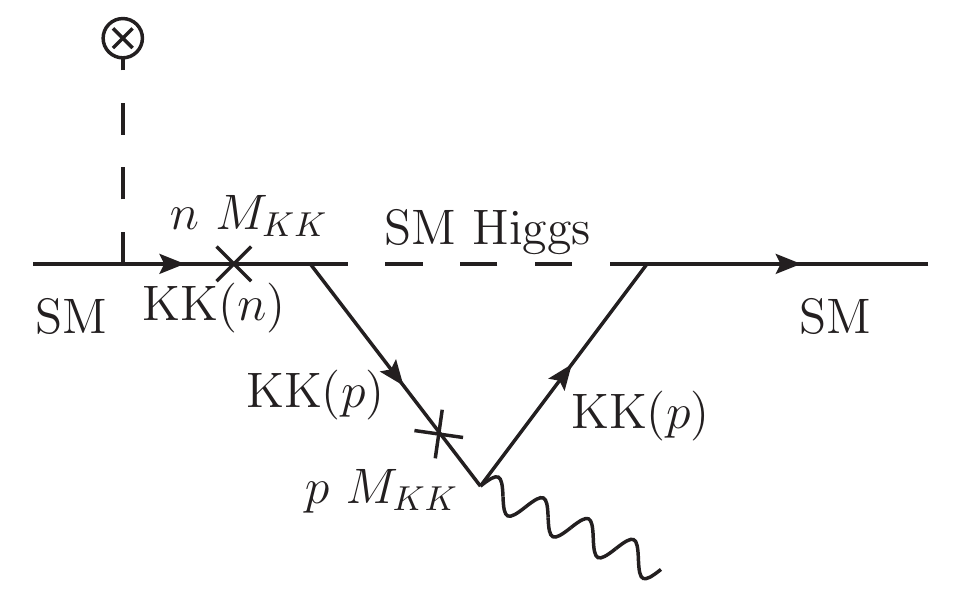}\quad
\caption{Wrong chirality with SM Higgs and Higgs VEV insertion outside the loop. The ``$\times$"'s on KK fermion line denote chirality flips.}
\label{fig:old_wrong2}
\end{center}
\end{figure}

The relevant diagram with a Higgs VEV insertion outside the loop is seen in Fig.~\ref{fig:old_wrong2}. In this case, the Higgs VEV insertion involves correct chirality, but one of the physical Higgs vertices (in the loop part of the diagram) comes with wrong chirality. However, it turns out to give the same combination of couplings as above, i.e., $( y^{ \rm light }_{ \rm SM \; KK } )^2 y^{ \rm light, -}_{ \rm KK \; KK } \sim y_{ \rm SM } \; y_{ \rm KK }^2 \left( n \; p \right) / \beta^2\ ( \tilde{Y}/Y )$. On the other hand, the dependence on KK fermion masses starts out looking different for fixed KK fermion modes,
\begin{eqnarray}\label{eq:NDAII:light:wrong:ext:np}
\l ( C^{ \rm light, \; wrong,\; ext}_{ \rm dipole } \r )_{(n, \; p)} & \sim & \frac{1}{ n } \frac{1}{ p } 
\left( \frac{ n \; p \; y_{ \rm KK }^2  }{ { \beta^2 } } \frac{ \tilde{Y} }{Y} \right)~,
\end{eqnarray}
where the superscript ``ext" denotes a Higgs VEV insertion outside the loop. In Eq.~\ref{eq:NDAII:light:wrong:ext:np}, we have already incorporated the estimate for the couplings which is same as for the case with a Higgs VEV insertion inside. Note that the 1st mass factor, $\frac{1}{n}$, in Eq.~\ref{eq:NDAII:light:wrong:ext:np} (again, $M_{ \rm KK }$ is already factored out in the definition of the dipole operator) comes from the external propagator (with chirality flip), while the 2nd one, $\frac{1}{p}$, is from the loop integral (where, as usual, we simply used dimensional analysis/power-counting). Once again, the exact loop function in Eq.~\ref{eq:outside} can be shown to match the above NDA estimate. However, the (double) KK sum gives similar estimate as for Higgs VEV insertion inside,
\begin{eqnarray}
\begin{split}
C^{ \rm light, \; wrong,\; ext}_{ \rm dipole } & \sim \quad \frac{ y_{ \rm KK }^2  }{ { \beta^2 } } 
\frac{ \tilde{Y} }{ Y } \sum^{ n \sim \beta }_{ n = 1 }  \sum^{ p \sim \beta }_{ p = 1 } 1~,\\
& \sim \quad y_{ \rm KK }^2 \frac{ \tilde{Y} }{ Y}~.
\end{split}
\end{eqnarray}

\subsection{KK Higgs in the Loop}
\label{KKhiggs}

This part leads to our new contribution, mainly driven by the different couplings and masses for the SM and KK Higgs. It follows a procedure similar to the above discussion. 

\begin{figure}
\begin{center}
\includegraphics[scale=0.75]{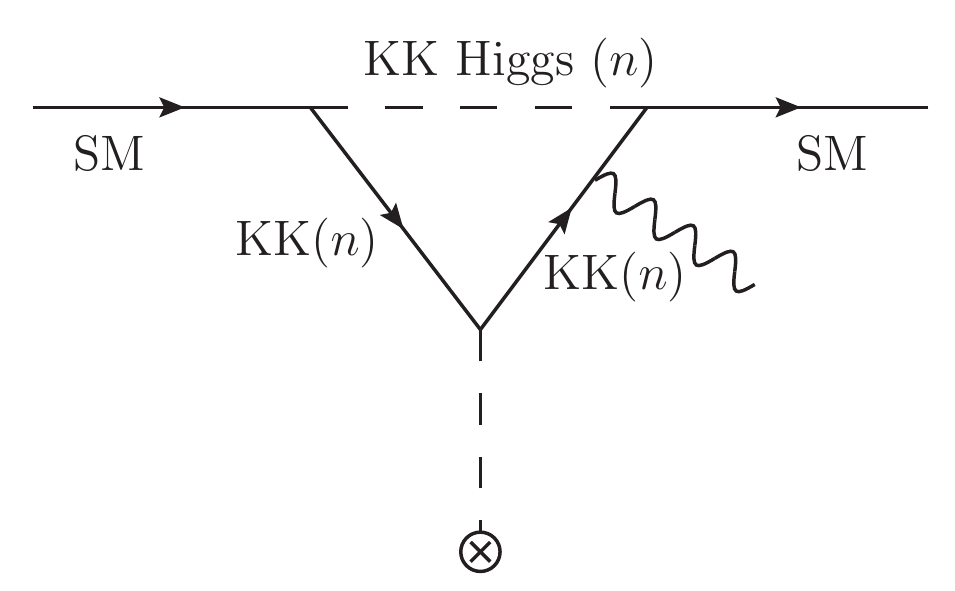}\quad
\includegraphics[scale=0.75]{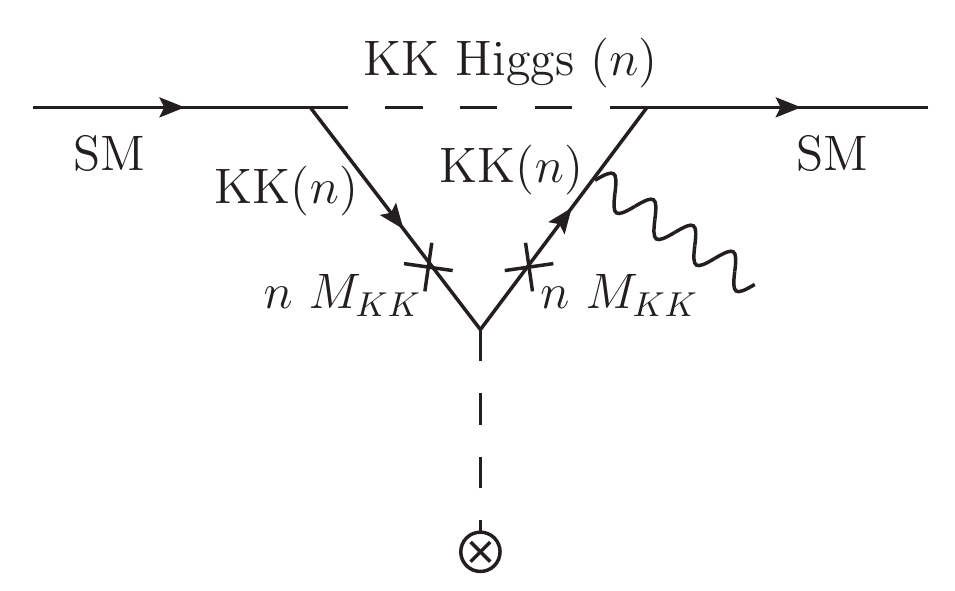}
\caption{Left: correct chirality with the KK Higgs. Right: wrong chirality with the KK Higgs and Higgs VEV insertion (denoted by ``$\otimes$'') inside loop. The ``$\times$''s on KK fermion line denote chirality flip.}
\label{fig:new_correctwrong}
\end{center}
\end{figure}

\subsubsection{Correct chirality}

\label{correctKKHiggs}

To begin with, we revisit the (purely) correct chirality diagram shown on the left of Fig.~\ref{fig:new_correctwrong}, but with a KK Higgs instead of a light Higgs. Based on the (accidental) suppression factor for the SM Higgs in the correct chirality contribution, as mentioned in Section~\ref{correctNDA}), we expect that there is no such suppression for KK Higgs modes, as $m_{h } \rightarrow (\hbox{at least}) \; M_{ \rm KK }$. Said another way, the suppression of the diagrams on the left side of Fig.~\ref{fig:old_correct:old_wrong1} was an artifact of neglecting the SM Higgs mass in the propagator (again, we are keeping track of only further suppressions here, i.e., beyond the 2 powers of KK mass from the loop integral). 

Furthermore, it is clear that the mode-number of the KK fermion has to roughly match that of the KK Higgs in this diagram to give an unsuppressed contribution. As already mentioned in Section~\ref{SMKKYukheavy}, this expectation is based on the profiles, in particular, their oscillations within the widths of their overlapping regions. In reality, a (small) range of KK fermion mode-numbers around the Higgs one contributes, but such an effect is within $O(1)$ here and so we simply equate the KK fermion and KK Higgs mode numbers for the NDA estimates we made. For the case of $n \ll \beta$, i.e., a (large) hierarchy between the KK Higgs and KK fermion masses, it is easy to estimate  that the contribution from loop momenta throughout this hierarchy gives the dominant effect, in the form of a logarithm factor of this hierarchy. 
This factor multiplies $\sim 1/  ( \beta + n )^2$ from the KK Higgs propagator inside the loop.
Whereas, loop momenta comparable to the KK Higgs mass  -- which are the only ones relevant for $n \sim \beta$ (i.e., KK fermion as heavy as KK Higgs)-- give a contribution with this $\log \rightarrow O(1)$. Combining these two cases, for fixed KK fermion and Higgs modes, $n \left( \lesssim\beta \right)$, we can then write
\begin{eqnarray}
\left ( C^{ \rm heavy, \; correct}_{ \rm dipole } \r )_{(n)} & \sim & 
\frac{1}{ \left( \beta + n \right)^2 } 
\left( \log \frac{ \beta }{n} + 1 
\right) 
\frac{ \left( y^{ \rm heavy }_{ \rm SM \; KK } \right)^2 y^{ \rm light, + }_{ \rm KK \; KK } }{ y_{ \rm SM } }~. 
\end{eqnarray}
This form of the estimate agrees with the exact loop function given in Eq.~\ref{eq:inside:correct}. Note that we get one factor of the SM Higgs Yukawa coupling due to a Higgs VEV insertion. It is clear here and similarly in the diagrams below that the mass scale suppression from the loop is dominated by that of the KK Higgs so that naively, the contribution is (highly) suppressed in the narrow bulk Higgs limit ($\beta \gg 1$), i.e., the KK Higgs decouples. However, we have the following two mitigating effects: as we saw in the previous section (see Eq.~\ref{SMKKYukapproxheavy}), 
\begin{itemize}

\item
the heavy Higgs coupling is larger than that of the SM Higgs, giving a partial compensation of the KK Higgs mass.
\end{itemize}
Thus, the above estimate is really
\begin{eqnarray}
\l ( C^{ \rm heavy, \; correct}_{ \rm dipole } \r )_{(n)} 
& \sim & \beta 
\left\{ \frac{ y_{ \rm KK }^2 }{ \left( \beta + n \right)^2 } \right\}
\left( \log \frac{ \beta }{n} + 1 
\right)~. 
\end{eqnarray}
Of course, naively, this is still vanishing in the narrow bulk Higgs limit (again, due to the heavy KK Higgs mass, in spite of its coupling being larger). However, we notice that the above contribution is (roughly) independent of KK mode-number, $n$ (up to $\sim \beta$), similar to the case of wrong chirality discussed above (and unlike the NDA estimate above). As a result, 
\begin{itemize}

\item

the KK fermion-Higgs (again, coordinated, i.e., not double) sum compensates the residual suppression due to the heaviness of the KK Higgs (in the brane-localized limit)

\end{itemize}
giving
\begin{eqnarray}
\begin{split}
C^{ \rm heavy,\; correct}_{ \rm dipole } & \sim \quad y_{ \rm KK }^2 \; \beta \sum^{ n \sim \beta }_{ n = 1 }  
\left( \log \frac{ \beta }{n} + 1 
\right) 
\left\{ \frac{ 1 }{ \left( \beta + n \right)^2 } \right\}~,
\\
& \sim \quad y_{ \rm KK }^2~.
\end{split}
\end{eqnarray}
Note that we do not get $\sim \log \beta$ in the end result after the KK sum, even though it was present at individual mode-level.

Note that each individual contribution is $1/\beta$-suppressed in the narrow Higgs limit. Adding up the log-independent contributions which are roughly comparable for $ \sim beta$ states gives a contribution which once again does not decouple with large $\beta$. The log contributions are different for the different states, so there is no log $\beta$ enhancement in the final answer. The KK Higgs degeneracy for the modes with $n \lesssim\beta$ is crucial in this argument for no suppression in the brane-localized limit. Recall that $y_{ \rm KK }$ is roughly held constant as we take $\beta \gg1$ by a rescaling of $Y$ (see discussion in Section~\ref{KKYuklight}). Such apparent ``non-decoupling'' of heavy KK modes is reminiscent of what was found for the wrong chirality effect above, but note that the particles which are more relevant are different, i.e., Higgs vs.~fermion, in the two cases and the couplings of the KK Higgs being enhanced compared to that of the SM Higgs played an equal role here. Once again, as $\beta \rightarrow$ 5D cutoff (in units of the curvature scale), we encounter UV sensitivity (even if there seems to be no divergence).

Just like for the wrong chirality effect, for a more spread-out Higgs, the KK Higgs (correct chirality) effect is clearly significant
even for the 1st KK level. For the sake of completeness, we mention that the diagrams with a Higgs VEV insertion outside of the loop (again, for correct chiirality) is suppressed for the KK Higgs just like for the SM Higgs case.

\subsubsection{Wrong chirality}
\label{wrong.in.out}
Finally, we consider wrong chirality couplings in diagrams involving the KK Higgs, again separating a Higgs VEV insertion inside and outside the loop. These are essentially the corresponding diagrams for the SM Higgs discussed above, but with the physical SM Higgs replaced by KK Higgs in the loop, while keeping the Higgs VEV insertion the same.\\

The corresponding Feynman diagram for a Higgs VEV insertion inside the loop is given on the right side of Fig.~\ref{fig:new_correctwrong}. As above, we use here approximate KK number conservation (at KK Higgs vertices); include factors from chirality flips ($\sim n$) in the numerator and the dimensional analysis/power-counting to obtain the denominator. We also consider the cases $ n \ll \beta$, i.e., KK Higgs much heavier than KK fermion vs.~$n \sim \beta$ (the two masses being comparable). The former loop integral is dominated by loop momenta comparable to the (much smaller) KK fermion mass, i.e., there is no logarithm here, unlike the case of correct chirality above, in such a way that the factors of $n$ from the chirality flip cancel against the same KK fermion masses from the loop integral. And, as before, the KK Higgs propagator simply gives $\sim 1 / ( \beta + n )^2$. Whereas in the $n \sim \beta$ case, loop momenta comparable to the KK Higgs mass are the relevant ones.

However, the chirality flip factors still (roughly) cancel the combination of KK Higgs and fermion masses from the loop integral, thus giving a similar estimate to the earlier one. Combining these two cases, it is straightforward to estimate this effect, starting with fixed KK modes:
\begin{eqnarray}
\l( C^{ \rm heavy, \; wrong, \; int}_{ \rm dipole }\r )_{(n)} & \sim & 
 \frac{ 1 }{ ( \beta + n )^2 } 
 \frac{ \left( y^{ \rm heavy }_{ \rm SM \; KK } \right)^2 y^{ \rm light, - }_{ \rm KK \; KK } }{ y_{ \rm SM } }~. 
\end{eqnarray}
(See Eq.~\ref{eq:inside:wrong} for the exact loop-function.) Next, we use the couplings estimated earlier: in particular, the wrong chirality SM Higgs coupling has a  suppression (compared to $y_{ \rm KK}$) for large $\beta$, but simultaneously an enhancement due to large mode-number, whereas there is a large $\beta$ enhancement for the correct chirality, KK Higgs coupling. So, the above estimate becomes
\begin{eqnarray}
\l ( C^{ \rm heavy, \; wrong, \; int}_{ \rm dipole }\r )_{(n)}
& \sim & \frac{ \tilde{Y} }{Y} \frac{ y_{ \rm KK }^2 }{ \beta } 
 \frac{ n^2 }{ ( \beta + n )^2 }~, 
 \end{eqnarray}
which up to the KK sum, gives an estimate similar to correct chirality one:
\begin{eqnarray}
\begin{split}
C^{ \rm heavy, \; wrong, \; int}_{ \rm dipole } & \sim \quad \frac{ \tilde{Y} }{Y} \frac{ y_{ \rm KK }^2 }{ \beta } 
\sum_{ n = 1 }^{ n \sim \beta } 
 \frac{ n^2 }{ ( \beta + n )^2 }~, 
 \\
& \sim  \quad  \frac{ \tilde{Y} }{Y } y_{ \rm KK }^2~.
\end{split}
\end{eqnarray}
in particular, it is unsuppressed even for $\beta \gg 1$.\\

\begin{figure}
\begin{center}
\includegraphics[scale=0.75]{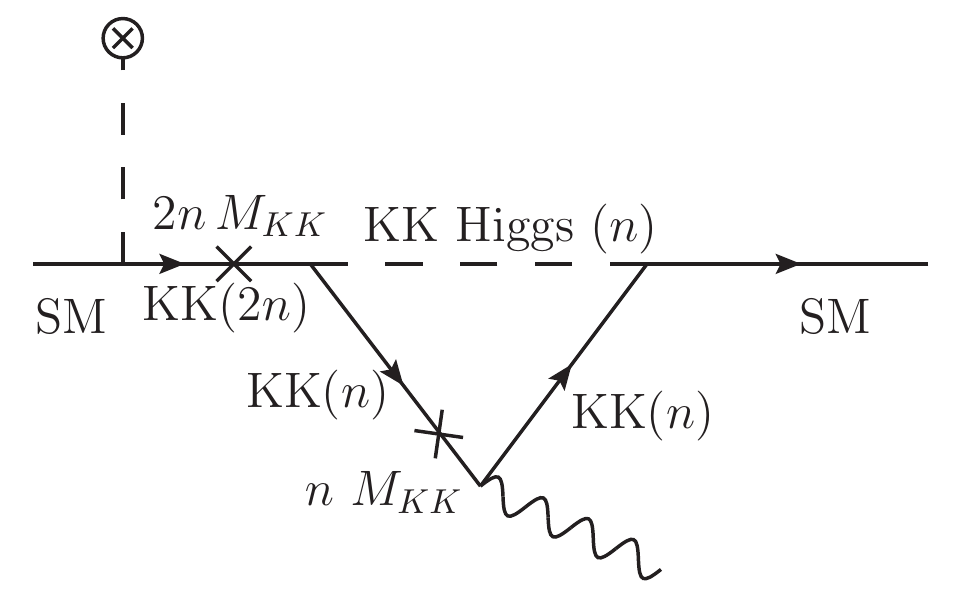}
\caption{Wrong chirality with the KK Higgs and Higgs VEV insertion (denoted by ``$\otimes$'') outside loop. The ``$\times$''s on the KK fermion line denote chirality flip.}
\label{fig:new_wrong2}
\end{center}
\end{figure}

For the case with a Higgs VEV insertion outside the loop shown in Fig.~\ref{fig:new_wrong2}, approximate KK number conservation at the KK Higgs vertices (but not for Higgs VEV insertion) implies that for Higgs mode-number $n$, the KK fermion inside the loop has the same mode-number, but the external KK fermion has mode-number $\sim 2 n$\footnote{This fermion is also allowed to be the zero-mode/SM by KK number conservation, but as already mentioned, we neglect such effects, since they involve suppressed Yukawa couplings}. 
In this case, we can write the estimate as%
\begin{eqnarray}
\l( C^{ \rm heavy, \; wrong, \; ext}_{ \rm dipole }\r )_{(n)} & \sim & 
\left\{
 \frac{1}{2n}\frac{n}{\l (\beta + n \r )^2}\l ( \log \frac{\beta}{n} + 1 \r )
 \right\}
 \frac{ y^{ \rm heavy}_{ \rm SM \; KK }\;  y^{ \rm heavy,\; - }_{ \rm KK \; KK }\; y^{ \rm light }_{ \rm SM \; KK }}{ y_{ \rm SM } }~. 
\end{eqnarray}
where as before we have used chirality flip factors in the numerator and dimensional analysis for the denominator, including the logarithm of the ratio of KK Higgs and KK fermion masses, like for the correct chirality contribution. This estimate is borne out by the exact loop function in Eq.~\ref{eq:outside}. We see that the dependence on KK masses (for fixed mode-number), and combination of couplings are  different from that for the diagram with a Higgs VEV insertion inside. The situation is different as well from the case of the SM Higgs in Section~\ref{SM.in.out} where the contributions from insertion inside and outside the loop were identical. The reason is that the wrong chirality is now in the coupling of the KK Higgs and it is unsuppressed even for $\beta \gg 1$, being actually enhanced compared to $y_{ \rm KK}$, just like for the correct chirality, KK Higgs coupling. Note that the Higgs VEV insertion (obviously of the SM Higgs) involves correct chirality.  Based on our earlier estimates of these couplings, it is easy to see that the combination of couplings for a Higgs VEV insertion outside is actually
parametrically different (it is larger for large $\beta$), giving for the above estimate:
\begin{eqnarray}
\l ( C^{ \rm heavy, \; wrong, \; ext}_{ \rm dipole } \r )_{(n)}
& \sim &  \frac{ \tilde{Y} }{Y} y_{ \rm KK }^2\; \beta\; 
\frac{1}{ 2 n } 
\frac{n}{ 
%
%
( \beta + n )^2  
%
%
}
\left( \log \frac{ \beta }{n} + 1 
\right)~. 
\end{eqnarray}
However, upon KK mode summation, the final estimate is the same as for a Higg VEV insertion inside (and thus not suppressed for $\beta \gg 1$):
\begin{eqnarray}
\begin{split}
C^{ \rm heavy, \; wrong,\; ext}_{ \rm dipole } & \sim \quad \frac{ \tilde{Y} }{Y} y_{ \rm KK }^2 \beta \sum_{ n = 1 }^{ n  \sim \beta }
\frac{ \left( \log \ds\frac{ \beta }{n} + 1 
\right) }{ 
\{ 
( \beta + n )^2 
%
%
}~,
%
%
\\
& \sim \quad 
\frac{ \tilde{Y} }{Y} y_{ \rm KK }^2~.
\end{split}
\end{eqnarray}
Finally, we also considered the potential contribution to the above effects from modes at the 5D cutoff scale $\Lambda$ running in the loops, which is found to be suppressed by $\left(\frac{\beta M_{\rm KK}}{\Lambda}\right)^2$. This is in contrast to the corresponding results in the case of a $\delta$-function brane-localized Higgs, where such an effect is significant  as found in \cite{Csaki:2010aj}, but is UV-sensitive: see Appendix~\ref{cornell} for details.
%

\section{Toward calculation in the 5D model}
\label{toy}
The above discussions in Sections~\ref{sec:NDAI} and ~\ref{sec:NDAII} involved only rough estimates. Here we add one more layer of semi-analytic estimate that we aim to capture actual calculations of $O(1)$ factors, while postponing the full numerical computation in a complete 5D model to Section~\ref{numerical}. The full computations of the dipole coefficients from loops require the precise spectrum of KK fermions and Higgs, their couplings, and the appropriate KK sum, in addition to the loop functions. The loop functions capture purely 4D factors which are more robust, whereas the other ingredients that capture more 5D effects are subject to the modifications due to brane-localized kinetic terms or the warp factor being modified from pure AdS near the TeV brane etc. Keeping track of these two effects separately will provide us with better insight on what we are dealing with. In this section, we focus on the calculation of the former contribution, namely, 4D loop functions. To this end, we consider the 4D effective field theory (what we call the 4D simplified model), describing the SM fields and just the first KK excitations of fermions and Higgs.

\subsection{Setting up 4D simplified model}
The 4D simplified model, where we only show what is relevant for a dipole operator for SM up-type quark for simplicity, is given by
\begin{eqnarray}
\begin{split}
{\cal L}^{4D}_{\rm Simplified} =&\ \  H^{ \rm light} \left( y^{u}_{ \rm SM } \overline{ q _L } u_R + y^{  \rm light, \; {\it u} }_{ \rm SM \; KK }  \overline{ q _L } U_R 
+ y^{\rm light, \; \it u }_{ \rm SM \; KK }\  \overline{ Q_L } u_R + y^{  \rm light, \; \it u+ }_{ \rm KK \; KK }\ \overline{ Q_L } U_R \right) + \rm h.c. \\
&  +  \tilde{H}^{ \rm light } \left(  y^{ \rm light, \; \it d }_{ \rm SM \; KK }\  \overline{ q_L } D_R + y^{ \rm light, \; \it d }_{ \rm KK \; KK }\  \overline{ Q_L } D_R \right)  + \rm h.c. \\
&  + H^{ \rm light } y^{ \rm light, \; \it u- }_{ \rm KK \; KK }\ \overline{ Q_R } U_L + \tilde{H}^{ \rm light} y^{ \rm light, \; \it d- }_{ \rm KK \; KK }\ \overline{ Q_R } D_L \rm h.c. \\
&  + H^{\rm heavy } \left( y^{ \rm heavy, \; \it u }_{ \rm SM \; KK }\ \overline{ q_L } U_R + y^{ \rm heavy, \; \it u }_{ \rm SM \; KK }\ \overline{ Q_L } u_R + y^{ \rm heavy, \; \it u + }_{ \rm KK \; KK }\  \overline{ Q_L } U_R  \right) +  \rm h.c. \\
&  + \tilde{H}^{\rm heavy } \left(  y^{ \rm heavy, \; \it d }_{ \rm SM \; KK }\ \overline{ q_L } D_R + y^{ \rm heavy, \; \it d + }_{ \rm KK \; KK }\  \overline{ Q_L } D_R  \right) +  \rm h.c. \\
&  + H^{ \rm heavy } y^{ \rm heavy, \; \it u - }_{ \rm KK \; KK }\ \overline{ Q_R } U_L + \tilde{H}^{ \rm heavy } y^{ \rm heavy, \; \it d  - }_{ \rm KK \; KK }\ \overline{ Q_R } D_L \rm h.c. \\
&  + M_Q \bar{Q} Q + M_D \bar{D} D + M_U \bar{U} U + M^2_H H^{ \rm heavy \; \dagger } H^{ \rm heavy }~.
\end{split}
\label{Ltoy}
\end{eqnarray}
Here, the superscript ``$\pm$" on the coupling denotes correct/wrong chirality. $q_L$ ($u_R$, $d_R$) are $SU(2)_L$ doublet (singlet) SM fermions. $Q$, $U$ and $D$ are vector-like KK fermions and their masses are denoted by $M_Q$, $M_U$, $M_D$. $H^{\rm light}$ corresponds to the (complex) SM Higgs doublet with mass $m_h$ (although it will be mostly neglected) whereas $H^{ \rm heavy}$ is a KK Higgs with the mass $M_H$. Even though we focus on the up-type quark dipole operator, we need down-type quark Yukawa couplings as well, which are (in general) different from that in the up-type quark sector and so the two are denoted by superscripts ``$u$" and ``$d$", respectively. The Higgs doublets for the down-type quark Yukawa couplings (for both light and KK modes) are given by the relation, $\tilde{H} = i \sigma_2 H^{ \ast }$. We will use the same notation for couplings as in above estimates: in particular, $L$ and $R$ chiralities of SM have same size of coupling, in turn, from the assumption of identical profiles in extra dimension (and similarly for all the correct chiralities of KK fermions and separately for all the wrong ones). These parameters are related to each other in the full 5D model, but this part of the calculation must be done numerically in order to do better than the $O(1)$ estimates of the previous section. We defer this step to the next section. Instead, here, for a semi-analytic calculation, we prefer to leave these couplings and masses as independent parameters (as far as we can afford to do so).

We calculate the coefficient of the chromomagnetic dipole operator using similar notation as in earlier Eq.~\ref{dipoledef},
\begin{eqnarray}
{\cal L} & \ni & m_{ \rm SM }  \frac{ g_{\rm S } C_{\rm dipole} }{ 16 \pi^2 M_{ \rm KK }^2 } \overline{ u_L } \sigma_{ \mu \nu } u_R G^{ \mu \nu }~. 
\label{chromodipoledef}
\end{eqnarray}
Note that in contrast to the electromagnetic (EM) dipole, we can attach gluons only to fermion lines, while photons can attach to either fermions or charged Higgses for EM dipoles, making the latter calculation a bit more involved (though equally straightforward). Here, $M_{ \rm KK }$ is the standard KK mass unit as defined earlier in Eq.~\ref{eq:mKK}.

This warm-up example will be a reasonable approximation to the full 5D model for the mass of the 5D Higgs field with $\beta \sim O(1)$ or smaller. Recall that in this case, the above estimates show that most of the KK effect comes from the lowest modes. The multiplicity of either fermion or Higgs fields is not really relevant here. In contrast, for the case of $\beta \gg 1$, the contributions from higher KK modes (up to mode number $n\sim\beta$) are crucial, and cannot be captured at all by our simplified model. Thus one needs to do the full 5D calculation (numerically). This will be done in Section~\ref{numerical}.

\subsection{SM Higgs in the loop}
\label{toylight}

The results and discussion in this section have a large overlap with recent work in~\cite{Delaunay:2012cz}. We adopt similar notations as ~\cite{Delaunay:2012cz}.

First, we consider the case where a Higgs VEV is attached to the internal quark lines only inside the loop (see Fig.~\ref{fig:insertioninside} which are detailed versions of Figs.~\ref{fig:old_correct:old_wrong1} and \ref{fig:new_correctwrong}). Irrespective of whether wrong or correct chirality coupling is involved, there is a cancellation in the neutral Higgs sector for this class of diagrams, namely, between the contributions of the physical Higgs boson and the would-be Nambu-Goldstone boson (which would be eaten by the longitudinal $Z$). Both the SM Higgs and $Z$-boson masses are negligible compared to the KK scale and we drop them in our calculation, (see Appendix B.2 of 1st reference in \cite{Csaki:2010aj} for the details). This cancellation can be understood as due to a Peccei-Quinn-like symmetry (see discussion above Eq.~(A3) in \cite{Delaunay:2012cz}). Therefore,  for this type of insertion we focus instead on the contribution from the (unphysical) charged Higgs (i.e., longitudinal $W$), which involves both up- and down-type Yukawa couplings. We drop its mass in our calculation. Similar to the procedure we followed for estimates of dipole operators in the previous section, we use couplings at vertices and masses in propagators as given in Eq.~\ref{Ltoy}, but calculating the loop integrals now. 

The resulting general formula for the dipole operator is then given by (see Appendix~\ref{app:loopfunction})
\begin{eqnarray}
\begin{split}
\frac{ C^{\phi, \; \rm int }_{ \; \rm dipole }}{ M_{ \rm KK }^2 } = &  
\left\{
\frac{ \left( y^{\phi, \; u }_{ \rm SM \; KK} \right) \left( y^{ {\rm light},  \; d + \; \ast }_{ \rm KK \; KK }  \right) \left( y^{\phi,  \; d }_{ \rm SM \; KK}  \right) }{ y^{ u }_{\rm SM } } 
\right\}
\left( \frac{ I^{ \rm int }_a + I^{ \rm int }_b }{2}  \right) \\
& + 
\left\{ 
\frac{  \left( y^{\phi,  \; u }_{ \rm SM \; KK }  \right) \left( y^{ {\rm light}, \; d- \; \ast }_{ \rm KK \; KK }  \right) \left( y^{\phi, \; d }_{ \rm SM \; KK }  \right) }{ y^{  u }_{\rm SM } }  
\right\}
\left( \frac{ J^{ \rm int }_a + J^{ \rm int }_b }{2}  \right)~,
\end{split}
\label{inside_general}
\end{eqnarray}
where the superscript $\phi$ collectively denotes both the SM and KK Higgses (i.e., $\phi=\{$``light'', ``heavy''$\}$), as both can propagate in the loop. The first and second terms in Eq.~\ref{inside_general} clearly correspond to the correct and wrong chirality coupling of KK fermions, respectively. Note that the middle factor in the Yukawa couplings corresponds to the Higgs VEV insertion and thus always involves a SM Higgs, i.e., regardless whether it is the heavy or light Higgs propagating in the loop. We drop the complex conjugate symbol in Yukawa couplings in the remainder of this section.  The detailed expression of the loop functions $I^{\rm int}_a$, $I^{\rm int}_b$ (for the correct chirality), and $J^{\rm int}_a$,$J^{\rm int}_b$ (for the wrong chirality) in Eq.~\ref{inside_general} are found in appendix~\ref{app:loopfunction}.

The result for the light Higgs in the loop is obtained by setting $M_H \rightarrow 0$ in the loop functions. As we mentioned before, the correct chirality contribution is negligible. They are suppressed by $\sim \left( m_{h} / M_{ \rm KK } \right)^2$ for an individual light Higgs (whether or not it is physical) in the loop, i.e., the suppression holds for each of the would-be Nambu-Goldstone boson contributions (charged and neutral) as well as the contribution from the physical (neutral) Higgs boson. We see this explicitly in our formula for loop-functions in the light Higgs limit (see Appendix~\ref{app:loopfunction} for more details), i.e., 
\begin{eqnarray}
I^{\rm int }_{ a, \; b } \left( M_H \rightarrow 0 \right)  & \sim & O \left( M_H^2 / M^4_{ Q, D } \right).
\label{Iaest}
\end{eqnarray}
Eq.~(\ref{Iaest}) actually tells us more than what we just mentioned above. It implies that the suppression holds separately for the loop functions where the gluon attaches to the right/left of Higgs VEV insertion (see discussion before Eq.~(15) in~\cite{Delaunay:2012cz} for a different approach).
We reiterate that this is independent of the above-mentioned cancellation within the neutral Higgs sector.
On the other hand, the loop-functions for the wrong chirality in the light Higgs limit become
\begin{eqnarray}
J^{ \rm int }_a + J^{ \rm int }_b \left( M_H \rightarrow 0 \right) 
\approx  \frac{1}{2} \frac{1}{ M_Q M_D }~.
\end{eqnarray}
Combining the above two features, we get 
\begin{eqnarray}
C^{ \rm light, \; int }_{ \rm dipole } & \approx & \frac{1}{4}  
\left\{ 
\frac{ \left( y^{ {\rm light},  \; u }_{ \rm SM \; KK } \right) \left( y^{ {\rm light}, \; d-}_{ \rm KK \; KK }  \right) \left( y^{ {\rm light}, \; d }_{ \rm SM \; KK }  \right) }{ y^{  u }_{\rm SM } }
\right\}
\left( \frac{ M_{ \rm KK }^2 }{ M_Q M_D }  \right)~,
\end{eqnarray}
which means that the contribution from the wrong chirality dominates (see Eq.~(A7) of \cite{Delaunay:2012cz} for similar discussion).\\

The Higgs VEV can also be attached to the external quark line outside the loop (see Fig.~\ref{fig:insertionoutside} which are more detailed versions of Figs.~\ref{fig:old_wrong2} and \ref{fig:new_wrong2}). In this case, both the neutral and charged Higgses contribute (i.e., the former does not encounter the cancellation of the earlier case and involves only up-type Yukawa couplings).
However, only the wrong chirality coupling is relevant here. The correct chirality effect is suppressed by the external KK fermion propagator between the Higgs VEV insertion and the loop, reducing to $\sim p / M^2_{ \rm KK }$ (where $p$ is the external quark momentum) by the requirement of no chirality flip (again, since only the correct chirality is chosen to couple). We emphasize that this suppression has nothing to do with the loop function unlike for the case of the SM Higgs contribution with the Higgs VEV insertion inside. 

The general formula for this case is (see Appendix~\ref{app:loopfunction} for the details)

\begin{eqnarray}
\begin{split}
\frac{ C^{\phi, \; \rm ext }_{ \rm dipole } }{ M_{ \rm KK }^2 }  = &  
\left\{
\frac{ \left( y^{ {\rm light}, \; u }_{ \rm SM \; KK }  \right) \left( y^{\phi, \; u -}_{ \rm KK \; KK }  \right) \left( y^{\phi, \; u }_{ \rm SM \; KK }  \right) }{  y^u_{ \rm SM }} 
\right\}
\left( \frac{ J^{ \rm ext }_{ u, \; e } + J^{ \rm ext }_{ u, \; f } }{2}  \right)
\\
& 
+ \left\{
\frac{ \left( y^{ {\rm light}, \; u }_{ \rm SM \; KK }  \right) \left( y^{\phi,  \; d- }_{ \rm KK \; KK } \right) \left( y^{\phi, \; d }_{ \rm SM \; KK } \right) }{  y^u_{ \rm SM }} 
\right\} 
\left( \frac{ J^{ \rm ext} _{ d, \; e } }{2}  \right)~,
\label{outside_general}
\end{split}
\end{eqnarray}
where $\phi$ again collectively denotes both the light SM and KK Higgs, $\phi=\{$``light'', ``heavy''$\}$. Note that only the wrong chirality coupling ($y^{\phi, \; u- \; \hbox{or} \; d- }$) enters in Eq.~\ref{outside_general}. The first factor in the Yukawa couplings corresponds to the Higgs VEV insertion and thus involves the SM Higgs (irrespective of whether the Higgs boson propagating in the loop is SM or KK). The details of these loop functions are given in Appendix~\ref{app:loopfunction}, where we see that the 1st term involving only up-type quark Yukawa couplings actually arises from both the charged and neutral Higgses, while the 2nd one only comes from the charged Higgs.

For the light SM Higgs, as before, this simplifies as:
\begin{eqnarray}
J^{ \rm ext }_{ u, \; e } 
\left( M_H \rightarrow 0 \right)  
\approx J^{ \rm ext }_{ u, \; f } \left( M_H \rightarrow 0 \right) &  \approx & \frac{1}{2 M_Q M_U }~,
\end{eqnarray}
and
\begin{eqnarray}
J^{ \rm ext }_{ d, \; e } \left( M_H \rightarrow 0 \right)  & \approx & \frac{1}{2 M_Q  M_D }~.
\end{eqnarray}
Therefore, Eq.~\ref{outside_general} leads to
\bea
\begin{split}
C^{ \rm light, \; ext }_{ \rm dipole } \approx &\  \frac{1}{2} 
\left\{
\frac{ \left( y^{ {\rm light}, \; u }_{ \rm SM \; KK } \right) \left( y^{ {\rm light}, \; u - }_{ \rm KK \; KK } \right)
\left( y^{ {\rm light}, \; u }_{ \rm SM \; KK } \right) }{ y^u_{ \rm SM } } 
\right\}
\left( \frac{ M_{ \rm KK }^2 }{ M_Q M_U } \right)\\
&  +
\frac{1}{4} 
\left\{
\frac{ \left( y^{ {\rm light}, \; u }_{ \rm SM \; KK } \right) \left( y^{ {\rm light}, \; d - }_{ \rm KK \; KK } \right)
\left( y^{ {\rm light}, \; d }_{ \rm SM \; KK } \right) }{ y^u_{ \rm SM } } 
\right\}
\left( \frac{ M_{ \rm KK }^2 }{ M_Q M_D } \right)~.
\end{split}
\eea

\subsection{Effects from KK Higgs modes in the loop}
\label{sec:toyheavy}

The above discussion involving the KK Higgs in the loops leads to our new results. 

First, consider the situation of the diagrams with internal Higgs VEV insertions in Fig.~\ref{fig:new_correctwrong} where the Higgs in the loop is one of the KK Higgs modes (instead of the SM Higgs). For the effect from the correct chirality coupling, the individual KK Higgs does not have any suppression, as opposed to the light Higgs which gives a suppressed effect, i.e., $\sim \l (m_{ W, \; Z, \; h } / M_{ \rm KK } \r )^2$. Nonetheless, just like for the SM Higgs, the neutral KK Higgs sector still has a cancellation between the real and imaginary KK Higgses: note that the latter is actually physical now, since the KK $Z$ boson (or any KK gauge boson in general) becomes massive by eating the $5^{\rm th}$ component of the corresponding 5D gauge field (instead of an imaginary scalar), whereas for SM modes the imaginary neutral Higgs boson becomes the longitudinal $Z$ boson. Note that (just like for light Higgs) this is irrespective of whether we consider the wrong or correct chirality couplings, i.e., holds for both cases (again, only for the internal Higgs VEV insertion that we are considering in this part). Of course, this cancellation in the neutral KK Higgs sector is not exact, since the real and imaginary KK Higgses are indeed split after EWSB, but the net effect is still suppressed by ratio of the splitting to $M_{ \rm KK }$ and so we simply neglect it here. Thus, this class of diagrams is dominated instead by the physical, charged KK Higgs.
This contribution is given by Eq.~\ref{inside_general} with $\phi =$ heavy.

For the case of the diagram in Fig.~\ref{fig:new_wrong2} with the Higgs VEV insertion outside the loop, we get the wrong chirality contribution for the KK Higgs by just setting $\phi =$ heavy in Eq.~\ref{outside_general}. 
The KK Higgs effect involving the correct chirality is suppressed for the same reason as for the SM Higgs as discussed in Section~\ref{toylight}, and it does not originate from the loop function.\\

In order to simplify the loop functions for a quick numerical estimate, we set all KK masses to be equal to $M_{\rm KK }$.
This roughly corresponds to the case where $\beta \sim O(1)$ or smaller in the complete 5D model. We keep track of symbols for wrong vs. correct chirality and light vs. heavy Higgs as these couplings can in general be different. It is only when we take various ratios of different contributions that we set these two sets of couplings equal. With the above assumption for KK masses, the loop functions for the KK Higgs are approximately
\begin{eqnarray}\label{eq:approxloopfunction:heavy}
\begin{split}
I^{\rm heavy, \; int} _{ a, \; b } M_{ \rm KK }^2 \approx - \ds\frac{1}{ 24},\quad & J^{\rm heavy, \; int }_{ a, \; b } M_{ \rm KK }^2 \approx  \ds\frac{1}{8}~, \\
J^{\rm heavy, \; ext} _{ d/u, \; e } M_{ \rm KK }^2 \approx \ds\frac{1}{3},\quad & J^{\rm heavy, \; ext}_{ u, \; f } M_{ \rm KK }^2 \approx \ds\frac{1}{3}~.
\end{split}
\end{eqnarray}
As expected, the loop functions $I_a$, $I_b$ in Eq.~\ref{eq:approxloopfunction:heavy}, involving the correct chirality Yukawa couplings, are not suppressed for the KK Higgs boson.
Also, note the negative sign in the 1st formula.

We focus on the terms involving both up and down-type quark Yukawa couplings (which come from the charged Higgs contribution) in all cases, for a fair comparison\footnote{Recall that there is a cancellation in the neutral Higgs sector between real and imaginary components of 
Higgs bosons, a subtlety we would like to avoid here, for simplicity.}.
We then get the contribution from the KK Higgs for the correct chirality  (internal Higgs VEV insertion only),
\begin{eqnarray}\label{eq:toy:heavy:correct}
C^{\rm heavy, \; correct }_{ \rm dipole } \approx  - \frac{1}{24} 
\frac{ \left( y^{{\rm heavy}, \; u}_{ \rm SM \; KK } \right) \left( y^{{\rm light}, \; d+}_{ \rm KK \; KK } \right) \left( y^{{\rm heavy}, \; d }_{ \rm SM \; KK } \right) }{ y^u_{ \rm SM } }~,
\end{eqnarray}
whereas the contribution from the KK Higgs for the wrong chirality,
\begin{eqnarray}\label{eq:toy:heavy:wrong}
\begin{split}
C^{\rm heavy, \; wrong }_{ \rm dipole } \approx  
\frac{1}{8} \frac{ \left( y^{{\rm heavy}, \; u }_{ \rm SM \;  KK } \right) \left( y^{{\rm light}, \; d - }_{ \rm KK \;  KK } \right) 
\left( y^{{\rm heavy}, \; d }_{ \rm SM \; KK } \right) }{ y^u_{ \rm SM } }
+
\frac{1}{6} \frac{ \left( y^{{\rm light}, \; u }_{ \rm SM \;  KK } \right) \left( y^{{\rm heavy}, \; d - }_{ \rm KK \;  KK } \right) 
\left( y^{{\rm heavy}, \; d }_{ \rm SM \; KK } \right) }{ y^u_{ \rm SM } }~, \\
\end{split}
\end{eqnarray}
where we included Higgs VEV insertions both inside and outside. We can take the ratio of the above two dipole coefficients, setting all couplings to be the same for simplicity:
\begin{eqnarray}
\frac{ C^{\rm heavy, \; correct }_{ \rm dipole } }{ C^{\rm heavy, \; wrong}_{ \rm dipole } } & \approx & - \frac{1}{7}~.
\label{toyratio1}
\end{eqnarray}
We see that correct chirality loop-function is smaller  than the wrong one by $3$, an $O(1)$ factor (considering
Higgs VEV insertions inside the loop for both) 
\footnote{Perhaps this is some sort of remnant of the cancellation that occurs for (individual) light Higgs contributions, i.e., between gluon attached to either side of the the Higgs VEV insertion. The point is that this cancellation is, of course, exact only for vanishing Higgs mass, which is a good approximation for the SM Higgs boson; while it is expected to be violated for the KK Higgs bosons, it might still result in an $O(1)$ factor suppression.}. In addition, the wrong chirality has a factor of $\sim 2$ enhancement from the Higgs VEV insertions inside and outside.

The total contribution from the KK Higgs which is the sum of Eq.~\ref{eq:toy:heavy:correct} and \ref{eq:toy:heavy:wrong}, is then (setting all couplings to be the same)
\begin{eqnarray}\label{eq:toy:heavy:total}
C^{ \rm heavy }_{ \rm dipole } & \approx & \frac{1}{4}
\frac{ \left( y^{{\rm heavy}, \; u }_{ \rm SM \;  KK } \right) \left( y^{{\rm light}, \; d }_{ \rm KK \; KK } \right) \left( y^{{\rm heavy}, \; d }_{ \rm SM \; KK } \right) }{ y^u_{ \rm SM } }~.
\end{eqnarray}
Similarly, the loop functions relevant for the SM Higgs boson, dominated by the wrong chirality, are roughly given by
\begin{eqnarray}
\begin{split}
I_{ a, \; b }^{ \rm light, \; int } M_{ \rm KK }^2 \sim O \left(  \frac{ m_{ h }^2 }{ M_{ \rm KK }^2 } \right) \sim 0,\quad & J^{\rm light, \; int}_{ a, \; b } M_{ \rm KK }^2 \approx  \ds\frac{1}{4}~, \\
J^{\rm light, \; ext} _{ d/u, \; e } M_{ \rm KK }^2 \approx \ds\frac{1}{2},\quad & J^{\rm light, \; ext }_{ u, \; f } M_{ \rm KK }^2 \approx \ds\frac{1}{2}~.
\end{split}
\end{eqnarray}
The contribution from the SM Higgs, combining Higgs VEV insertions outside and inside the loop, is given by
\begin{eqnarray}\label{eq:toy:light:wrong}
C^{\rm light, \; wrong }_{ \rm dipole } & \approx & \frac{1}{2} 
\frac{ \left( y^{{\rm light}, \; u }_{ \rm SM \; KK } \right) \left( y^{{\rm light}, \; d - }_{ \rm KK \; KK } \right) \left( y^{{\rm light}, \; d }_{ \rm SM \; KK } \right) }{ y^u_{ \rm SM } }~.
\end{eqnarray}

The comparison of the two wrong chirality effects from KK Higgs bosons in Eq.~\ref{eq:toy:heavy:wrong} and the SM Higgs in Eq.~\ref{eq:toy:light:wrong} gives a measure of how much suppression is from all particles in loop being heavy vs.~the Higgs being light (the form of the loop-function is the same here, whereas the masses are different):
\begin{eqnarray}
\frac{ C^{\rm heavy, \; wrong }_{ \rm dipole } }{ C^{ \rm light, \; wrong }_{ \rm dipole } } & \approx & \frac{7}{12}~,
\label{toyratio2}
\end{eqnarray}
where we set light and heavy Higgs couplings to be the same\footnote{This is the case for $\beta \sim O(1)$ or smaller in the 5D model: see estimates done earlier or actual calculations later on}. We see that the heavy Higgs loop is $\sim 2$ (still $O(1)$) smaller than the SM Higgs (as expected, based on 
masses of particles in the loop).

To get an idea of how much contribution from KK Higgs modes was missed in the earlier literature, we can further take the ratio of the two effects in Eq.~\ref{eq:toy:heavy:total} and Eq.~\ref{eq:toy:light:wrong},
\begin{eqnarray}\label{toyratio3}
\frac{ C^{ \rm heavy }_{ \rm dipole } }{ C^{ \rm light, \; wrong }_{ \rm dipole } } & \approx & \frac{1}{2}~.
\end{eqnarray}
Eq.~\ref{toyratio3} implies that the KK Higgs boson is comparable (even numerically) to the SM Higgs boson.

Finally, we compare the two net chirality effects by taking ratio of total correct chirality effect (dominated by the KK Higgs) to the total wrong chirality one (with contributions from both the SM and the KK Higgses):
\begin{eqnarray}
\frac{ C^{\rm heavy, \; correct }_{ \rm dipole } }{ C^{\rm light, \; wrong }_{ \rm dipole } + C^{ \rm heavy, \; wrong }_{ \rm dipole } } & \approx & - \frac{1}{19}~,
\end{eqnarray}
i.e., even when we do the calculation consistently including the KK Higgses, the sizes of the two chiralities are not quite comparable, with the correct chirality effect being smaller by $\approx 20$. However, it boils down to $O(1)$ factors from the evaluations of loop-functions, and one might still say parametrically they are on similar footing.

\section{Numerical evaluation in a complete 5D model}
\label{numerical}
In this section, we carry out full numerical 5D calculations of the dipole operator. The goal of these exact calculations is to validate the qualitative results presented in the previous sections. 
We will report them in terms of the coefficients ($C$) of the dipole operator defined in Eq.~\ref{chromodipoledef}. To be consistent with the discussion in Section~\ref{toy} we focus on the chromomagnetic operator of the up-type SM quark and on the terms which depend on both the up and down-type Yukawa couplings that scale like $y_U y_D^2$.

The procedure for doing the full computation in a complete 5D model is straightforward. As already hinted above, we can simply re-use the above calculations (of 4D loops) in the simplified model. First, we plug in exact couplings and masses (listed in Appendices~\ref{higgsRS}-\ref{app:loopfunction}) in the dipole operator coefficients, given in Eqs.~\ref{inside_general} and~\ref{outside_general}, in order to obtain the contribution from each KK level. Then, we perform the KK sum over both fermion mode numbers (denoted by $n_{ F_{1,2} }$) and Higgs mode number (denoted by $n_H$). That is,
\bea
\begin{split}
\label{eq:exactdipoles5D}
 &\frac{C^{\rm light, \; wrong, \; int}_{\rm dipole}}{M_{KK}^2}= \sum_{n_{F_1},n_{F_2}}\frac{y^{u}_{(n_{F_1},0,0)}\; y^{d,\,-\; *}_{(n_{F_1},n_{F_2},v)}\; y^{d}_{(0,n_{F_2},0)}}{y_{SM}}\times \frac{(J_a+J_b)(m_{n_{F_1}},m_{n_{F_2}},m_h)}{2}~,\\
 &\frac{C^{\rm light,\; wrong,\; ext}_{\rm dipole}}{M_{KK}^2}= \sum_{n_{F_1},n_{F_2}}\frac{y^{u}_{(n_{F_1},0,v)}\; y^{d,\, -\; *}_{(n_{F_1},n_{F_2},0)}\; y^{d}_{{(0,n_{F_2},0)}}}{y_{SM}} \times \frac{J_{d,e}(m_{n_{F_1}},m_{n_{F_2}},m_h)}{2}~,\\
  &\frac{C^{\rm heavy,\; wrong,\; int}_{\rm dipole}}{M_{KK}^2}= \sum_{n_{F_1},n_{F_2},n_H}\frac{y^{u}_{(n_{F_1},0,n_H)}\; y^{d,\, -\; *}_{(n_{F_1},n_{F_2},v)}\;  y^{d}_{(0,n_{F_2},n_H)}}{y_{SM}} \times \frac{(J_a+J_b)(m_{n_{F_1}},m_{n_{F_2}},m_{n_H})}{2}~,\\
 &\frac{C^{\rm heavy,\; wrong,\; ext}_{\rm dipole}}{M_{KK}^2}= \sum_{n_{F_1},n_{F_2},n_H}\frac{y^{u}_{(n_{F_1},0,v)}\; y^{d,\, -\; *}_{(n_{F_1},n_{F_2},n_H)}\; y^{d}_{{(0,n_{F_2},n_H)}}}{y_{SM}} \times \frac{J_{d,e}(m_{n_{F_1}},m_{n_{F_2}},m_{n_H})}{2}~,\\
 &\frac{C^{\rm heavy,\; correct,\;  int}_{\rm dipole}}{M_{KK}^2}= \sum_{n_{F_1},n_{F_2},n_H}\frac{y^{u}_{(n_{F_1},0,n_H)}\; y^{d,\, +\; *}_{(n_{F_1},n_{F_2},v)}\; y^{d}_{(0,n_{F_2},n_H)}}{y_{SM}} \times \frac{(I_a+I_b)(m_{n_{F_1}},m_{n_{F_2}},m_{n_H})}{2}~,
\end{split}
\eea
where $y$'s are the Yukawa couplings, obtained by integrating the 5D Yukawa couplings with the wave function profiles over the fifth dimension. The first two subscripts in $y$ are reserved for KK fermion numbers $n_{F_1}$, $n_{F_2}$ and the zeroth mode SM fermion (explicitly written as $0$). The last subscript denotes either the KK Higgs number $n_H$ or the light SM Higgs (explicitly written as $0$, and it is replaced with $v$ in the case of the Higgs VEV insertion). The exact definitons of the Yukawa couplings $y$'s and the complete forms of loop functions $I$'s, $J$'s, are given in Appendices \ref{app:couplings} and~\ref{app:loopfunction}.

We take various combinations of the above individual dipole coefficients in Eq.~\ref{eq:exactdipoles5D}. To this end, we also define 
some summed effects:
\bea
\begin{split}\label{eq:exactdipoles5D:comb}
 C^{\rm light,\; wrong}_{\rm dipole} &\equiv\ C^{\rm light,\; wrong,\; int}_{\rm dipole}+C^{\rm light,\; wrong,\; ext}_{\rm dipole}~,\\
 C^{\rm heavy}_{\rm dipole}     &\equiv\ C^{\rm heavy,\, correct}_{\rm dipole}+C^{\rm heavy,\, wrong}_{\rm dipole}~,\\
 C^{\rm heavy,\; correct}_{\rm dipole} &\equiv\ C^{\rm heavy,\; correct,\; int}_{\rm dipole}~,\\
 C^{\rm heavy,\; wrong}_{\rm dipole} &\equiv\ C^{\rm heavy,\; wrong,\; int}_{\rm dipole}+C^{\rm heavy,\; wrong,\; ext}_{\rm dipole}~.
\end{split}
\eea
However, before presenting the actual results for the dipole operators, we first check that the patterns of exact couplings and masses are in accord with expectations in Section~\ref{sec:NDAI}.
In particular, we will be interested in the $\beta \gg 1$ limit, where key ingredients were estimated as follows: 
\vspace{0.05in}

\begin{itemize}
\item the wrong chirality light Higgs coupling is suppressed by $1/\beta^2$, but grows with KK fermion mode
number, as intuitively shown in Eq.~\ref{KKYukapproxwronglight}~;

\item the KK Higgs coupling is enhanced compared to that of the SM Higgs,
as shown schematically in Eqs.~\ref{SMKKYukapproxheavy} and~\ref{KKYukapproxheavy}~;

\item the KK Higgs coupling exhibits approximate KK mode number conservation, as mentioned below Eqs.~\ref{SMKKYukapproxheavy} and \ref{KKYukapproxheavy}~;

\item the KK Higgs spectrum has a region of quasi-degenerate modes as sketched in Eq.~\ref{KKHiggsmassapprox}.
\end{itemize}

\begin{figure}
\centering
\includegraphics[scale=0.34]{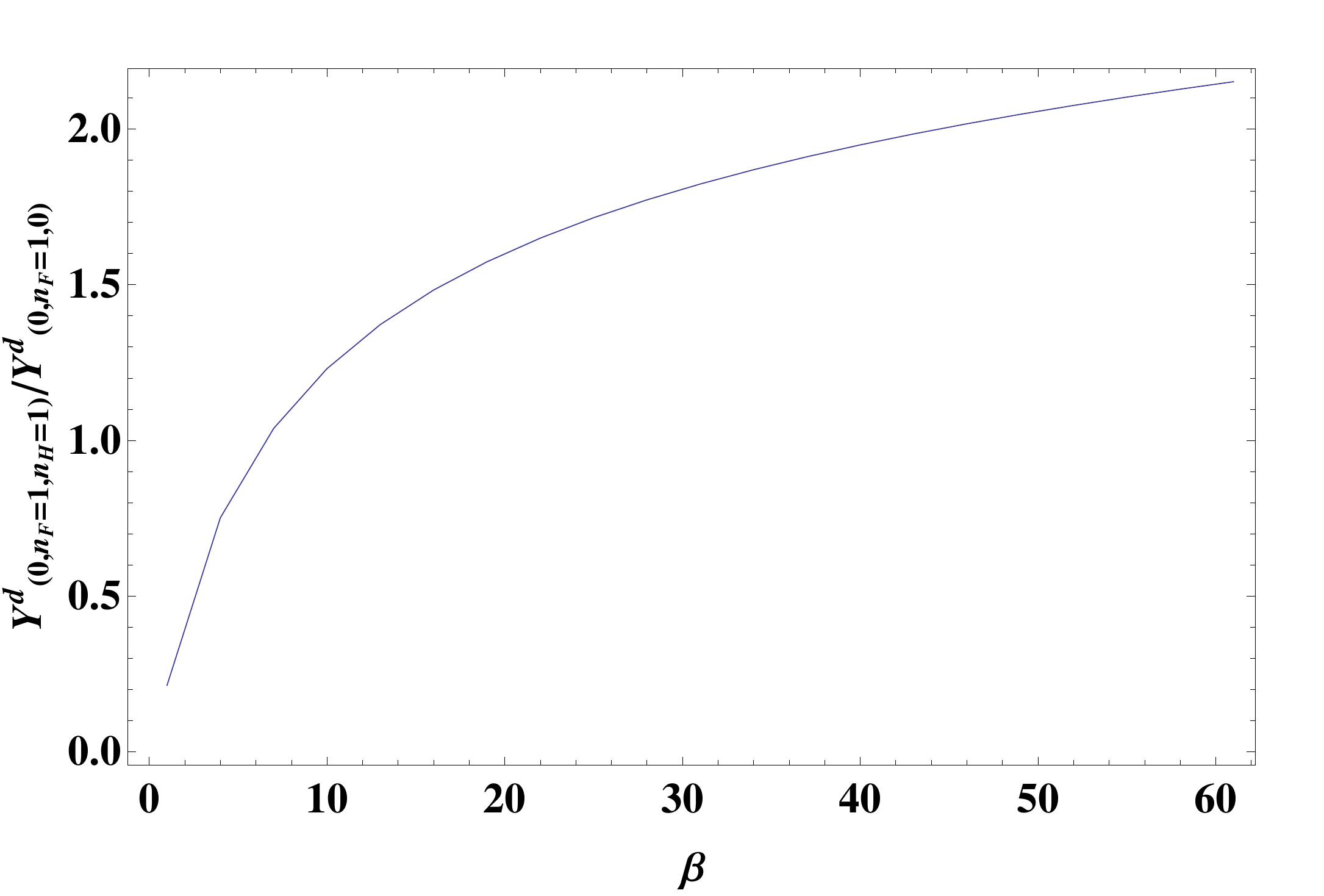}\quad
\includegraphics[scale=0.395]{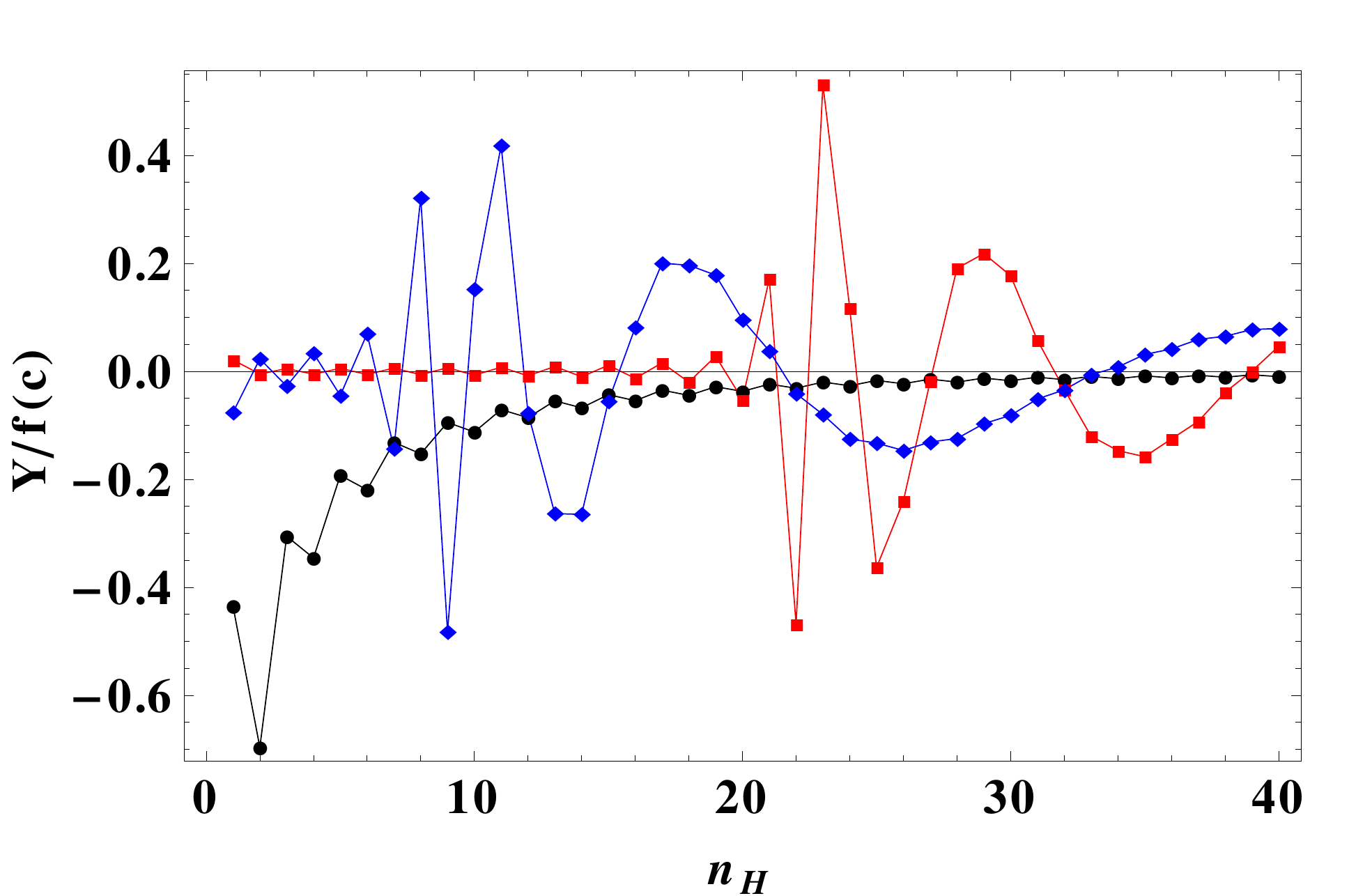}
\caption{Left: The ratio of two Yukawa couplings (between the SM fermion, KK fermion and Higgs), differing by whether the Higgs is SM- or KK-type, as a function of the bulk mass parameter $\beta$; for the KK fermion and KK Higgs, we selected the 1st level mode, $n_F = 1$ and $n_H=1$. Right: the effective Yukawa couplings of the SM fermion, KK fermion, and KK Higgs; three lines correspond to three different KK fermion modes: $n_F =$ 1 (black), 15 (blue) and 30 (red), $n_H$ on $x$-axis is KK Higgs mode number. The Yukawa coupling was normalized to $f(c)$, which is the SM fermion profile on the IR brane.}
\label{fig:yukratio}
\end{figure}

The first bullet point has already been discussed in~\cite{Azatov:2009na, Delaunay:2012cz} and so we refer the reader to those discussions. The second bullet point is illustrated in the left panel of Fig~\ref{fig:yukratio}. It shows that the couplings of the SM Higgs have an additional suppression of $1/\sqrt{ \beta }$ compared to the couplings of the KK Higgses. In detail, the ratio plotted is
\bea
\frac{ y^{d}_{(0,n_F = 1,n_H = 1)} }{ y^{d}_{(0,n_F = 1,0)}}~,
\eea
where the numerator (denominator) corresponds to the Yukawa coupling of the SM Higgs (the 1st KK Higgs) with the 1st KK and SM fermions (see Eq.~\ref{SMKKYukexact1} or \ref{SMKKYukexact2} for the exact definitions). The third bullet point, which states the approximate KK number conservation is illustrated in the right panel of Fig.~\ref{fig:yukratio} where we plot the Yukawa coupling between the SM fermion, KK fermion and KK Higgs (see Eq.~\ref{SMKKYukexact1} for the definition), normalized to $f(c)$, which is the value of the fermion zero mode wave function on the IR brane. The right panel of Fig.~\ref{fig:yukratio} shows the coupling as a function of Higgs mode number $n_H$ (with $\beta=20$ chosen) for three different values of KK fermion mode number, $n_F =$1, 15, 30. For the numerical illustration, we set the 5D mass parameters of the light quarks to the values, $c_q=-c_d=-c_u=0.6$ and the 5D Yukawa couplings to $Y^u_{5D}=Y^d_{5D}=1/\sqrt{k}$ (these will be our default values for all numerical studies, unless otherwise specified). One can see the approximate KK number conservation: the coupling vanishes once we go to the values of the Higgs KK number, $n_H$, that are very different from the fermion KK numbers, $n_F$. 
In more detail, for high KK numbers the wave functions become approximate trigonometric functions and the overlap integrals follow approximate orthogonality relations. Finally, the degeneracy in the KK Higgs spectrum mentioned in the fourth bullet point is clearly seen in (the more exact) Eq.~\ref{KKHiggsmassexact}.\\

\begin{figure}
\centering
\includegraphics[scale=0.46]{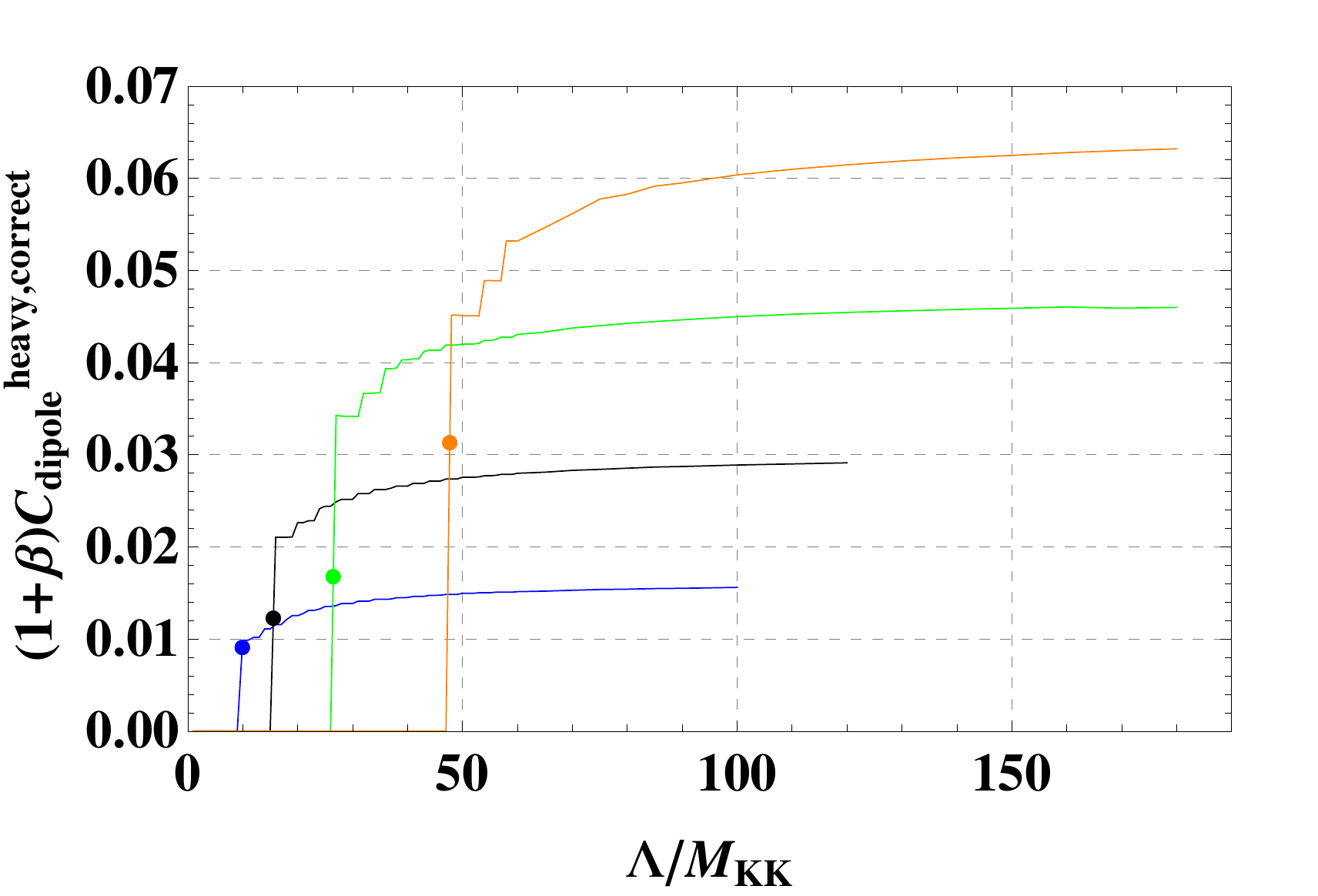}
\includegraphics[scale=0.48]{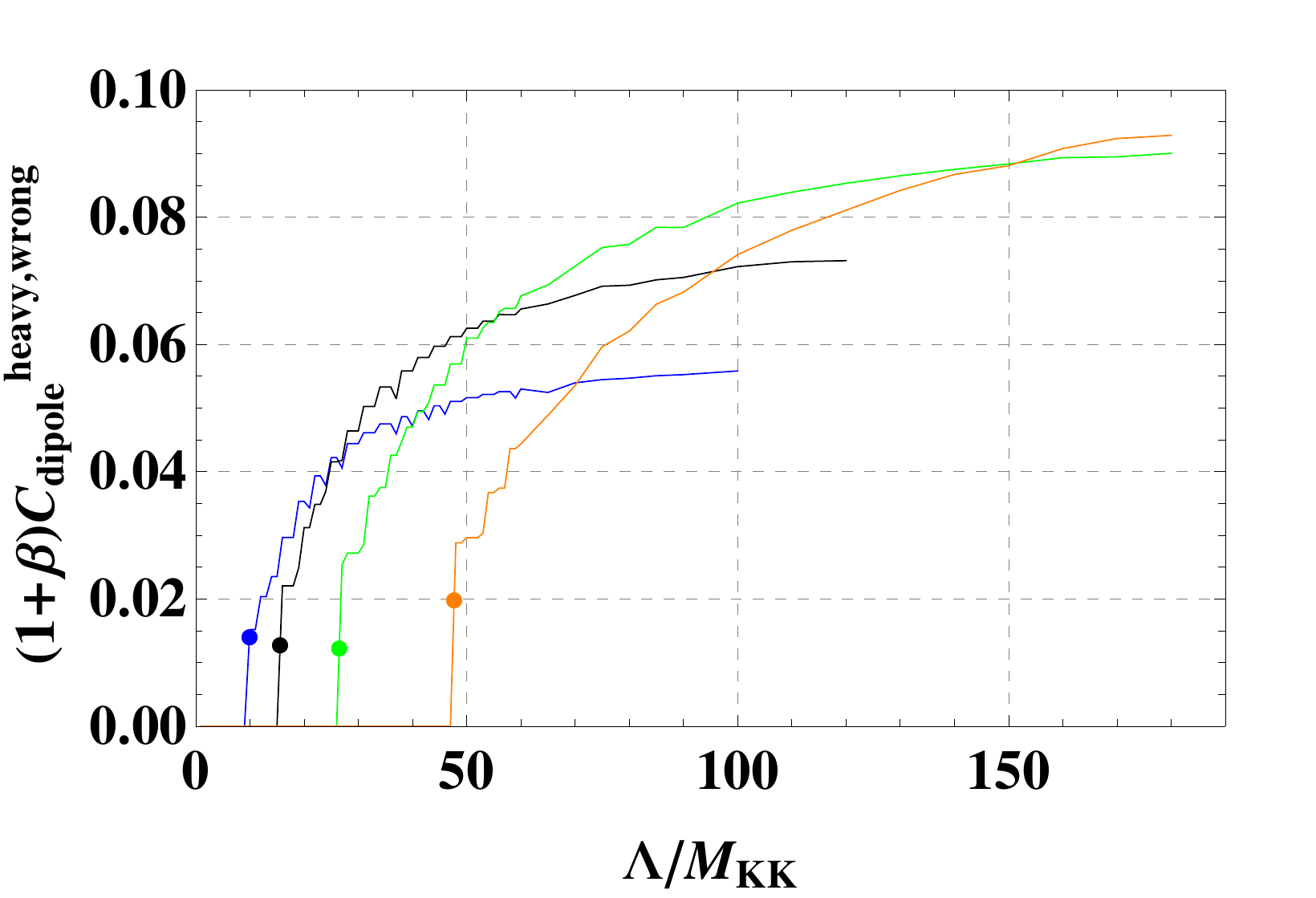}\\
\includegraphics[scale=0.47]{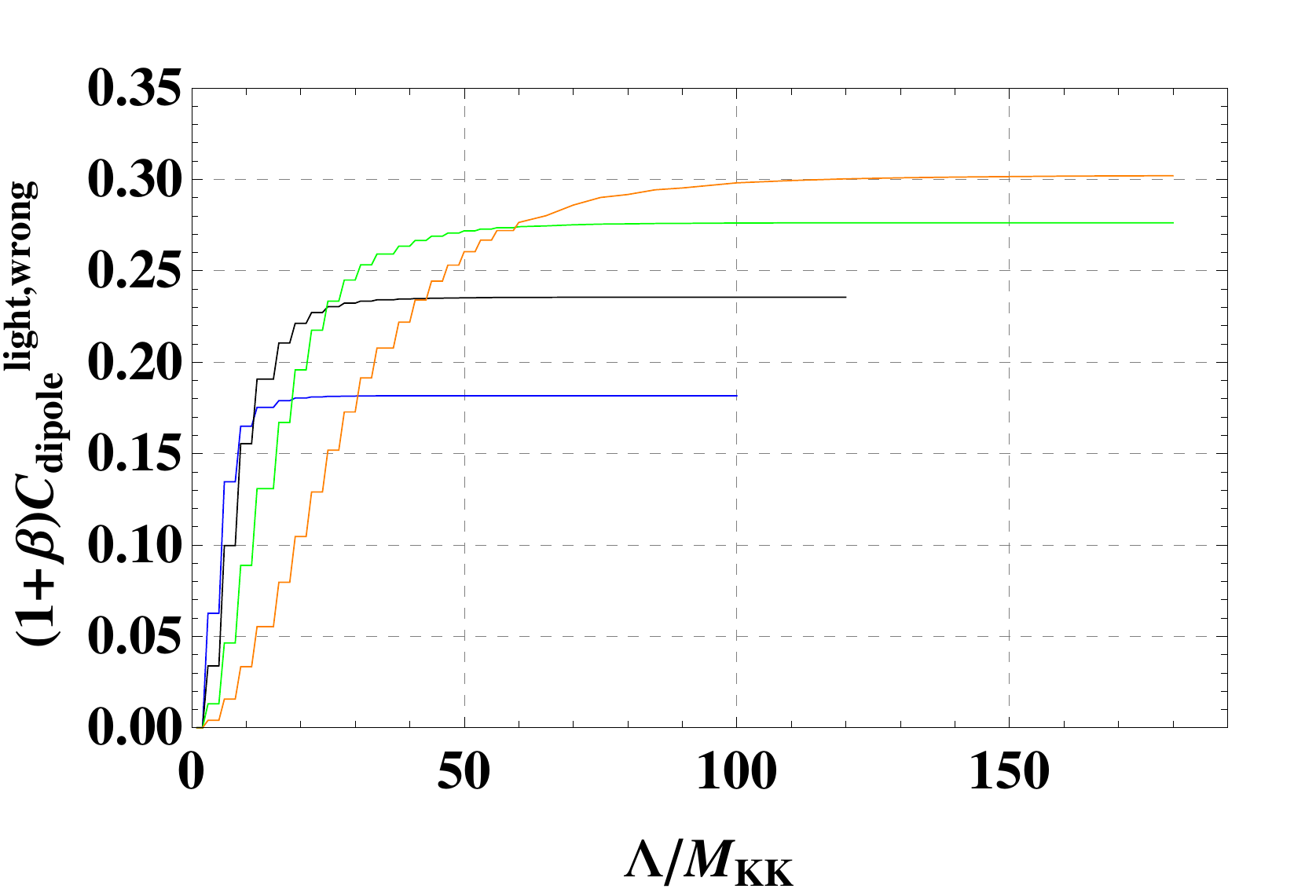}
\includegraphics[scale=0.41]{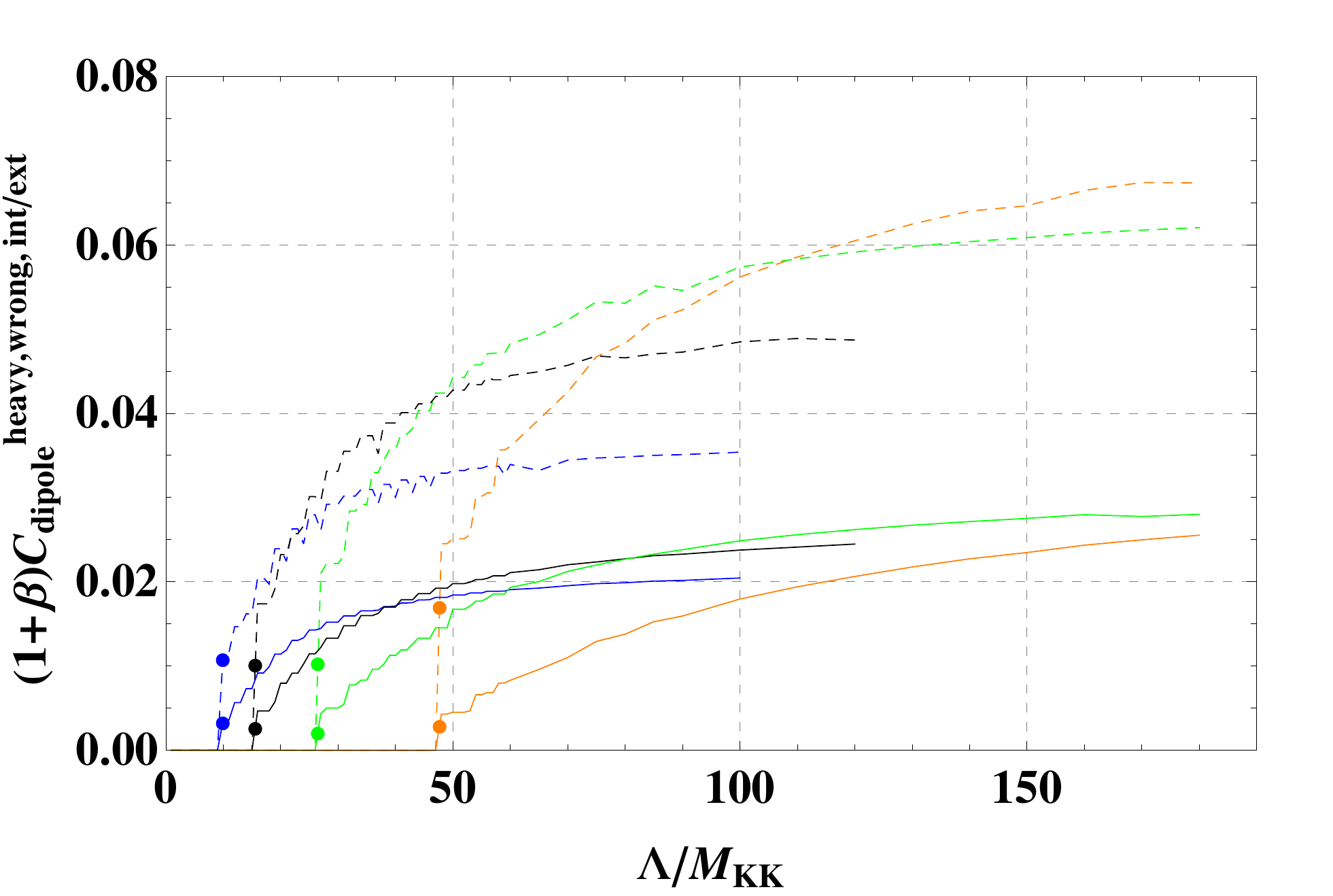}
\caption{\label{kkdip} Our numerical estimates of three coefficients, $C^{\rm heavy,\; correct}_{\rm dipole}$ (upper-left), $C^{\rm heavy,\; wrong}_{\rm dipole}$ (upper-right), and $C^{\rm light,\; wrong}_{\rm dipole}$ (bottom-left). The four lines correspond to $\beta=5$ (blue), 10 (black), 20 (green), 40 (orange). $\Lambda$ on the $x$-axis is the cutoff for the mass of all the modes. The circle indicates the mass of the first KK Higgs and it's contribution. The dipole coefficients are appropriately rescaled by $1+\beta$. 
The $C^{\rm heavy,\; wrong}_{\rm dipole}$ is subdivided into two individual contributions (bottom-right): $C^{\rm heavy,\; wrong,\; int}_{\rm dipole}$ (solid) and $C^{\rm heavy,\; wrong,\; ext}_{\rm dipole}$ (dot-dashed).  The definitions of the dipole coefficients are given in Eqs.~\ref{eq:exactdipoles5D} and~\ref{eq:exactdipoles5D:comb}.}
\end{figure}

Based on the above checks, we expect the results of our full numerical calculation of dipole operators to roughly agree with the earlier estimates in Sections~\ref{sec:NDAI},~\ref{sec:NDAII}. These dipole coefficients $C$, multiplied by the factor $(1+\beta)$ (for the reason explained in Section~\ref{SMYukawa}) are shown in Fig.~\ref{kkdip}. The two plots in the upper panel of Fig.~\ref{kkdip} are dipole coefficients from the KK Higgs in the loop for the correct and wrong chiralities for four different choices of $\beta$. The bottom-left panel shows the SM Higgs loop effect, where only the wrong chirality is significant. Finally, the bottom-right plot of Fig.~\ref{kkdip} separate the wrong chirality KK Higgs contributions depending on whether a Higgs VEV insertion is inside or outside the loop, while the correct chirality has only the former effect. The dipole coefficients are shown as a function of the cutoff scale $\Lambda$ (in units of $M_{KK}$), which is defined as follows: the KK sum includes only KK fermion and KK Higgs modes whose masses are below $\Lambda$. Note that the numbers of KK fermion modes and KK Higgs modes that are below $\Lambda$ actually vary with $\beta$, recalling that the KK Higgs masses are roughly $\sim (\beta + n)\; M_{KK}$ whereas KK fermion masses are $\sim  n\; M_{KK}$. In particular, there is no contribution from loops of KK Higgs modes as long as the cutoff $\Lambda$ is below the first KK Higgs mass, roughly given by $\sim \beta M_{KK}$ (up to $O(1)$ difference from the exact values). This explains in Fig.~\ref{kkdip} the difference he starting point on the $x$-axis of the curves (i.e., what value of $\Lambda$ does dipole contribution kick-in) between the two cases with the SM Higgs and KK Higgs, as far as t is concerned.

We clearly see in Fig.~\ref{kkdip} that in the case of the KK Higgses, the dipole effect saturates only after summing over modes with masses up to $\sim \hbox{a few} \times \beta M_{KK}$. The saturation also means that the result becomes insensitive to the modes much beyond $\beta M_{KK}$  (demonstrating the UV-insensitivity). The underlying reason for this saturation was already discussed in Section~\ref{sec:NDAII}. A similar saturation is observed for the case with the SM Higgs in the loop, as seen in the bottom-left plot of Fig.~\ref{kkdip}. In this case, the saturation is reached while summing over KK fermion modes (this result was first calculated in~\cite{Delaunay:2012cz}). We see that the KK Higgs effects (both wrong and correct chirality) are indeed roughly comparable to the SM Higgs one (see further discussion on this point below). Also, the asymptotic values are roughly independent of $\beta$, up to a small, $O(1)$ growth with $\beta$. This observation corresponds to our central result, i.e., it clearly indicates an ``apparent'' non-decoupling behavior, against our naive expectation of the KK Higgs effect dropping with increasing $\beta$. From the bottom-right panel, it appears that the two sub-contributions within the wrong chirality KK Higgs effect are of the same order, again in agreement with the semi-quantitative discussion in Sections~\ref{SM.in.out} and \ref{wrong.in.out}

\begin{figure}
\centering
\includegraphics[scale=0.8]{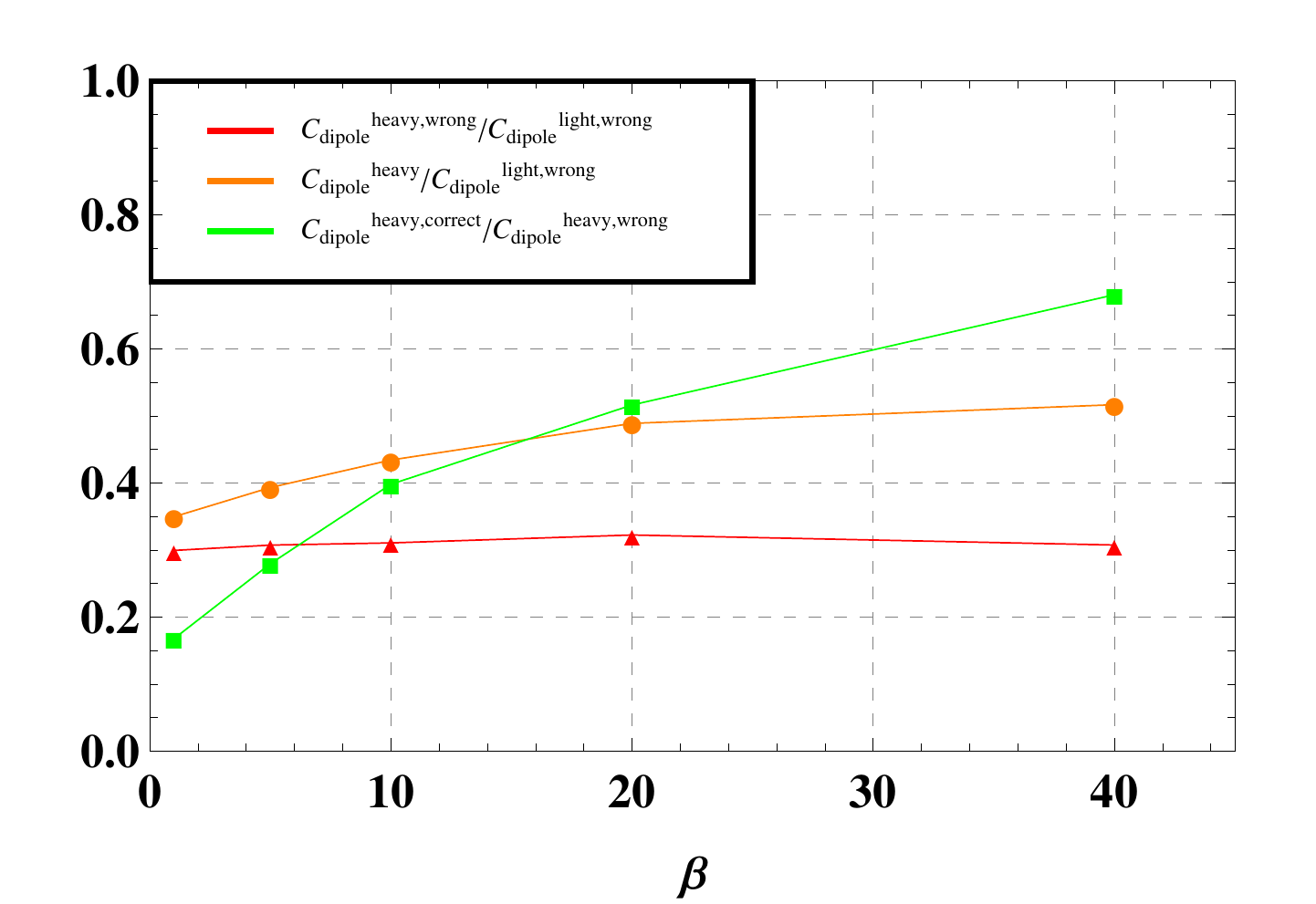}
\caption{\label{kkdipR1} 
Various ratios of the dipole coefficients as a function of $\beta$ (5D Higgs mass parameter). The dipole coefficients in the legend are the same as those in Fig.~\ref{kkdip} (see Eqs.~\ref{eq:exactdipoles5D} and~\ref{eq:exactdipoles5D:comb} for the exact definitions).
}
\end{figure}
%

%
Our finding of the KK Higgs effect is more pronounced in Fig.~\ref{kkdipR1} where we take ratios of the various saturated values of the dipole coefficients in Fig.~\ref{kkdip} (i.e., differing by their main contributors),
%
%
revealing a 
different perspective on our results.
We do this for $\beta=$5, 10, 20, 40.
These ratios are also expected to qualitatively agree with our NDA-type estimates done in Sections~\ref{sec:NDAI},~\ref{sec:NDAII}. 
In detail, the red line in Fig.~\ref{kkdipR1} represents the relative size of the wrong chirality effect between the KK Higgs modes and the SM Higgs, and it is expected to match our NDA-type estimate in Eq.~\ref{toyratio2}. This ratio is somewhat smaller than 1 (although within an order of magnitude), probably reflecting the loop function being smaller for the former, due to all particles in the loop being heavy. 
The green line measures the relative size between the correct and wrong chirality effect for the KK Higgs. The ratio is somewhat smaller than 1, partly because of the difference in the corresponding loop functions with different chirality (both containing heavy particles) which were discussed in Section~\ref{sec:toyheavy} (also see Eq.~\ref{toyratio1}).
The orange curve in Fig.~\ref{kkdipR1} indicates what fraction of the total effect was missed in the earlier literature, namely the ratio of total KK Higgs effect to the SM one. It is seen to be significant (see Eq.~\ref{toyratio3} for a related NDA-type estimate).

Finally, we note that the NDA estimates \cite{Agashe:2004cp, Agashe:2006iy} give a log-divergence in the brane-localized Higgs case,  which corresponds to the $\beta \rightarrow \Lambda/k$ (or $\infty$) limit of the bulk Higgs. This expectation should have shown up as a $\log \beta$ dependence in the KK summed result for the bulk Higgs. However, as already mentioned, our semi-analytic estimates (shown in Section~\ref{KKhiggs}), which are based on exact KK number conservation, do not have such a factor (of course, we do see a UV-sensitivity in this limit, in agreement with NDA estimates). Our numerical results do include the (small) violation of KK number conservation present in the model (see the right panel of Fig.~\ref{fig:yukratio}); nonetheless, we do not see a clear $\log \beta$ dependence here either (see Fig.\ref{kkdip}).

In summary, the full 5D calculation does agree with estimates of Section~\ref{sec:NDAII} and the calculation in the simplified model of Section~\ref{toy}. The KK Higgs contribution to dipole operators is important and interesting. The correct chirality's effect is significant, and both chirality effects are unsuppressed even as we take the brane Higgs limit, i.e., make the KK Higgs heavy.%

\section{Conclusions}
\label{conclude}
\indent

In this paper, we have calculated for the first time the contribution from KK Higgs bosons (along with KK fermions in the loop diagrams) to dipole operators of SM fermions in the framework of warped extra dimension models with SM fields, including the Higgs boson, propagating in the bulk. The previous work on such dipole operators involved only the SM Higgs boson in the loop diagrams. We found that the KK Higgs effect is in fact comparable to that from the SM Higgs. Therefore while the new result does not change the associated phenomenology by more than an $O(1)$ factor, it is clearly important to include KK Higgses for the sake of completeness in warped 5D models.

In addition, the KK Higgs effect is interesting on several theoretical fronts: first of all, it is necessary to include for
consistency with 5D covariance. Furthermore, it receives sizeable contribution from the SM-like (what we call ``correct") chirality couplings of the Higgs boson, as opposed to the SM Higgs effect which is dominated by the wrong chirality couplings. Finally, when the mass of the 5D Higgs field becomes much larger than the AdS curvature scale ($\beta \gg1$ in our notation) (as needed for localizing the zero-mode Higgs 
(very) close to the
IR brane), the summed KK Higgs modes contribution features an ``apparent'' non-decoupling effect: it is unsuppressed even as the mass of the 1st KK Higgs increases, due to the quasi-degeneracy of the KK Higgs spectrum up to mode number $\sim\beta$, as well as the enhanced coupling of the KK Higgs relative to the SM Higgs. Ultimately, we demonstrated the above features of the KK Higgs effect with a numerical analysis in the full 5D model. To build intuition, we also performed semi-analytic NDA estimates and an analytic calculation in a simplified model, both of which agree qualitatively with the numerical computation. The chosen simplified model mimics the lowest-level KK sector of the 5D model with $\beta \sim O(1)$ or smaller, where the 1st KK Higgs boson mass is at the typical KK scale.

As an aside, we mention another model which is often employed in order to analyze various effects in 
this 5D framework, namely the ``two-site" model (\cite{Contino:2006nn} and its variations), which is related to the simplified model that we studied here. The two-site model is based on the deconstruction of the 5D model, combined with the AdS/CFT correspondence. It consists of elementary and composite sectors which mix (even before EWSB), and the resultant eigenstates (i.e., after diagonalizing this mixing) would roughly correspond to the particles in our simplified model, including SM/zero and the 1st KK modes\footnote{These can further mix due to EWSB, but this is a sub-leading effect.}. However, in the existing two-site models, there is only the SM Higgs boson on the composite site. Thus, in this model, clearly the dipole operator arises only from the wrong chirality couplings \footnote{References \cite{Agashe:2008uz, Konig:2014iqa} studied dipole operators in such a framework.}. On the other hand, in this two-site model it is straightforward to ``model" the KK Higgs/correct chirality effect that we calculated in this paper, by simply adding a heavy Higgs boson on the composite site, with a mass comparable to the gauge and fermion composites there, and with Yukawa couplings similar to the SM Higgs boson. Basically, this modified 2-site model will then be even more similar to (if not the same as) the simplified model that we studied.

We close with some remarks on possible directions for follow-up studies related to this topic. In this paper we focused on general aspects of the dipole calculations. Corresponding detailed analyses for specific dipole observables have more direct phenomenological implications, which we leave for future work. Another interesting avenue to pursue is the application of the AdS/CFT correspondence to this KK Higgs effect. In particular, the limit of $\beta \gg 1$ is dual (on the CFT side) to the scaling dimension of the Higgs operator becoming large: what then is the dual interpretation of the ``apparent'' non-decoupling effect seen on the 5D side for such $\beta$? In this context, references \cite{Pappadopulo:2012jk} might be relevant.

Finally and curiously, as far as we know, this is the first time that the effect of the KK Higgs on low-energy observables has been found to be significant. Of course, merely detecting a signal for such a dipole operator will not constitute ``evidence" for the KK Higgs, since not only is it an indirect effect that can be mimicked by other types of new physics, but the SM Higgs can also give similar effects within this model. Moreover, one cannot distinguish $\beta \lesssim O(1)$ from $\beta \gg 1$ simply based on observables originating from dipole operators, since they are of similar size in both cases. Clearly, we need a direct signal for the KK Higgs, i.e., KK Higgs production and detection at colliders, particularly, production via gluon fusion and decay to $t\bar{t}$ pairs. Of course, the cross-section for such a KK Higgs signal is expected to be small at the 14 TeV LHC, given the loop-level production channel, and a few TeV mass for the KK Higgs, based on direct and indirect bounds on other KK particles (and the relation between all of these masses). On the other hand, recently the possibility of a 100 TeV hadron collider has been widely discussed, which would allow a better probe of multi-TeV KK Higgs bosons. Such collider searches can allow distinction between large and small $\beta$ as well, since the KK Higgs is similar in mass to other KK particles for small values of $\beta$, whereas it is much heavier for larger $\beta$. Thus, it would be timely to further study collider phenomenology of such KK Higgs bosons, which are ``must-have" in warped extra-dimensional models, and yet have been overlooked so far.

\section*{Acknowledgements}

The work of K.~A. and Y.~C.~was supported in part by NSF Grant No.~PHY-1315155 and by the Maryland Center for Fundamental Physics.
Y.~C.'s research was supported in part by the National Science Foundation under Grant No. PHY11-25915T and by Perimeter Institute for Theoretical Physics. Research at Perimeter Institute is supported by the Government of Canada through Industry Canada and by the Province of Ontario through the Ministry of Research and Innovation. 
The work of LR was supported in part by NSF grant PHY-0855591 and PHY-1216270 
and the Fundamental Laws Initiative of the Harvard Center for the Fundamental Laws of Nature.
We would like thank Csaba Csaki, Yuval Grossman, Gilad Perez, Raman Sundrum, Philip Tanedo, and Luca Vecchi for discussions.
K.~A., Y.~C. and L.~R. thank the Aspen Center for Physics (where part of the work was done) for hospitality.


\appendix

\section{Solutions for bulk Higgs}
\label{higgsRS}

The model, including the 5D Lagrangians, was already outlined in Section~\ref{model}.
Here, we present details of the KK decompositions, starting in this Appendix with the Higgs field.
The bulk Higgs construction was first suggested in the \cite{Luty:2004ye} (see~\cite{Archer:2012qa} for a recent discussion of the bulk Higgs, especially its massive modes). The five dimensional Lagrangian is given by
\bea
{\cal L}_{\rm Higgs} = \int dz d^4x \left(\frac{R}{z}\right)^3
\left[ |{\cal{D}}_M H|^2 - \frac{\mu^2}{z^2} |H|^2  \right] -
V_{UV}(H)\delta(z-R) - V_{IR}(H)\delta (z-R')~,
\label{HiggsLag}
\eea
where the $V_{UV}$, $V_{IR}$ are the potentials on the UV and IR branes. 
The equation of motion is derived from Eq.~\ref{HiggsLag} which is
\bea
 \d_z\l(\frac{1}{z^3}\d_z\r)H + \frac{p^2}{z^3}H - \frac{\mu^2}{z^5}H=0~,
\label{5DEOMHiggs}
\eea
and its boundary conditions at the UV and IR branes are given by
\bea\label{eq:5DEOMHiggs:BC}
\begin{split}
 \d_z H-\frac{\d}{\d H^*}V_{UV}~ &= 0 \quad {\rm for}\quad z=R~,\\
 -\l(\frac{R}{R'}\r)^3\d_z H - \frac{\d }{\d H^*}V_{IR}~ &=0 \quad {\rm for}\quad z=R'~.
\end{split}
\eea
Solving the equation of motion in Eq.~\ref{5DEOMHiggs} for massless mode, $p^2 = 0$, we obtain the profile of the Higgs VEV along the fifth dimension,
\bea\label{eq:5Dvev}
v(z)\sim a z^{2+\beta}+b z^{2-\beta}~,
\eea
where $\beta = \sqrt{4+\mu^2}$. The second term in the Higgs VEV in Eq.~\ref{eq:5Dvev} can be removed by an appropriate boundary condition at the UV boundary,
\bea
V_{UV}=m_{UV}|H|^2,~~m_{UV} = \frac{2+\beta}{R}~,
\label{VUV}
\eea
leaving only $z^{2+\beta}$ term shown in Eq.~\ref{v(z)}. The coefficient $a$ in Eq.~\ref{eq:5Dvev} is still an $x_\mu$-dependent field and we introduce the usual ``Mexican hat"-type potential of the Higgs at the IR brane,
\bea
V_{IR}=\l(\frac{R}{R'}\r)^4\frac{\lambda R^2}{2}\l( H^2-\frac{v_{IR}^2}{2}\r)^2~,
\label{VTeV}
\eea
to develop a VEV:
\bea\label{eq:5Dvev:fixed}
 v(z)=V(\beta)\ z^{2+\beta}~,
\eea
where $V(\beta)$ is defined as
\bea
V(\beta)=\sqrt{\l(v_{IR}^2-\frac{2(2+\beta)}{\lambda R^3}\r)}\frac{1}{(R')^{2+\beta}}~.
\eea
$v_{IR}$ in the above equation can be replaced by the 4D VEV by the relation
\beq
v_4^2 = \int_{R}^{R'} dz \left ( \frac{R}{z} \right )^3 v^2(z)~.
\eeq
That is, $V(\beta)$ in terms of $v_4$ is rewritten as 
\bea
\begin{split}
&V(\beta) = \sqrt{\frac{2(1+\beta)}{R^3 (1-(R/R')^{2+2\beta})}}\frac{v_4}{(R')^{1+\beta}}~,
\end{split}
\eea
where the $v_4=246$ GeV is the usual four dimensional Higgs VEV. Note that we need to fine-tune
the bulk mass $\beta$ against the IR-brane localized mass term $v_{IR}$ in order to obtain $v_4 \ll 1 / R^{ \prime} \equiv M_{ \rm KK}$.

Next, we consider the fluctuations around the VEV, i.e., modes contained in the real component
of the neutral (but of course still complex) 5D Higgs field, equivalently, the tower of the CP-even Higgs
bosons. After plugging the parameterization of the field around the VEV in Eq.~\ref{eq:5Dvev:fixed} into the Lagrangian in Eq.~\ref{HiggsLag}
(and Eqs.~\ref{VUV} and \ref{VTeV} for the brane-localized potentials), the equation of
motion for a mode with $p^2=m^2$ is given by
\bea\label{eq:hEOM:aroundvev}
\l(z^3\d_z\frac{1}{z^3}\d_z+m^2-\frac{\mu^2}{z^2}\r)h=0~.
\eea
along with the boundary conditions at UV and IR branes:
\bea
\begin{split}
\d_ z h-\frac{2+\beta}{R} h &= 0~\quad {\rm for}\quad z=R~,\\
\d_z h+\frac{R}{R'}m_{TeV}h   &= 0~\quad {\rm for}\quad z=R'~,
\end{split}
\label{eq:hEOM:aroundvev:BC}
\eea
where the effective mass term at the IR brane is $m_{TeV} R=\lambda R^3 v^2(R')-(2+\beta)$.
The Higgs profile that solves Eq.~\ref{eq:hEOM:aroundvev} with the boundary conditions in Eq.~\ref{eq:hEOM:aroundvev:BC} is a basically Bessel function:
\bea
h(z)=A z^2 \l[
J_\beta(m z)+b_\beta(m) Y_\beta(mz)\r]~,
\label{Higgsprofile}
\eea
where the coefficient $b_\beta(m)$ is fixed by the boundary condition at the UV brane to be 
\bea
b_\beta(m)=-\frac{J_{\beta+1}(m R)}{Y_{\beta+1}(m R)}~.
\eea
The KK spectrum is determined by the boundary condition at the IR brane, that is
\bea
\label{neutralm}
R' m \l[J_{\beta+1}(m R')+b_\beta(m) Y_{\beta+1}(mR')\r]-\lambda R^3 v^2(R')\l[J_\beta(m R')+b_\beta(m) Y_{\beta}(mR')\r]=0~.
\eea

We can then study the approximate profiles and masses for $v_4 \ll 1 / R^{ \prime }=M_{ \rm KK}$. Setting 
$m^2 = 0$ in Eq.~\ref{eq:hEOM:aroundvev} and simply neglecting the $v^2$ term present in Eq.~\ref{eq:hEOM:aroundvev:BC}, we see that it is the same sets of equations as in Eq.~\ref{5DEOMHiggs} with $p^2 = 0$ (dropping consistently $v^2$ term in there as well). Thus, in
this approximation, we get a zero-mode Higgs boson, whose profile is the same as that of the VEV given in Eq.~\ref{eq:5Dvev:fixed}. Similarly, spectrum of KK Higgs bosons becomes (neglecting $v^2$ term in Eq.~\ref{neutralm}):
\bea
\label{chargedm}
\l[J_{\beta+1}(m R')+b_\beta(m) Y_{\beta+1}(mR')\r] = 0~.
\eea
Furthermore, in the limit when $m^{ (n) } R^{ \prime } \gg \beta$, i.e., the arguments of Bessel functions are much larger
than their indices, we can then approximate the Bessel functions by the trigonometric functions and the
following approximate relation for the KK spectrum can be derived:
\bea
m^{(n)} \simeq \l(\frac{3}{4}\pi +\pi n+\frac{1}{2}\beta \pi \r)\frac{1}{R'}~.
\label{KKHiggsmassexact}
\eea
Note that this is neglecting $O\left( v^2 / M^2_{ \rm KK } \right)$ effects. Even though, strictly speaking, the above formula
cannot be shown analytically to be valid for the Higgs modes with $n \lesssim\beta$ (recall these are the ones
relevant for the dipole calculation), we have checked numerically that it is actually also good enough for these masses.

Including the $O \left( v^2 \right)$ terms in the equations of motion above, it is straightforward to show that the
zero-mode is ``lifted", but it still remains (much) lighter than $M_{\rm KK}$. This is to be identified with
the SM-like 126 GeV Higgs boson, but clearly it now has a small admixture of the above KK Higgs
modes (and vice versa for the much heavier Higgs bosons). In particular, the SM Higgs profile
is then shifted from that of the VEV. We have checked (both
semi-analytically) and
numerically that the differences between these two profiles scale as $\sim m^2_H R^{ \prime \; 2 } / 
\beta$ \cite{Azatov:2009na} ($m_H$ is
the mass of the SM Higgs) which becomes negligible in the limit of an IR localized bulk Higgs.
Similarly, the masses of the heavy Higgs receive corrections from Eqs.~\ref{chargedm} and~\ref{KKHiggsmassexact}, i.e., we have to use
the exact Eq.~\ref{neutralm}. Comparing the equations Eq.~\ref{neutralm} and Eq.~\ref{chargedm}, we can see that their solutions will be
approximately the same and by expanding the Bessel functions for the small SM Higgs mass we can find that  the difference between the two mass eigenvalues  will scale as $ \Delta m^{ (n) } 
\sim m^2_H /(m^{(n)} \beta) $, which tends
to be zero for the heavy KK Higgs bosons.

We now move on to similar analysis of charged (and imaginary, neutral components of 5D Higgs field, i.e.,
CP-odd Higgs bosons). The equation of motion is the same as in Eq.~\ref{eq:hEOM:aroundvev}, except that the $v^2$ term is
absent in the last line. Setting $m^2 = 0$ here, it is easy to see that these are the same as Eq.~\ref{5DEOMHiggs} with $p^2 = 0$, at
all orders in $v^2$ (i.e, without having to neglect $v^2$ terms). Specifically, there is then a zero-mode for the charged (and imaginary, neutral) Higgs bosons. And,
clearly the KK modes masses and profiles satisfy Eqs.~\ref{chargedm},~\ref{KKHiggsmassexact} and~\ref{Higgsprofile} (again, no $v^2$ correction here, unlike
for real, neutral Higgs bosons above). So, both these properties for the charged and imaginary, neutral Higgs
bosons are similar to real, neutral Higgs modes, up to $O \left(
v^2 / M^2_{\rm KK } \right)$
effects.

All of the above discussion did not include EW gauging, which we now consider. At 
leading order in $v^2 / M^2_{\rm KK}$, it is zero-modes of the charged (and similarly imaginary, neutral) Higgs bosons which
are ``eaten" by the zero-mode $W/Z$ in order to become massive (but obviously they are still lighter than the
KK scale), whereas the KK $W/Z$ basically use $A_5$ (extra-dimensional component of gauge field) for this
purpose. 
However, at higher order, the zero and KK $W/Z$ modes undergo (mass) mixing, i.e., 
the longitudinal components of heavy $W/Z$
actually have (small)
admixtures of the above zero and KK charged (or imaginary, neutral) Higgs modes;
similarly the longitudinal components of the SM W/Z now contain bit of $A_5$.
In turn, the spectrum (and profiles) of
heavy charged (and imaginary, neutral) mass eigenstate are corrected by powers of $\left ( M^2_{ W/Z } / M^2_{\rm KK} \right )$ relative to Eqs.~(\ref{chargedm}) and (\ref{Higgsprofile}) in this process. On the other hand, the properties of the real, neutral unchanged
by this gauging.

Note that, in general, in our calculations of dipole operator, we are neglecting all the effects of
$O \left( v^2 / M^2_{ \rm KK } \right)$, including differences in the masses and profiles of the charged (or imaginary, neutral) Higgs bosons vs. real, neutral
Higgs bosons
(which is present even before the above gauging) and the shift from the gauging.


\section{Solutions for bulk fermions}
\label{fermionRS}

Next, we consider the 5D fermion.
The five dimensional action for the fermions is given by
\bea
S_{\rm Fermion} & = & \int d^4 x d z \l(\frac{R}{z}\r)^5 \Big [\frac{i}{2}\l(\bar Q \Gamma^A {\cal D}_A Q-{\cal D}_A \bar Q \Gamma^A Q \r)+\frac{c_q}{R}\bar{Q} Q +(Q,\; c_q \Leftrightarrow U,\; c_u \ {\rm and}\ D,\; c_u)  \nonumber \\
& & \hspace{3cm} + Y^{ u }_{5D} \bar Q H U + Y^{ d } _{5D} \bar Q H D \Big ]~.
\eea
Performing the KK decomposition for the field $Q=\sum Q^{(n)}(x) q^{(n)}(z)$ (again, neglecting the Higgs VEV
in 2nd line above) we get 
\bea
\begin{split}
-m^{(n)}q_L^{(n)}-{q_R^{(n)}} '+\frac{c_q+2}{z}q_R &=0~,\\
m^{(n)}q_R^{(n)}+{q_L^{(n)}} '+\frac{c_q-2}{z}q_L &=0~.\\
\end{split}
\eea
The solution is given by the Bessel functions
\bea
&&q_{L}^{(n)}=N_{n} z^{5/2}\l( J_{1/2+c_q}(m^{(n)} z)+b_n Y_{1/2+c_q}(m^{(n)} z)\r)~,
\label{correctprofile}
\eea
for the correct chirality
and
\bea
&&q_{R}^{(n)}=N_{n} z^{5/2}\l( J_{-1/2+c_q}(m^{(n)} z)+b_n Y_{-1/2+c_q}(m^{(n)} z)\r)~,
\label{wrongprofile2}
\eea
for the wrong chirality, 
where the coefficient $b_n$ is fixed by the boundary conditions; for example for the 
$q_L$ with $(++)$ boundary conditions (i.e., Neumann on both branes) it is given by 
\bea
b_n=-\frac{ J_{-1/2+c_q}(m^{(n)} z_{UV})}{ Y_{-1/2+c_q}(m^{(n)} z_{UV})}=-\frac{ J_{-1/2+c_q}(m^{(n)} z_{IR})}{ Y_{-1/2+c_q}(m^{(n)} z_{IR})}~.
\label{bn}
\eea
The KK masses, $m^{ (n) }$ are determined by solving the 2nd of Eq.~\ref{bn} and (for $ m^{ (n) } R^{\prime }
\gg1$) are given approximately by
\bea
m^{ (n) } & \approx & \pi \left( n - \frac{1}{2} + 2 \; c_q \right) \ds\frac{1}{R^{\prime }}~.
\label{KKfermionmass}
\eea
The normalization $N_n$  is fixed by requiring 
\bea
\int \l(\frac{R}{z}\r)^4|q_L|^2=\int \l(\frac{R}{z}\r)^4|q_R|^2=1~.
\eea
In the case where the fermion $q_L$  has $(++)$ boundary conditions,
there is a zero mode in the spectrum with profile given by:
\bea
\label{qL0}
q_L^0(z) &=& f(c_q)\frac{{R'}^{-\frac{1}{2}+c_q}}{ R^{2}} z^{2-c_q}~,
\eea
where the $f(c)=\sqrt{\frac{1-2c}{1-(R/R')^{1-2c}}}$ is proportional to the value of the wave function of the zero mode fermion at the IR brane. Similarly if the fermion $u_L$ (or $d_L$) has $(--)$ boundary conditions  then there will be a right handed zero mode with the profile,
\bea
\label{uR0}
u_R^0(z), d_R^0(z) &=& f(-c_u)\frac{{R'}^{-\frac{1}{2}-c_u}}{ R^{2}} z^{2+c_u \; \hbox{or} \; c_d}.
\eea
i.e., same as for $q_L$, but with $c_q \rightarrow - c_{ u, d }$
And, the corresponding KK profiles are given by 
Eqs.~\ref{correctprofile}, \ref{wrongprofile2} and \ref{bn}, with appropriate changes in $c$ parameters.


\section{Relevant couplings}
\label{app:couplings}

As already mentioned, the relevant couplings for our work are the Yukawa ones, i.e., between the Higgs and fermion modes (again, the gauge modes are not used here).
These are to be
obtained from the corresponding 5D Yukawa coupling, multiplied by overlap of profiles.
In turn, these couplings are of various types, depending on which the mode of the Higgs and fermion is involved here.
A ``master" formula was schematically given in Eq.~\ref{mastercoupling}
and estimates for individual couplings were also given in that section.
Here, we would like to present the exact formulae for each of these types of couplings, using the profiles from the earlier appendices.

We begin with the SM Yukawa coupling between the two fermion zero-modes and the SM Higgs. The up- and down-type SM Yukawa couplings are given by (we collectively call them $y_{SM}$ as in Eq.~\ref{SMYukapprox})
\bea
\label{SMYukexact}
\begin{split}
 y^u_{SM} &=\quad \frac{Y_{5D}^u}{v_{SM}}\int_R^{R'}dz\l(\frac{R}{z}\r)^5 u_R^{0}(z)v(z)q_L^{0}(z)~,\\
 y^d_{SM} &=\quad \frac{Y_{5D}^d}{v_{SM}}\int_R^{R'}dz\l(\frac{R}{z}\r)^5 d_R^{0}(z)v(z)q_L^{0}(z)~,
\end{split}
\eea 
where $v(z)$ is the VEV profile given in Eq.~\ref{eq:5Dvev:fixed} and $u_R^{0}(z)$, $d_R^{0}(z)$, $q_L^{0}(z)$ are the fermion zero-modes, given in Eqs.~\ref{qL0} and \ref{uR0}. In Eq.~\ref{SMYukexact}, we used the property that the SM Higgs profile is (approximately) the same as the profile of VEV. The difference between the Higgs zero-mode and the Higgs VEV profiles is almost marginal, roughly of order $\sim v^2 /(\beta\; M_{ \rm KK }^2)$.

Next,  we consider the coupling between one SM, one KK fermion and a Higgs (either the SM Higgs for $n_H = 0$ or KK Higgs modes for $n_{H} \geq 1$). For the Yukawa coupling with the KK Higgs, $n_H \geq 1$ we get
\bea
\label{SMKKYukexact1}
y^{d}_{(0,\; n_F,\; n_H)}=  Y_{5D}^d\int_R^{R'}dz\l(\frac{R}{z}\r)^5 q_L^{0}(z)h^{(n_H)}(z)d_R^{(n_F)}(z)~,
\eea
and
 \bea
 \label{SMKKYukexact2}
 y^{u}_{(n_F,\; 0,\; n_H)}= Y_{5D}^u\int_R^{R'}dz\l(\frac{R}{z}\r)^5 q_L^{(n_F)}(z) h^{(n_H)}(z) u_R^{0}(z)~.
 \eea
where the wavefunctions, $q_L^{(n_F) }(z)$ and $d_R^{(n_F)}(z)$ with the correct chirality, are found in Eqs.~\ref{correctprofile} and the KK Higgs wave function $h^{(n_H)}(z)$ is given in Eq.~\ref{Higgsprofile}. Similarly for the Yukawa couplings involving $u_R^{(n_F)}(z)$. For the SM Higgs field, corresponding to $n_H=0$, one can replace the KK Higgs wave function in Eqs.~\ref{SMKKYukexact1},~\ref{SMKKYukexact2} with the Higgs VEV profile, namely $v(z)/v_{SM}$ and they are denoted by $y^{d}_{(0,\; n_F,\; 0)}$ and $ y^{u}_{(n_F,\; 0,\; 0)}$. The Yukawa couplings with the Higgs VEV insertion involving one SM and one KK fermion are same as those for the SM Higgs with SM- and KK fermion, and they are 
\bea
\label{SMKKYukexactlight}
\begin{split}
y^{d}_{(0,\; n_F,\; v)} =\frac{Y_{5D}^d}{v_{SM}}\int_R^{R'}dz\l(\frac{R}{z}\r)^5 q_L^{0}(z)v(z)d_R^{(n_F)}(z)~,\\
y^{u}_{(n_F,\; 0,\; v)} =\frac{Y_{5D}^u}{v_{SM}}\int_R^{R'}dz\l(\frac{R}{z}\r)^5 q_L^{(n_F)}(z)v(z) u_R^{0}(z)~.
\end{split}
\eea
Even though (as already mentioned) the profiles of the SM Higgs and its VEV are almost identical (and thus so are the above two sets of overlaps integrals), we will still differentiate between the Higgs VEV insertions and the Yukawa couplings in order to make explicit the correspondence with the Feynman diagrams.
The NDA-type estimates for these couplings were given in Eqs.~\ref{SMKKYukapproxlight} and \ref{SMKKYukapproxheavy}.

Finally, we present the couplings of two KK fermions to Higgs (whether SM or KK), where we have to distinguish between wrong $(-)$ and correct $(+)$ chiralities. For KK Higgs modes with $n_H \geq 1$ they are given by
\bea
\label{KKYukexact}
\begin{split}
 y^{d,\, -}_{(n_{F_1},\; n_{F_2},\; n_H)} &=\quad Y_{5D}^d\int_R^{R'}dz\l(\frac{R}{z}\r)^5 q_R^{(n_{F_1})}(z)h^{(n_H)}(z)d_L^{(n_{F_2})}(z)~,\\
 y^{d,\, +}_{(n_{F_1},\; n_{F_2},\; n_H)} &=\quad Y_{5D}^d\int_R^{R'}dz\l(\frac{R}{z}\r)^5 q_L^{(n_{F_1})}(z)h^{(n_H)}(z)d_R^{(n_{F_2})}(z)~,
\end{split}
\eea
where the KK fermion profile with the wrong chirality, $q_R^{(n_{F_1})}(z)$ (similarly for $d_L^{(n_{F_2})}(z)$), is given in Eq.~\ref{wrongprofile2}. As before, we replace the KK Higgs wave function in Eq.~\ref{KKYukexact} with Higgs VEV profile $v(z)/v_{SM}$ for the case of the Higgs VEV insertions (or SM Higgs for $n_H=0$) involving two KK fermions. They are denoted by $y^{d,\, \pm}_{(n_{F_1},\; n_{F_2},\; 0\ {\rm or}\ v)}$. These correspond to the our NDA-type estimates in Eqs.~\ref{KKYukapproxcorrectlight}, \ref{KKYukapproxwronglight} and \ref{KKYukapproxheavy}.
%

%
Let us now look now at the brane-localized limit of the bulk Higgs, i.e., $\beta \gg 1$. In this limit the SM Yukawa coupling can be expressed in terms of the five dimensional parameters as
\begin{eqnarray}
y_{SM}^u=\frac{\sqrt{2(1+\beta)}}{(2-c_q+c_u+\beta)}\frac{Y^u_{5D}}{\sqrt{R}} f(c_q)f(-c_u)~,
\end{eqnarray}
where the $f(c)$ is defined in Eq.~\ref{fc}.
For a fixed $Y_{5D}^u$ (5D Yukawa coupling), we can see that  in the $\beta\ra \infty$ limit the SM coupling will have an additional suppression of $1/\sqrt{1+\beta}$.(One can understand this as originating from the different normalizations of the bulk and brane Higgs VEVs.) One way to cure this behavior (i.e., in order to have well defined ``brane" limit of the bulk Higgs for fixed 
fermion profiles) is to rescale 5Dl Yukawa coupling with the factor $\sim\sqrt{\beta}$. Nonetheless, in our calculations we keep this additional rescaling factor explicit.


\section{Loop functions}
\label{app:loopfunction}

\begin{figure}
\begin{center}
\includegraphics[scale=0.6]{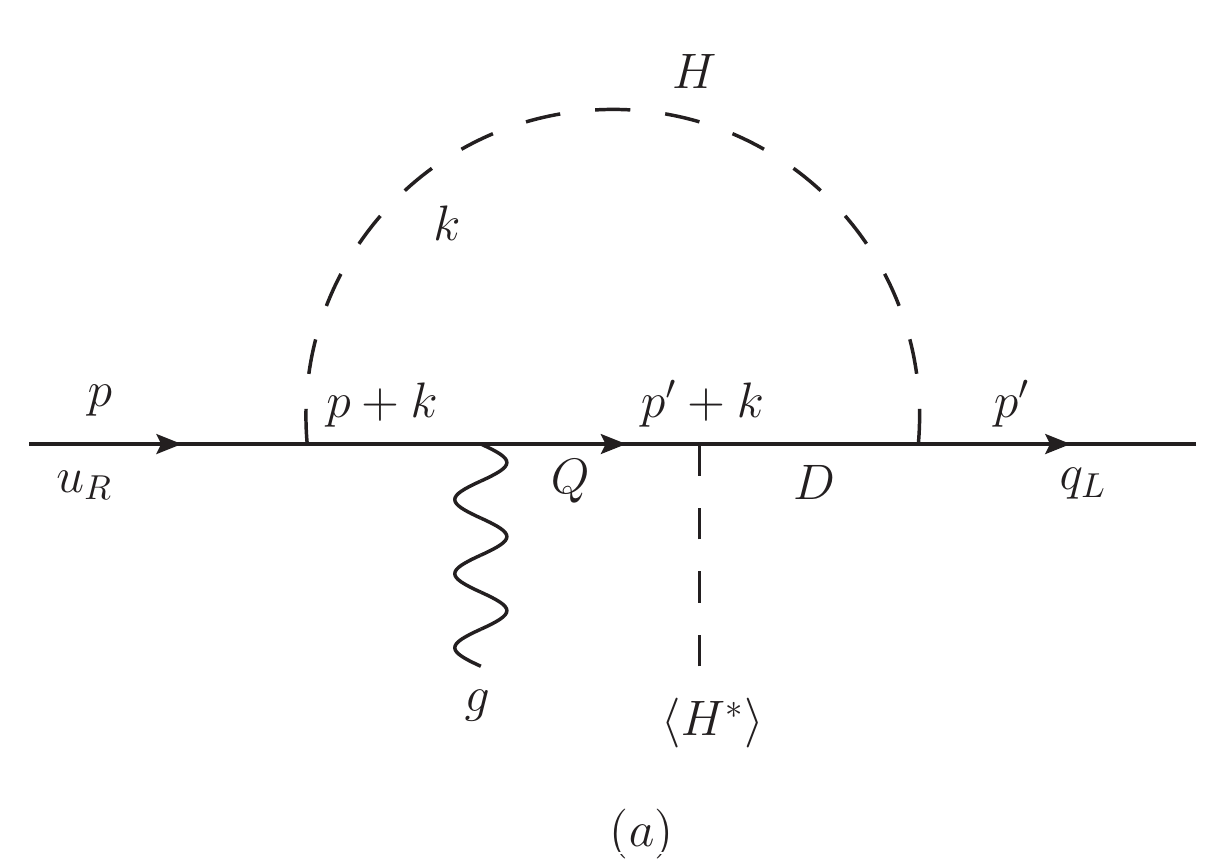}\quad
\includegraphics[scale=0.6]{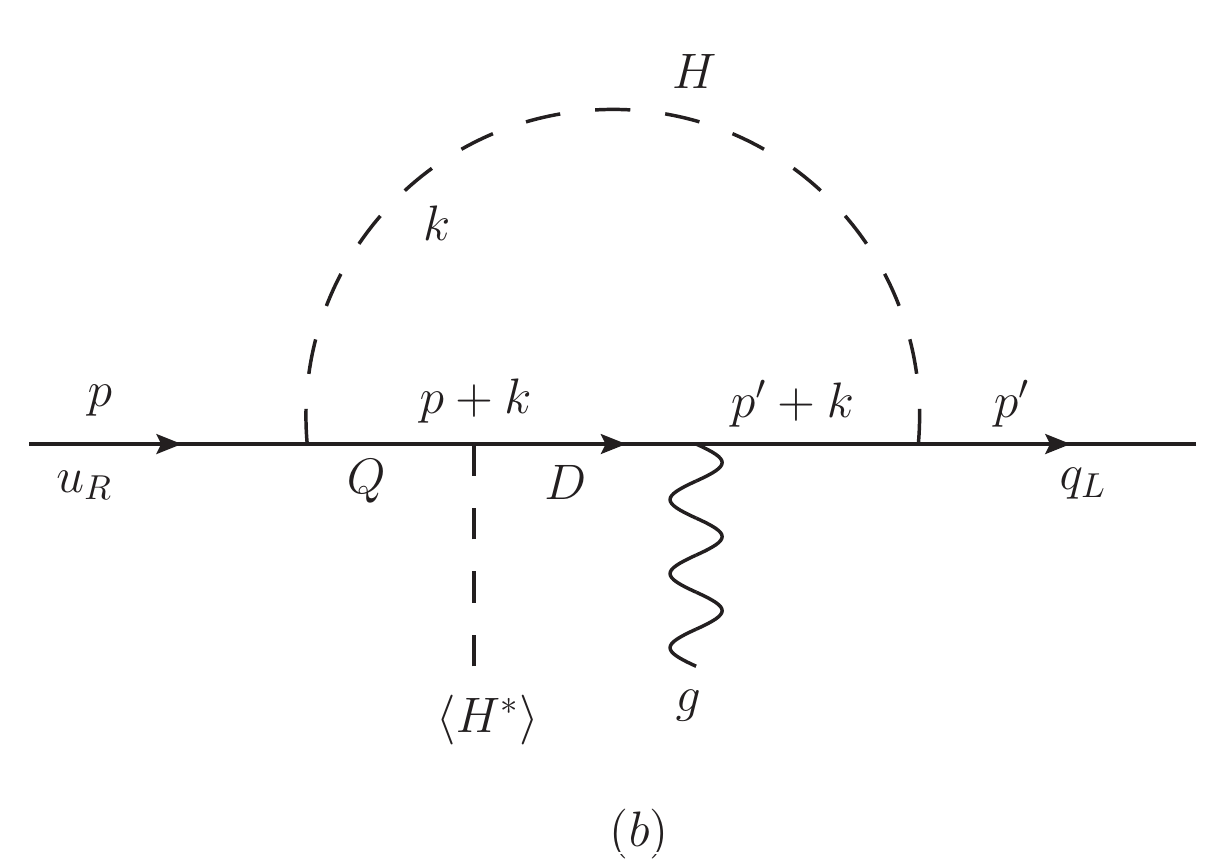}
\caption{The diagrams contributing to the dipole operators (for SM up-type quark) in the weak basis with Higgs VEV attached to the internal quark-lines only, i.e., inside loop. The internal quark lines (here, all are inside the loop) correspond to the heavy KK fermions whereas $H$ line in the loop can be either SM light Higgs or heavy KK Higgs modes.}
\label{fig:insertioninside}
\end{center}
\end{figure}

Recall that in the simplified model calculation of Section~\ref{toy}, the result for the dipole operator coefficient (for up-type quark) was given in terms of certain loop functions. In this section we provide the complete derivation of these loop functions in the weak basis: this is essentially a cross-check of a similar calculation done in appendix A of \cite{Delaunay:2012cz}, but we still show it for the sake of completeness. We sub-divide them according to whether the Higgs VEV is attached inside the loop (i.e., to internal quark lines only) or to the external quark line (i.e., outside the loop). 

We begin with the Higgs VEV attached to the internal quark lines only (i.e., inside the loop). There are actually two diagrams shown in Fig.~\ref{fig:insertioninside} (labelled $a$ and $b$), depending on which side of the Higgs VEV is the gluon attached to. As noted earlier (see discussion just above Eq.~\ref{inside_general}), we consider only the charged Higgs contribution here, since the neutral sector has a cancellation. Thus, these diagrams have only down-type quarks in the loop (again, incoming on-shell lines are up-type quark) and are proportional to down-type Yukawa coupling (in addition to up-type Yukawa coupling).
The corresponding amplitudes are given by
\beq
\begin{split} \label{eq:amp:insertioninside}
%
%
 i{\mathcal M}_a &= \int \frac{d^4k}{(2\pi)^4} \bar{u}(p') (i y^d_{\rm SM\ KK} P_R )\frac{i(\slashed{k}+\slashed{p'}+M_D)}{(k+p')^2-M^2_D} \left (i y^{{\rm light},\, d+\; *}_{\rm KK\ KK} P_L + i y^{{\rm light},\, d-\; *}_{\rm KK\ KK} P_R\right )\\
&\times \frac{i(\slashed{k}+\slashed{p'}+M_Q)}{(k+p')^2-M^2_Q} \left ( i g_s T^a \slashed{\epsilon}^* \right ) \frac{i(\slashed{k}+\slashed{p}+M_Q)}{(k+p)^2-M^2_Q} (i y^u_{\rm SM\ KK} P_R)u(p)\frac{i}{k^2-M_H^2}~, \\
%
%
 i{\mathcal M}_b &= \int \frac{d^4k}{(2\pi)^4} \bar{u}(p') (i y^d_{\rm SM\ KK} P_R )\frac{i(\slashed{k}+\slashed{p'}+M_D)}{(k+p')^2-M^2_D}  \left ( i g_s T^s \slashed{\epsilon}^* \right ) \frac{i(\slashed{k}+\slashed{p}+M_D)}{(k+p)^2-M^2_D} \\
&\times \left (i y^{{\rm light},\, d+\; *}_{\rm KK\ KK} P_L + i y^{{\rm light},\, d-\; *}_{\rm KK\ KK} P_R\right ) \frac{i(\slashed{k}+\slashed{p}+M_Q)}{(k+p)^2-M^2_Q} (i y^u_{\rm SM\ KK} P_R)u(p)\frac{i}{k^2-M_H^2}~,
\end{split}
\eeq
where $P_{L/R}$ are the projection operators of each chirality fermion and the Yukawa couplings to both the light Higgs and the heavy KK Higgses are allowed unless explicitly specified, for instance, $y^d_{\rm SM\ KK} = \left \{ y^{{\rm light},\, d}_{\rm SM\ KK},\, y^{{\rm heavy},\, d}_{\rm SM\ KK} \right \}$ (similarly for other types of Yukawa couplings as well). The $\epsilon_\mu$ denotes the polarization four vector of the gluon. 
\begin{figure}
\begin{center}
\includegraphics[scale=0.6]{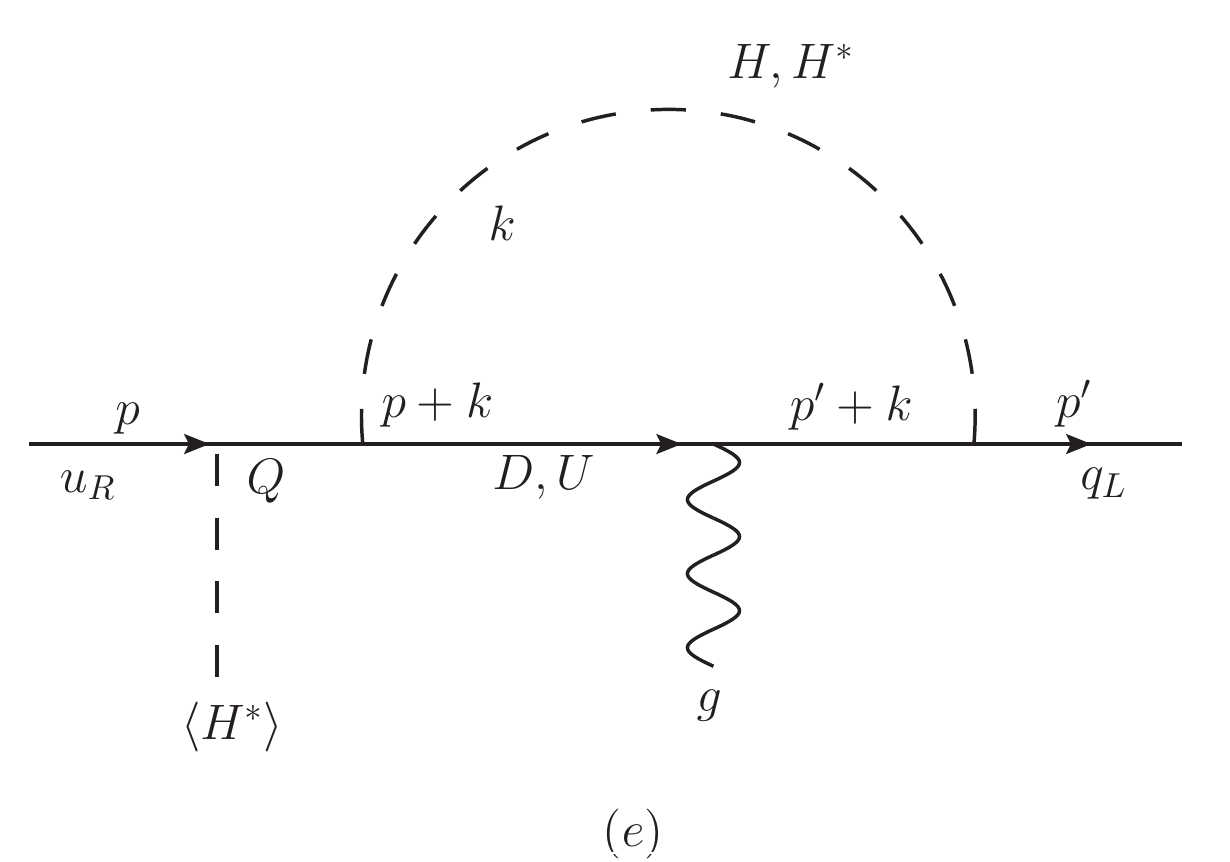}\quad
\includegraphics[scale=0.6]{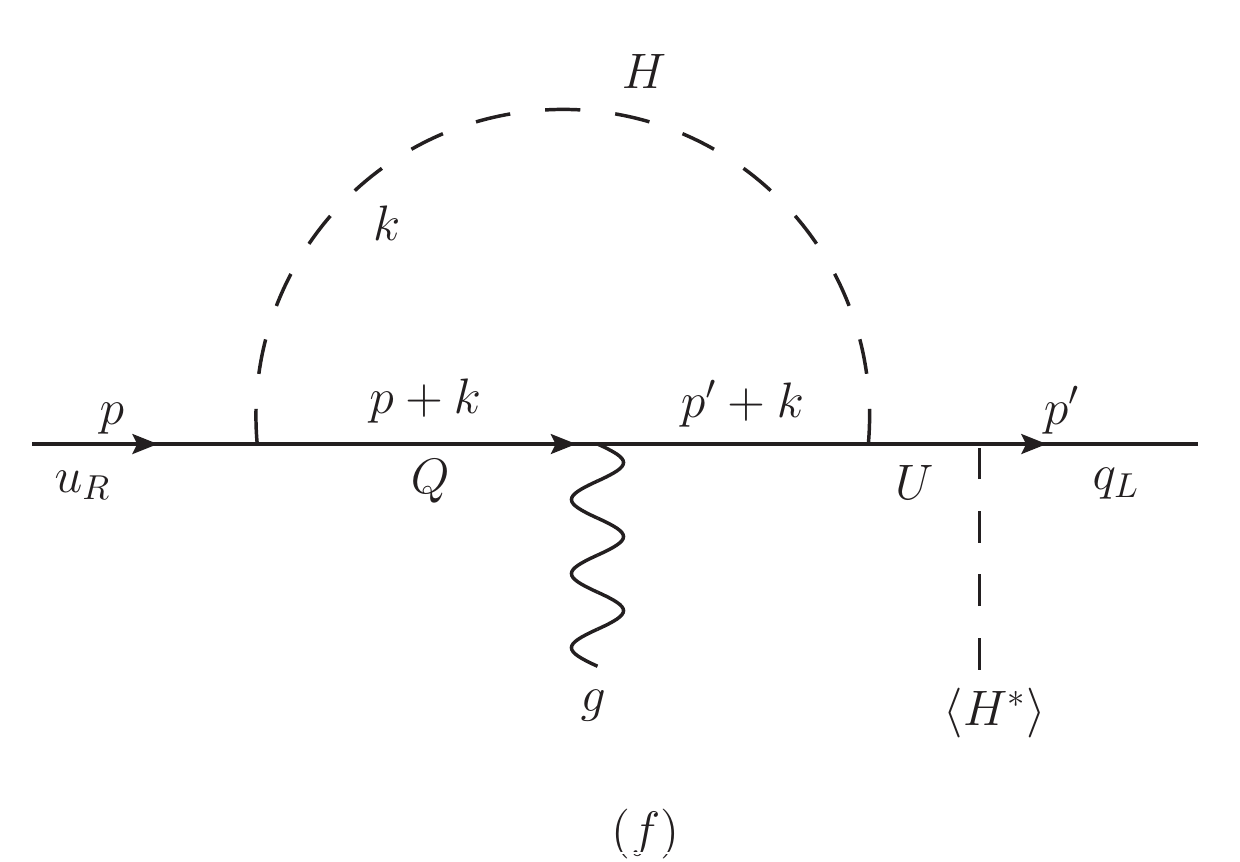}
\caption{The diagrams contributing to the dipole operator (for SM up-type quark) in the weak basis with the Higgs VEV attached to the external quark (on-shell) line, i.e., outside the loop. The internal quark lines (including the one between VEV insertion and the loop) correspond to the heavy KK fermions whereas $H$ line in the loop can be either the SM light Higgs or heavy KK Higgs modes.}
\label{fig:insertionoutside}
\end{center}
\end{figure}
Note that these amplitudes simultaneously include two distinctive contributions from the correct (denoted by superscript ``$+$'') Yukawa and the wrong Yukawa couplings (denoted by ``$-$'').

On the other hand, the diagrams with the Higgs VEV attached to the external quark line (i.e., outside the loop) are shown in Fig.~\ref{fig:insertionoutside}.
Once again, there are two types (labelled $e$ and $f$ for notational clarity), but now depending on whether the Higgs VEV is on the incoming (i.e., on-shell)
$u_R$ or $u_L$ line.
In the first case, we can easily work out that the heavy quarks inside the loop are $SU(2)_L $ singlets and 
can be either down or up-type, corresponding to the Higgs inside loop being charged or neutral and the Yukawa couplings involved being both up and down-type or only up-type.
In particular, the neutral Higgs contribution does not encounter the cancellation (unlike
for insertion inside the loop mentioned above).
Whereas, the second diagram involves $SU(2)_L$ doublet quark inside loop, i.e., either up and down-type here (along with neutral or charged Higgs), but involves only up-type Yukawa couplings.
The corresponding amplitudes are
\beq
\begin{split}\label{eq:amp:insertionoutside}
%
%
 i{\mathcal M}_e &= \int \frac{d^4k}{(2\pi)^4} \bar{u}(p') (i y^{d,u}_{\rm SM\ KK} P_R) \frac{i(\slashed{k}+\slashed{p'}+M_{D,U})}{{k+p'}^2-M^2_{D,U}} \left ( i g_s T^a \slashed{\epsilon}^* \right ) \frac{i(\slashed{k}+\slashed{p}+M_{D,U})}{(k+p)^2-M^2_{D,U}}  \\
&\times \left (i y^{d +,u +\; *}_{\rm KK\ KK} P_L + i y^{d-,u-\; *}_{\rm KK\ KK} P_R\right )  \frac{i(\slashed{p}+M_Q)}{p^2-M^2_Q} (i y^{{\rm light},\, u}_{\rm SM\ KK} P_R)u(p)\frac{i}{k^2-M_H^2}~,\\
%
%
 i{\mathcal M}_f &= \int \frac{d^4k}{(2\pi)^4} \bar{u}(p') (i y^{{\rm light},\, u}_{\rm SM\ KK} P_R) \frac{i(\slashed{p'}+M_U)}{{p'}^2-M^2_U} \left (i y^{u +\; *}_{\rm KK\ KK} P_L + i y^{u-\; *}_{\rm KK\ KK} P_R\right ) \\
&\times \frac{i(\slashed{k}+\slashed{p'}+M_Q)}{(k+p')^2-M^2_Q} \left ( i g_s T^a \slashed{\epsilon}^* \right ) \frac{i(\slashed{k}+\slashed{p}+M_Q)}{(k+p)^2-M^2_Q} (i y^{u}_{\rm SM\ KK} P_R)u(p)\frac{i}{k^2-M_H^2}~.
\end{split}
\eeq
Once again, in principle, this includes both the correct and wrong chirality coupling contributions. In practice, only the wrong chirality is significant in this case, since as mentioned earlier (see discussion
just above Eq.~\ref{outside_general}), the correct chirality is suppressed due to the form of (heavy) quark
propagator in-between Higgs VEV insertion and the loop.
Note that  for simplicity in Eqs.~\ref{eq:amp:insertioninside},~\ref{eq:amp:insertionoutside}, the indices for the KK mode numbers in the Yukawa couplings and masses are not shown.
%

%
%
Based on the Dirac structure and the various types of Yukawa couplings involved, the amplitudes in Eqs.~\ref{eq:amp:insertioninside} and
\ref{eq:amp:insertionoutside} can be clearly decomposed as
%
%
\beq
\begin{split} \label{eq:amp:insertion:ab:secondstep}
i {\mathcal M}_{a+b} =&\
-\frac{i}{8 \pi^2}\, \left [ y^{d}_{\rm SM\ KK}\, y^{{\rm light},\, d\; *}_{\rm KK\ KK}\, y^{u}_{\rm SM\ KK}\ (I^{\rm int}_a+I^{\rm int}_b) + y^{d}_{\rm SM\ KK}\, y^{{\rm light},\, d-\; *}_{\rm KK\ KK}\, y^{u}_{\rm SM\ KK}\ (J^{\rm int}_a+J^{\rm int}_b) \right ]\\
& \hspace{1.5cm} \times \bar u(p')\, g_s T^a (\epsilon^*\cdot p)P_R u(p)~ +...,\\
\end{split}
\eeq
and 
\beq
\begin{split} \label{eq:amp:insertion:ef:secondstep}
i {\mathcal M}_{e+f} =&\
-\frac{i}{8 \pi^2}\, \left [ y^{d}_{\rm SM\ KK}\, y^{d-\; *}_{\rm KK\ KK}\, y^{{\rm light},\, u}_{\rm SM\ KK}\ J^{\rm ext}_{d,e} + y^{u}_{\rm SM\ KK}\, y^{u-\; *}_{\rm KK\ KK}\, y^{{\rm light},\, u}_{\rm SM\ KK}\ J^{\rm ext}_{u,e} \right . \\
 & \left . \hspace{3.5cm} +\ 2\ y^{{\rm light},\, u}_{\rm SM\ KK}\, y^{u-\; *}_{\rm KK\ KK}\, y^{u}_{\rm SM\ KK}\ J^{\rm ext}_{u,f} \right ]\   \bar u(p')\, g_s T^a (\epsilon^*\cdot p)P_R u(p)~ +...,\\
\end{split}
\eeq
where ``..." in above amplitudes denotes other Dirac structures, namely, $\gamma_{ \mu }$-type. The $J/I$'s are loop functions (whose actual expressions are given below).
The familar form of the dipole amplitude can then be obtained by two replacements, $(\epsilon^*\cdot p) \rightarrow i/2\ (\sigma^{\mu\nu}\epsilon^*_\mu q_\nu)$ and  $i\, \sigma^{\mu\nu} \epsilon^*_\mu q_\nu \rightarrow -(1/2)\, \sigma^{\mu\nu}G_{\mu\nu}$ (the gluon polarization four vector replaced with $G_\mu$). 
The factor of 2 in the third term of Eq.~\ref{eq:amp:insertion:ef:secondstep} is due to $SU(2)$ multiplicity of the doublet KK fermion $Q$ inside the loop (see diagram $(f)$ in Fig.~\ref{fig:insertionoutside}), in turn, corresponding to that of the Higgs bosons (i.e., charged and neutral) in the same loop. It is absent in the case of diagram $(e)$ (1st and 2nd terms above), where only one type of quark (up or down $SU(2)$ singlet)
propagates in the loop, which is 
accompanied 
by either charged Higgs or neutral Higgs.
Matching the above form of the amplitudes to the Lagrangian shown in Eq.~\ref{chromodipoledef} gives us the coefficient of the dipole operator ($C_{ \rm dipole }$).

This is how we obtained Eqs.~\ref{inside_general} and \ref{outside_general} in Section~\ref{toy}.
In other words, the various loop functions ($J$'s and $I$'s) appearing in Eqs.~\ref{inside_general} and \ref{outside_general} --
which were not specified in that section -- simply correspond to the relevant parts of the amplitudes
in Eqs.~\ref{eq:amp:insertioninside} and~\ref{eq:amp:insertionoutside}, respectively. \\

In more detail,
we  first introduce Feynman parameters in order to combine the denominators of the propagators in Eqs.~\ref{eq:amp:insertioninside} and 
\ref{eq:amp:insertionoutside}. Then,
we perform the loop momentum integrals, thus leaving the loop functions as
integrations over the Feynman parameters.
%
%
In this way,
the loop functions for insertion inside lthe oop (i.e., 
from Eq.~\ref{eq:amp:insertioninside}) with the wrong chirality couplings (i.e., 
$y^-$ terms) are given by  
\begin{eqnarray}
J_a^{ \rm int } & = & M_Q M_D \int_{0}^{1} d x  \int_{0}^{ 1 - x } d y  \int_{0}^{ 1 - x - y } d z \frac{ ( x + y + z ) }{ \Delta^2}~,
\end{eqnarray}
where the denominator is given by 
\begin{eqnarray}
\Delta & = & M_Q^2 ( y + z ) + x M_D^2 + M_H^2 ( 1 - x - y - z )~.  
\end{eqnarray}
On the other hand, for the correct chirality contribution (with insertion still being inside loop), i.e., considering
terms without $y^-$, we get
\begin{eqnarray}
I_a^{ \rm int } & = & \int_{0}^{1} d x  \int_{0}^{ 1 - x } d y  \int_{0}^{ 1 - x - y } d z \Bigg[  \frac{1 - 3 ( x + y ) }{ \Delta } +  \frac{ z M_Q^2 }{ \Delta^2 } \Bigg]~.
\label{Ia}
\end{eqnarray}
The $I_b$, $J_b$ are easily obtained from $I_a$ and $J_a$ by replacing $\left\{ x, M_D \right\} \leftrightarrow \left\{ z, M_Q \right\}$.

Similarly, we can obtain the expressions for the loop functions for insertion outside loop, i.e., from eq.~\ref{eq:amp:insertionoutside} 
(again, only wrong chirality
contributions, i.e., terms with $y^-$, are significant here). 
We get from the contributions involving (only) up-type Yukawa couplings:
\begin{eqnarray}
J^{ \rm ext }_{ u, \; e } & = & \frac{ M_U }{ M_Q } \int^{1}_{0} d x \int^{ 1 - x }_{0} d y \frac{ ( x + y ) }{ \Delta_U }~,
\end{eqnarray}
and
\begin{eqnarray}
J^{ \rm ext }_{ u, \; f } & = & \frac{ M_Q }{ M_U } \int^{1}_{0} d x \int^{ 1 - x }_{0} d y \frac{ ( x + y ) }{ \Delta_Q }~,
\end{eqnarray}
where the denominators are given by
\begin{eqnarray}
\Delta_X & = & M_X^2 ( x + y ) + M_H^2 ( 1 - x - y )~.
\end{eqnarray}
Finally, the loop function involving the down type Yukawa coupling (only from part of the first amplitude in Eq.~\ref{eq:amp:insertionoutside}, corresponding to the down-type heavy quark in the loop) is simply $J^{ \rm ext }_{ d, \; e } = J^{ \rm ext }_{ u, \; e }$ with $M_U \rightarrow M_D$. In our study, we focus only on the contribution involving both up and down-type quark Yukawa couplings and thus do not need to consider the loop function $J^{ \rm ext }_{ u \; e \; \hbox{or} \; f }$.
The integrations over Feynman parameters, are straightforward, giving
\bea
\label{eq:outside}
J^{\rm ext}_{d,e}=\frac{M_D\left( M_D^4-4 M_D^2 M_H^2 + 4 M_H^4 \log
   \left(M_D/M_H\right)+3 M_H^4\right)}
   {2 M_Q \left(M_D^2-M_H^2\right)^3}~,
\eea
which corresponds to the 1st diagram in the Fig \ref{fig:insertionoutside} with the $D$ quark in the loop. 

For the diagrams with the insertion inside the loop we get the following: for the correct chirality
\bea
\begin{split}\label{eq:inside:correct}
I^{\rm int}_a+I^{\rm int}_b =&
\ \Bigg\{ 4 M_H^2 M_Q^2(M_H^2-M_D^2)^3 \log \left(M_Q/M_D \right)\\
&\ +(M_Q^2-M_D^2)\bigg [ (M_H^2-M_Q^2) \Big (3 M_H^6 - (4 M_D^2 + M_Q^2) M_H^4 + M_D^4 M_H^2 + M_D^4 M_Q^2 \Big ) \\
&\  + 4 M_H^2 \Big (M_H^6 -3 M_Q^2 M_D^2 M_H^2 + M_D^4 M_Q^2 + M_D^2 M_Q^4 \Big ) \log \left ( M_D/M_H \right ) \bigg ] \Bigg\}\\
&\times \frac{M_H^2}{2 ( M_Q^2-M_D^2 ) ( M_H^2-M_Q^2 )^3 ( M_H^2-M_D^2 )^3}~,
\end{split}
\eea
and, for the wrong chirality
\bea
\begin{split}\label{eq:inside:wrong}
&J^{\rm int}_a+J^{\rm int}_b \\
&\hspace{0.5cm}= \ \Bigg\{
 \ M_Q M_D \left (M_Q^2-M_D^2 \right ) \left (M_H^2-M_Q^2 \right ) \left (M_H^2-M_D^2 \right )\bigg [ 5 M_H^4 - 3 M_H^2 \left ( M_Q^2+M_D^2 \right )+M_Q^2 M_D^2 \bigg ] \\
&\hspace{1cm}\ + 4\, M_H^4 M_Q M_D \bigg [ 3 M_H^4 \left ( M_Q^2-M_D^2 \right ) - 3 M_H^2 \left ( M_Q^4-M_D^4 \right ) +  M_Q^6-M_D^6 \bigg ] \log \left (M_D/M_H \right ) \\
&\hspace{1cm}\ - 4\, M_H^4 M_Q M_D \left ( M_H^2-M_D^2 \right )^3 \log \left (M_D/M_Q \right ) \Bigg\} \\
&\hspace{1cm}\ \times \frac{1}{2 (M_Q^2-M_D^2) ( M_H^2-M_Q^2 )^3 (M_H^2-M_D^2)^3}~.
\end{split}
\eea
The above formulae in Eqs.~\ref{eq:outside},~\ref{eq:inside:correct},~\ref{eq:inside:wrong} were not explicitly given in \cite{Delaunay:2012cz}. We see explicitly in Eq.~\ref{eq:inside:correct} that the loop function involving the correct chirality is proportional to $M_H^2$. It causes the suppression by $\sim (m_h/M_{KK})^2$ for the SM Higgs loop where $M_H$ corresponds to the SM Higgs mass, $m_h$. While we do not show the loop functions of $I_a$ and $I_b$ separately, we emphasize that the individual loop function $I_a$ (similary $I_b$) is proportional to $M^2_H$.

In the light Higgs limit where we take $M_H \rightarrow 0$ (as in Section~\ref{toylight}), we get
\bea
 J^{\rm ext}_{d,e} =\frac{1}{2 M_D M_Q}~,
\eea
and 
\bea
I^{\rm int}_a+I^{\rm int}_b =-\frac{M_H^2}{2 M_Q^2 M_D^2}~,\quad J^{\rm int}_a+J^{\rm int}_b =\frac{1}{2 M_Q M_D}~,\\
\eea
On the other hand, for the case of universal KK masses, namely $M_Q=M_D=M_H=M_{KK}$ (this choice was made in Section~\ref{sec:toyheavy}) we get
\bea
J^{\rm ext}_{d,e}=\frac{1}{3 M_{KK}^2}~,\quad J^{\rm int}_a+J^{\rm int}_b=\frac{1}{4 M_{KK}^2}~,\quad I^{\rm int}_a+I^{\rm int}_b=-\frac{1}{12 M_{KK}^2}~.
\eea
The above limiting values were also mentioned in Section~\ref{toy}.

\section{Cutoff Contribution: Comparison to Models with (strictly) Brane-Localized Higgs }
\label{cornell}

Here, we re-consider some of the above dipole effects from the Higgs boson modes, but from a somewhat different angle. As mentioned in the introduction, the references in \cite{Csaki:2010aj} start and stay with the strictly brane-localized ({\em aka} $\delta$-function) Higgs and thus they only have the correct chirality coupling at disposal in the loop diagrams. There is really no KK Higgses (or they are infinitely heavy) in this case and only the SM Higgs appears in loop.
As discussed in Section~\ref{correctNDA}, if we set the 4D loop momentum cutoff $\rightarrow \infty$, i.e.,
$M_{ \rm KK}$ is the only scale in the loop, then such a contribution vanishes (again, we neglect effects suppressed by $\sim m_{h }^2 / M_{ \rm KK }^2$). Nonetheless, these references showed that respecting 5D covariance implies that we do get a sizeable contribution (as follows) to the dipole operator.
Their point is that in the KK approach taken by earlier literature one should ``coordinate'' the 4D loop momentum cutoff with the KK sum cutoff, denoting both by an appropriately warped-down of $\Lambda$, the scale at which 5D effective field theory (EFT) description breaks down and the physics of UV completion of 5D model comes in.\footnote{For example, in order to accomplish this in a 5D gauge-invariant way, one could use the 5D Pauli-Villars (PV) regularization. In this case, 4D modes of the PV field (with 5D mass $\sim \Lambda$) will also appear in loops. }
Therefore, for a finite yet large $\Lambda$ and for fixed KK levels ($n, \; p \lesssim\Lambda / k$), we 
expect to 
get instead%
\begin{eqnarray}
\begin{split}
\l (C^{ \rm covariant}_{ \rm dipole } \r )_{(n,\; p)} & \sim \quad \frac{ M_{ \rm KK }^2 }{ \Lambda^2 } \frac{ \left( y^{ \rm light}_{ \rm SM \; KK } \right)^2 y^{ \rm light, + }_{ \rm KK \; KK } }{ y_{ \rm SM } }~, \\
& \sim \quad \frac{  M_{ \rm KK }^2 }{ \Lambda^2 } y_{ \rm KK }^2~,
\end{split}
\end{eqnarray}
i.e., schematically, we get an additional contribution where it is the finite 4D loop momentum cutoff (being larger than most of the KK masses) that sets the mass scale for the loop integral.
Strictly speaking, we have not shown that one actually gets such a term, but just that it is allowed: for example, at the least, it matches the earlier finding of zero contribution
for $\Lambda \rightarrow \infty$.
(See Section 6.6 of the 1st reference in~\cite{Csaki:2010aj} and Appendix D of the 2nd reference therein for more discussion
about this issue: the effect we sketch here is similar in spirit)

For a fixed KK level, the contribution seems suppressed by the cutoff $\Lambda$, but the crucial 
point 
is that upon KK double sum (again, up to the same $\sim \Lambda$), we get 
\begin{eqnarray}
\begin{split}
C^{ \rm covariant}_{ \rm dipole } & \sim \quad  \frac{  M_{ \rm KK }^2 }{ \Lambda^2 } \sum_{ n, p = 1 }^{ n, p \sim \frac{ \Lambda }{k} } y_{ \rm KK }^2~, \\
& \sim \quad y_{ \rm KK }^2~. 
\end{split}
\label{braneCornell}
\end{eqnarray}
Note that the brane-localized Higgs coupling does not conserve KK number, no matter how high, as expected
from the $\beta \rightarrow \infty$ limit of the bulk Higgs, and consequently the double-sum persists up to the cutoff $\Lambda$.
The subtlety is that the above effect is 
missed completely if one first takes the 4D loop momentum cutoff to infinity for a fixed KK level and only
%
%
try do KK sum afterwards, because the latter cannot catch-up with 4D loop cutoff, given this order of operations!
Alternatively, one can use the 5D propagators, as was done for the actual calculation in the references \cite{Csaki:2010aj}, which
of course should be equivalent (since 5D propagators entail an implicit sum over KK modes after all).
In passing, let us note that even though this contribution is finite (i.e., $\Lambda$'s cancel above), it is still UV-sensitive because KK modes all the way up to the cutoff 
are relevant. 
This is perhaps not surprising given the NDA estimate of the dipole operator being UV-sensitive for a brane-localized Higgs.

The above cutoff contribution is present even for a Higgs field being in the bulk: here we focus on the case where only the SM Higgs (i.e., not the KK Higgs) and the KK fermions with the correct chirality appear in the loops, since that
effect seems suppressed if the 4D loop momentum cutoff $\rightarrow \infty$ to start with, i.e., there is a chance of missed, subtle effect here as before.
However, as discussed earlier, the KK fermions with mode-numbers above $\sim \beta$, i.e., oscillate within the SM Higgs width, thus respect
KK number conservation and result in a suppressed, single sum.
As a result, the double sum which appeared in Eq.~\ref{braneCornell} for this case 
is instead effectively cutoff by the inverse of the Higgs width:
\begin{eqnarray}
\begin{split}
C^{ \rm covariant, \; bulk }_{ \rm dipole } & \sim \quad   \frac{  M_{ \rm KK }^2 }{ \Lambda^2 } \sum_{ n, p = 1 }^{ n, p \sim \beta } y_{ \rm KK }^2~, \\
& \sim \quad y_{ \rm KK }^2 \left( \frac{ \beta \; M_{ \rm KK } }{ \Lambda } \right)^2~.
\end{split}
\label{bulkCornell}
\end{eqnarray}
Note that this truncation of KK sum happens automatically, that too within 5D EFT, for the case of a bulk Higgs. On the other hand, in the earlier case of a brane-localized Higgs, it had to be done ``by-hand'', i.e., via the considerations of going beyond 5D EFT. Thus, this cutoff effect is now much smaller than the NDA estimate (again, for $\beta 
\ll
\Lambda / k$, even if $\beta 
\gg 1$).

In contrast, the KK Higgs effect that we calculated in the main text in this paper is (roughly speaking) independent of the Higgs profile width, and so clearly is different from the effect discussed above (even though in the brane-localized limit, they look similar).

Finally, we note that the same 5D covariance principle (i.e., cutting off of the 4D loop momentum and KK sum hand-in-hand) applies to previous sections' bulk Higgs calculations, i.e., the wrong chirality SM Higgs or the correct/wrong chirality effect for the KK Higgs. 
There we took the 4D loop momentum cutoff $\rightarrow \infty$ to start with,
even so finding unsuppressed contributions (cf. the SM Higgs, correct chirality contribution above).
However, \`a la 2010 references~\cite{Csaki:2010aj}, strictly speaking
we should
have allowed the 4D loop momentum and the KK sum to go only till
$\Lambda$.\footnote{Again, we need to take this into account whether we take the brane-localized limit or not.}
Fortunately, as we already saw, the Higgs profile width provides an effective cutoff on the KK sum in those calculations also, and thus these suppressed cutoff contributions do not effectively affect our main results. Furthermore, note that this implies that it is the 4D loop momenta up to inverse of Higgs width (i.e., $\lesssim\beta k$) which gave the dominant effect.\footnote{There is in a sense an intrinsic
``coordination" of  these two ``cutoffs" within 5D EFT here, so that no extra 

carefulness is needed here (cf. 
the case of brane-localized Higgs as above)}, while
the contribution from the 4D loop momenta above $\sim \Lambda$ (which were unnecessarily included earlier) is 
suppressed by powers of $\sim \beta k / \Lambda$.
In other words, the correction to those results from the actual 5D cutoff effects of the above type 
is (very) small provided $\beta \ll \Lambda / k$, like in 
Eq.~\ref{bulkCornell}:
it was indeed justified then 
-- at least {\em a posteriori} -- to take the 4D loop momentum cutoff
to infinity first.
These estimates are 
consistent with the NDA expectation of UV-insensitivity for the case with a bulk Higgs.



\end{document}